# Data-Driven Review and Machine Learning Prediction of Diamond Vacancy Center Synthesis

Zhi Jiang,[1,2] Marco Peres,[3,4,5] Carlo Bradac,[6] and Gil Gonçalves[1,2*]

1. Centre for Mechanical Technology and Automation (TEMA), Mechanical Engineering Department, University of Aveiro, 3810-193 Aveiro, Portugal
2. Intelligent Systems Associate Laboratory (LASI), 4800-058 Guimarães, Portugal
3. Instituto de Engenharia de Sistemas e Computadores - Microssistemas e Nanotecnologia, Rua Alves Redol 9, 1000-029 Lisboa, Portugal
4. IPFN, Instituto Superior Técnico, University of Lisbon, Av. Rovisco Pais 1, 1049 001 Lisbon, Portugal
5. DECN, Instituto Superior Técnico, University of Lisbon, Estrada Nacional 10 (km 139.7), 2695 066 Bobadela, Portugal
6. Department of Physics and Astronomy, Trent University, 1600 West Bank Drive, Peterborough, ON, K9L 0G2, Canada

* Corresponding author: Gil Gonçalves, ggoncalves@ua.pt

**Keywords:** High Pressure High Temperature (HPHT), Microwave Plasma Chemical Vapor Deposition (MPCVD), Ion Implantation, Electron/Ion Irradiation, Diamond Color Centers, Machine Learning, Decision Tree Regression (DTR), Extreme Gradient Boosting (XGB), Shapley Values.

## Abstract

Diamond and diamond color centers have become prime hardware candidates for solid state-based technologies in quantum information and computing, optics, photonics and (bio)sensing. The synthesis of diamond materials with specific characteristics and the precise control of the hosted color centers is thus essential to meet the demands of advanced applications. Yet, challenges remain in improving the concentration, uniform distribution and quality of these centers. Here, we perform a review and meta-analysis of some of the main diamond synthesis methods and their parameters for the synthesis of N-, Si-, Ge- and Sn-vacancy color-centers. We extract quantitative data from over 60 experimental papers and organize it in a large database (170 data sets and 1692 entries). We then use the database to train two machine learning algorithms to make robust predictions about the fabrication of diamond materials with specific properties from careful combinations of synthesis parameters. We use traditional statistical indicators to benchmark the performance of the algorithms and show that they are powerful and resource-efficient tools for researchers and material scientists working with diamond color centers and their applications.

**Table of contents**

## 1. Introduction

Diamond synthesis can be carried out through several fabrication processes each involving multiple variables such as pressure, temperature, chemical precursors, substrates, run time, etc. that can display complex nonlinear relations.[1] In addition, diamond can be synthesized to host extrinsic defects of foreign atoms that display unique spin-optical properties and have been successfully employed in a multitude of both fundamental and practical applications including in quantum communication[2–5] and computation,[6–8] quantum simulation,[9,10] and quantum sensing.[11–13] However, traditional orthogonal synthesis methods often struggle to effectively capture and single out the optimal experimental parameters that lead to the synthesis of diamond materials and their dopant concentrations with desired characteristics. This limitation opens the opportunity for alternative approaches such as machine learning (ML), which shows incredible promise in contexts concerning the effective extraction of patterns from complex data and the capture of difficult-to-model physical mechanisms akin to the ones involved in material synthesis.[14,15] Models based on ML algorithms can, for instance, quickly identify the key parameters of fabrication processes[16–18], and can be used to tailor and design specific material properties,[19] heterostructures,[20,21] and devices.[22–25]

Currently, in material science the focus is on the synthesis of diamond hosting extrinsic dopants, specifically on controlling the inclusion of diamond vacancy color centers (DVCCs). These are atom(s)-vacancy(ies) complexes replacing carbon atoms in the diamond crystalline matrix: the most prominent examples are the nitrogen-vacancy (NV) center and the group IV diamond color centers (SiV, GeV, SnV and PbV).[26,27] Notably, achieving precise control of these vacancy color centers, whilst fabricating high-quality diamond films and nanomaterials, remains a significant challenge under current technological capabilities.[28–32]

In this study, we propose to use ML to address this challenge. Our approach is based on two ML methods: Decision Tree Regression (DTR) and Extreme Gradient Boosting (XGB).[33–35] Briefly, we perform an extensive meta-analysis of the existing literature (60 studies) and create a large database (170 data sets and 1692 entries) containing the synthesis parameters and color centers yield for each of the identified diamond fabrication techniques. We then use this database to train-and-test the DTR and XGB algorithms and perform Shapley value analyses to identify the optimal combinations of synthesis parameters that produce diamond materials with desired characteristics and high-quality color centers. Specifically, we train and optimize the DTR and XGB algorithms with the constructed database to predict the *target value* of certain variables of interest from the values of all the other *input features*. We demonstrate the effectiveness of this approach for each synthesis technique by performing two *Tasks*. *Task I* (§ 3.1) is specific to each synthesis method: the target variable and input features are chosen amongst the relevant synthesis parameters of the method. For example, we use the DTR/XGB algorithms to predict the size (target variable) of diamond materials synthesized by high pressure high temperature (HPHT) from the value of the chamber temperature, chamber pressure and run time (input features) of the process. *Task II* (§ 3.2) is more general and comparative: for all techniques, we select the full width at half maximum (FWHM) of the DVCCs' zero-phonon line (ZPL) as the target variable, while the input features are chosen amongst the relevant synthesis parameters of each method. This allows us to directly compare all synthesis techniques. For *Task II*, we use the FWHM as a proxy for the quality of the DVCCs. This is motivated by a series of observations: *i)* the need to base our analysis on a *measurable* physical observable that is *ii) common* to all synthesis techniques, and *iii)* that is *physically meaningful* (many of the technological quantum realization involving DVCCs heavily rely on their optical and spectral properties, for which the FWHM can be a useful indicator).

The objective of this study is to develop a data-driven and practical framework for diamond synthesis and design, aimed at maximizing the yield of color centers with optimized optical properties and performance. This combined focus on material synthesis and color center incorporation complements ongoing machine learning efforts that independently target the optimization of material fabrication,[36,37] device engineering,[38] and measurement efficiency.[39,40]

## 2. Methods and theoretical models

Our analysis focuses on four different processes for synthesizing diamond materials and controlling the inclusion of extrinsic defects forming DVCCs. We concentrate on high pressure high temperature (HPHT) and chemical vapor deposition (CVD) for material synthesis, and on ion implantation and electron/ion irradiation for the control and inclusion of target dopants. These techniques are briefly reviewed in §§ 2.1.1–2.1.4 to establish context, help identify the critical parameters of each technique, and detail the implementation of *Tasks I* and *II* for each method. The construction of the database is discussed in § 2.2 (more details can be found in the Supplementary Information, SI, § S1). The methods and ML algorithms employed to extract the parameters for the optimal synthesis of diamond and color centers are discussed in § 2.3.

### 2.1 Diamond synthesis and dopants inclusion techniques

For our analysis, we start by selecting a set of techniques for diamond synthesis and inclusion of dopants in the fabricated materials. These methods are reviewed in §§ 2.1.1–2.1.4, where we highlight for each technique the main experimental variables that affect the yield of material and of DVCC inclusion. We use these variables in the database to train and test the DTR and XGB algorithms. To compile the database, we reviewed over 60 published papers, gathering a total of 170 datasets, which are summarized and organized in Tables ST2–ST5 (see SI, § S2). Figure 1 summarizes the workflow of our analysis—from data collection and organization, through training and validation of the ML models, to Shapley value-based feature interpretation.

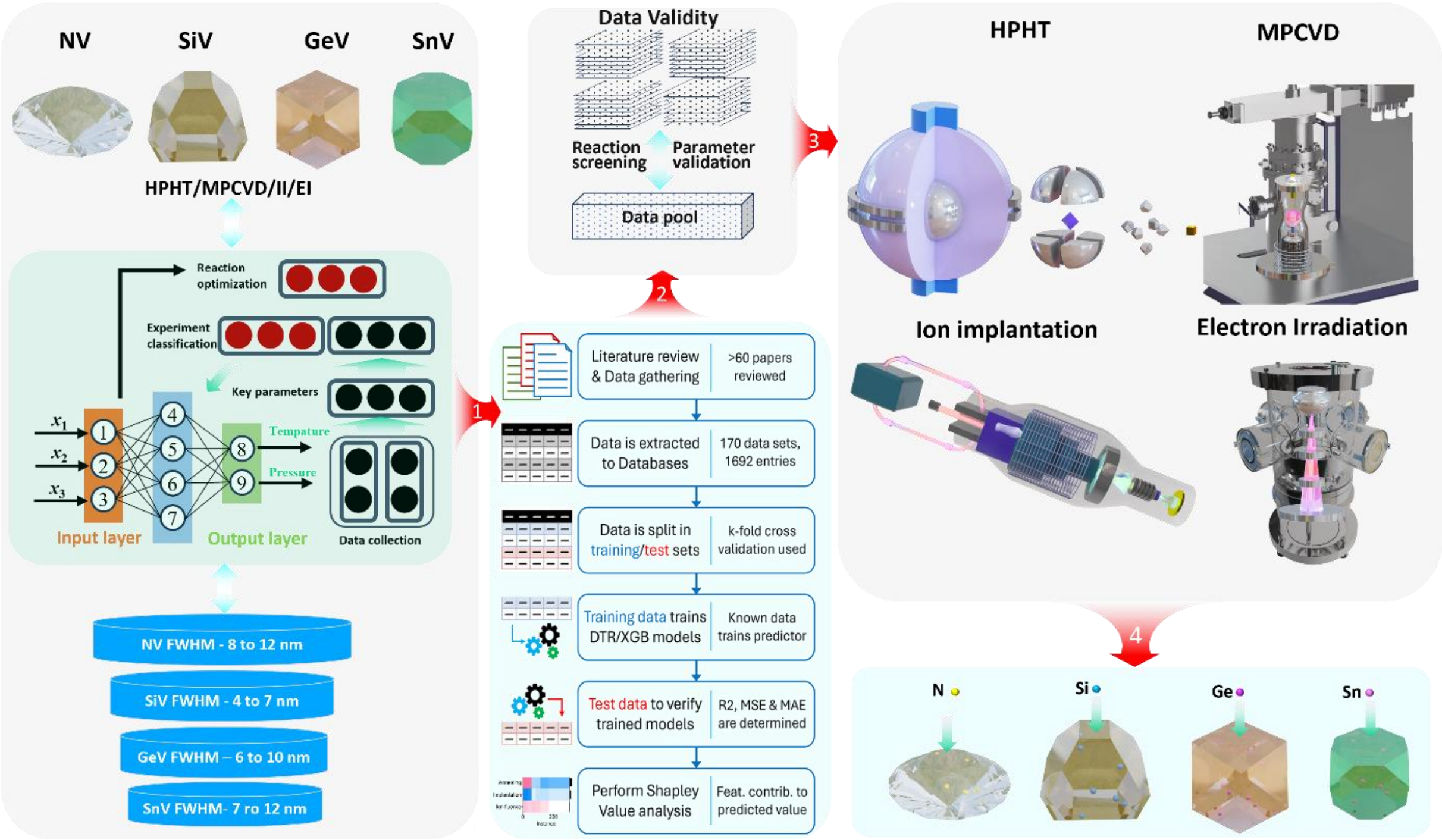

**Figure 1**. Schematic representation of the workflow used in this study: data extraction and preprocessing, training of the ML model and $k$-fold validation, Shapley value-based feature interpretation. For the analysis, we gathered datasets from the literature (~60 papers, 170 sets and 1700 entries) for HPHT, MPCVD, ion implantation, and electron/ion irradiation processes. The underlying aim is to identify how synthesis parameters influence diamond and color-center characteristics.

#### *2.1.1. High Pressure High Temperature (HPHT)*

The synthesis of artificial diamond by high pressure high temperature (HPHT) finds its origin in the mid-1950s.[41] The process mimics the geological formation of natural diamond: carbon-based precursor materials are subjected to high temperature (above 1500 °C) and high pressure (above 5 GPa) to induce the phase transition of graphite-like carbon to diamond. This technique usually involves high process and equipment costs, high energy consumption, and limitations in large-scale production,[42] but it reliably yields high-quality diamond material and large-grained, single-crystalline micro/nanoparticles (M/NPs) hosting color centers. The HPHT process is attractive for the inclusion of DVCCs directly during synthesis. For instance, the employment of inorganic or organic compounds high in nitrogen content, such as $NaN_3$, $C_3H_6N_6$, $Ba(N_3)_2$ and $Fe_3N$, can lead to the relatively high incorporation of nitrogen-vacancy (NV) centers directly during material fabrication.[43–49]

The characteristics of synthesized HPHT diamond depend on its fabrication parameters. For our analysis, we identify five key variables that prominently determine the yield and properties of HPHT diamond and its color centers: chamber pressure and temperature, run time, diameter of the synthesized diamonds and full width at half maximum (FWHM) of the hosted DVCCs' zero-phonon line (ZPL). With relevance to the *Tasks* introduced in § 1, we have the following for our analysis (see SI, § S1, Table ST1). For *Task I*, the target variable is the average diameter of the synthesized diamond NPs, and the input features are the chamber pressure and temperature, and the run time. For *Task II*, the target variable is the FWHM of the DVCCs' ZPL, and the input features are the chamber pressure, chamber temperature, run time, and average diameter of the synthesized diamond NPs.

### *2.1.2. Chemical vapor deposition (CVD)*

Chemical vapor deposition (CVD) and microwave plasma chemical vapor deposition (MPCVD) are diamond synthesis techniques that consist in the atom-by-atom growth of diamond on a heated substrate from a chemically reacting hydrocarbon gas mixture. Usually, the growth process is carried out inside a vacuum chamber where temperature, pressure and, in the case of MPCVD, microwave (MW) frequency and power are precisely controlled. Research on the growth of artificial diamonds from the gas phase started in the early 1950s,[50] with other major milestones in the development of the technique following in the late 1960s,[51] 1970s[52] and 1980s.[53–56] Diamond synthesis via CVD is an attractive method as it can yield materials with a certain flexibility in terms of size and properties, whilst also offering practical advantages. These include the lack of need for extremely high pressures (which can be limiting for HPHT synthesis) and, in the case of MPCVD, the fact that the material synthesis process takes place at relatively high growth rates, efficient carbon source utilization, lower deposition temperatures and low contamination.[57–59] Similar to the HPHT process, CVD growth also allows for the inclusion of color centers in the diamond material directly during synthesis by injection into the chamber of gas mixtures containing specific dopants. The incorporation of dopants during growth can however be inefficient,[60–63] and requires careful control over experimental variables such as chemical composition of the reaction atmosphere (e.g. via the use of inert gases such as $N_2$ and Ar, and the reduction of $H_2$), pressure, substrate temperature, and power of the MW radiation.[64]

The synthesis of diamond by CVD and MPCVD involves a complex and large set of inter-dependent physical mechanisms and chemical reactions. To build our database and train our algorithms to make predictions about quality, rate, morphology, doping, and DVCC inclusion of the synthesized diamond, we use the following parameters: chamber pressure, substrate temperature, MW plasma power, and FWHM of the hosted DVCCs' ZPL. With relevance to the *Tasks* introduced in § 1, we have the following for our analysis (see SI, § S1, Table ST1). For *Task I*, the target variable is the MW plasma power, and the input features are the chamber pressure and substrate temperature. For *Task II*, the target variable is the FWHM of the DVCCs' ZPL, and the input features are the chamber pressure, substrate temperature, and MW power.

*2.1.3. Ion implantation*

Ion implantation is a process for the inclusion of foreign atoms in the diamond crystalline matrix after synthesis.[65] In this process, high-energy ions of specific elements are accelerated towards the target diamond and are embedded into it.[66–68] Ion implantation is advantageous for DVCC inclusion because it has a relatively high spatial resolution, both transversal and longitudinal, and allows for the fabrication of isolated color centers. The deterministic implantation is desirable for quantum applications as these often rely on the ability to isolate individual single-photon sources.[69–72] Further control over the creation of DVCCs can be achieved through low-energy carbon ion implantation, which can induce the formation of atom-vacancy complexes in the shallow surface layer of diamond, without introducing additional impurities.[73–76]

Ion implantation is technically challenging. The implantation parameters for each of the different dopant species forming DVCCs (i.e. N, Si, Ge, Sn, Pb) must be controlled with high precision as they are specific to and different for each element, and also depend on the concentration of impurities created during synthesis.[43,77–82] Additionally, the stochastic nature of the implantation and the straggle of the ions require careful control over their implantation kinetic energies and fluence (i.e. the number of striking atoms per unit area) to minimize lateral displacement while maximizing DVCCs yield.[81,83] Ion implantation usually involves at least one annealing step (at temperatures ≥850 °C). The reason is twofold: to repair the damage to the diamond $sp^3$ bonds caused by the implanted ions,[71,76,79,80] and to activate the mobilization of vacancies and thus promote the formation of DVCCs in the diamond material.[84,85]

Ion implantation is markedly non-trivial, as the combinatorics of the values for the synthesis parameters is considerably large. To build our database and train our algorithms, we use the following parameters: implantation energy, fluence, annealing temperature, annealing time and FWHM of the hosted DVCCs' ZPL. With relevance to the *Tasks* introduced in § 1, we have the following for our analysis (see SI, § S1, Table ST1). For *Task I*, the target variable is the annealing time, and the input features are the implantation energy, fluence, and annealing temperature. For *Task II*, the target variable is the FWHM of the DVCCs' ZPL, and the input features are the implantation energy, fluence, annealing temperature, and annealing time.

*2.1.4. Electron/ion irradiation*

Electron/ion irradiation is a common approach to induce the formation of DVCCs in diamond. A focused beam of electrons (or ions) is used to irradiate the target diamond material and create vacancies in the crystalline structure. Subsequent annealing at temperatures ≥850 °C promotes the migration of the vacancies, which become trapped by the foreign atoms already present in the diamond lattice, forming stable DVCCs.[86–88] The annealing step also helps repair the damage caused by the irradiation process. This is particularly relevant for this study as the optical and spectroscopic properties of the DVCCs are affected by strain, which can alter the photon purity,[89,90] lifetime[91] and coherence[92] of the DVCCs. Ad-hoc annealing conditions and advanced annealing strategies (e.g. laser and pulsed annealing[93,94]) can strongly affect the number and quality of the DVCCs and are taken into consideration in our models.

To build our database and train our algorithms, we use the following parameters: electron/ion irradiation energy, fluence, annealing temperature, annealing time, and FWHM of the hosted DVCCs' ZPL. With relevance to the *Tasks* introduced in § 1, we have the following for our analysis (see SI, § S1, Table ST1). For *Task I*, the target variable is the annealing time, and the input features are the electron/ion irradiation energy, fluence, and annealing temperature. For *Task II*, the target variable is the FWHM of the DVCCs' ZPL, and the input features are the irradiation energy, fluence, annealing temperature, and annealing time.

## 2.2 Database

The next step in our analysis is the construction of the database. We use the database to train and test the DTR and XGB algorithms and, for each synthesis technique, make predictions about the selected target variables from the input features. The database is compiled from 60 experimental studies and has a total of 1692 entries, organized into 170 datasets and four sub-databases covering each one of the four synthesis methods; this information is displayed and organized in Tables ST2–ST5 (see SI, § S2). Each table is arranged by color center: NV, SiV, GeV, and SnV, and shows the values and specifications for the critical synthesis parameters discussed in secs. 2.1.1–2.1.4, with regard to *Tasks I* and *II*.

## 2.3 Machine learning models and algorithms

In this work, we use two machine learning algorithms to make predictions about target values from input features: Decision Tree Regressor (DTR) and Extreme Gradient Boosting (XGB). We also implement Shapley value analysis to quantify the contributions of the various input features to the target variables. These methods are briefly reviewed below from a theoretical standpoint; their application to the collected data is discussed in §§ 3.1–3.3.

### *2.3.1. Decision Tree Regression (DTR)*

Decision Tree Regression (DTR) is a supervised learning algorithm that models the relation between a set of input features and a continuous target variable.[95] The model constructs a decision tree by recursively partitioning the data into subsets based on the values assumed by the input features, aiming to minimize the prediction error within each subset. In our case the data is organized as follows. Each one of the four sub-databases has a total number of observations, $n$. These are the rows in the sub-database (SI, § S2, Tables ST2–ST5) constructed from the meta-analysis of the literature. Each diamond synthesis method has a number of input features, $d$ (summarized for each technique and for each *Task* in Table ST1), and a target variable, $y$. The target variable is the observable whose value we are interested in predicting and is specified for each technique and for each *Task* in the SI, § S1, Table ST1.

The data has form $(\boldsymbol{X}, \boldsymbol{y})$ where $\boldsymbol{X} = [\boldsymbol{x}_1, \boldsymbol{x}_2, \ldots, \boldsymbol{x}_n]^{\mathrm{T}} \in \mathbb{R}^{n \times d}$ is the feature matrix and $\boldsymbol{y} = [y_1, y_2, \ldots, y_n]^{\mathrm{T}} \in \mathbb{R}^n$ is the target vector, such that each $i$-th observation $\boldsymbol{x}_i = [x_{i1}, x_{i2}, \ldots, x_{id}] \in \mathbb{R}^d$ is a row of input features with corresponding output $y_i \in \mathbb{R}$. For each synthesis method, the algorithm starts from the entire dataset considering all the features and proceeds in steps such that, at each steps, it finds the best feature $j$ and corresponding threshold $s$ with respect to which the data should be split in left, L: $\{\boldsymbol{x}_i \mid x_{ij} \leq s\}$, and right, R: $\{\boldsymbol{x}_i \mid x_{ij} > s\}$. The best split is defined with respect to the criterion of minimizing the sum of squared errors (SSE) in each L and R partition:

$$\mathrm{SSE}_{\mathrm{split}} = \sum_{i \in \mathrm{L}} (y_i - \bar{y}_{\mathrm{L}})^2 + \sum_{i \in \mathrm{R}} (y_i - \bar{y}_{\mathrm{R}})^2 \quad (1)$$

Where $\bar{y}_{\mathrm{L}}$ and $\bar{y}_{\mathrm{R}}$ are the mean target values of the L and R partitions, respectively. The algorithm proceeds by splitting the data in each subset recursively until a stopping criterion is met (e.g., minimum number of samples, or no gain in error reduction). In the final tree, each leaf node predicts a constant value, which is the mean of all $y_i$ values in that region.

Once the regressor tree is constructed, if an unknown input $\boldsymbol{x}_u = [x_{u1}, x_{u2}, \ldots, x_{ud}] \in \mathbb{R}^d$ is passed to it, the tree routes it through a set of decisions until it reaches a leaf node of output:

$$\hat{y}_u = \frac{1}{|S|} \sum_{i \in S} y_i \quad (2)$$

Where $S$ is the set of training samples in that leaf region.

### *2.3.2. Extreme Gradient Boosting (XGB)*

Extreme Gradient Boosting (XGB) is a supervised learning algorithm that combines the predictions of many decision trees to produce a more accurate model than a single tree (see DTR above).[96] It uses an ensemble technique called gradient boosting, which builds trees sequentially, with each tree learning to correct the residual errors of the previous ones. As for the case of DTR, the input features and a target variable are identified for each set of data of interest (in our case, for each synthesis technique and each *Task*).

The data is formatted in the same way as per the DTR case (see § 2.3.1), in the form $(\boldsymbol{X}, \boldsymbol{y})$, where $\boldsymbol{X} = [\boldsymbol{x}_1, \boldsymbol{x}_2, \ldots, \boldsymbol{x}_n]^{\mathrm{T}} \in \mathbb{R}^{n \times d}$ is the feature matrix and $\boldsymbol{y} = [y_1, y_2, \ldots, y_n]^{\mathrm{T}} \in \mathbb{R}^n$ is the target vector, such that each $i$-th observation $\boldsymbol{x}_i = [x_{i1}, x_{i2}, \ldots, x_{id}] \in \mathbb{R}^d$ is a row of input features with corresponding output $y_i \in \mathbb{R}$. As per the DTR case, $n$ is the number of observations, $d$ is the number of features and $y$ is the target variable. The XGB algorithm builds the model as a sum of many regression trees:

$$\hat{y}_i = \sum_{k=1}^{K} f_k(\boldsymbol{x}_i), \quad f_k \in \mathcal{F} \tag{3}$$

Where $\hat{y}_i$ is the prediction for the $i$-th sample, $f_k$ is the $k$-th tree function (decision tree), $\mathcal{F}$ is the space of regression trees and $K$ is the number of boosting rounds (trees). Each new tree is added to minimize a loss function, which measures the error between predicted and actual values. Specifically, the algorithm aims at minimizing the object:

$$\mathcal{L}^{(t)} = \sum_{i=1}^{n} \ell\left(y_i, \hat{y}_i^{(t-1)} + f_i(\boldsymbol{x}_i)\right) + \Omega(f_t) \tag{4}$$

Where $\ell$ is the loss function (e.g., the squared error $\ell(y, \hat{y}) = (y - \hat{y})^2$), the term $\hat{y}_i^{(t-1)}$ is the current prediction, $f_t$ is the new tree and $\Omega(f_t)$ is the regularization term, i.e. a function that represents a penalty for the complexity of a tree and discourages overly complex trees (briefly, trees that have more leaves and are deeper can model data very closely, but run into the risk of overfitting the data, which this regularization term helps avoid). To fit the new tree, XGB uses Taylor expansion of the loss function (gradient boosting). Unlike DTR, which builds a single deep tree, XGB builds shallower trees and adds them up. Each tree is trained to focus on the residuals (errors) from the previous model. After all $K$ trees are trained, any unknown input $\boldsymbol{x}_u = [x_{u1}, x_{u2}, \dots, x_{ud}] \in \mathbb{R}^d$ passed to the algorithm produces the target variable prediction:

$$\hat{y}_u = \sum_{k=1}^{K} f_k(\boldsymbol{x}_u) \tag{5}$$

Which is usually a more robust, less overfitted model than using one tree (DTR).

To train and test our DTR and XGB algorithms we implemented $k$-fold cross validation. That is, the database entries are split into $k$ equally sized subsets, or *folds*. The model is trained on $k-1$ of these folds and tested on the remaining one, and this process is repeated $k$ times so that each fold serves as the test set once. The final performance is then computed as the average across all $k$ runs. The errors are calculated on the test data and are given as the differences between the true experimental values in the database and the values produced by the DTR/XGB predictors (see § 3 for details). A schematic implementation of the XGB algorithm for the synthesis of diamond material via HPHT, used as an example, is shown in the SI, § S1, Figure SF1. The list of hyperparameters used in our models and the range of their values can be found in the SI, § S3, Table ST6.

We note that in our analysis, we excluded certain *categorical* parameters for the optimization of the models, such as the carbon source for HPHT and CVD/MPCVD, the dopant source and substrate for CVD/MPCVD, the ion source for Ion implantation and Electron/ion irradiation. These parameters are known to affect color-center formation kinetics and stability. However, we exclude them from our analysis for practical, defensible reasons. These variables are difficult to include in our models as they require qualitative, categorical encoding, i.e., converting non-numeric, qualitative data into numeric form. In many cases, even when quantitative values are applicable, the studies do not report exact numbers (e.g. for the ratios of the precursors in CVD/MPCVD). Finally—and more importantly—a post-facto analysis of our results shows that already with the input features listed in Table ST1 (SI, § S1), the optimized DTR/XGB models have good predictive power (high coefficient of determination $R^2$, and relatively small mean squared error (MSE) and mean absolute error (MAE), see §§ 3.1, 3.2), without the need to include these categorical features. (Additional details about the handling of categorical variables in our models can be found in the SI, § S4, Figs. SF2-5).

*2.3.3. Shapley value analysis*

In addition to using DTR and XGB to make predictions, we use Shapley value analysis to establish the relevance of each parameter in determining the characteristics of the synthesized diamond materials and DVCCs.[97] Shapley value analysis is a method developed from cooperative game theory that helps explain the contribution of each feature to a target value prediction made by a (machine learning) model, through fair, additive, and consistent feature attributions.

As per our discussion in §§ 2.3.1 and 2.3.2, the data is in the form $(\boldsymbol{X}, \boldsymbol{y})$, where $\boldsymbol{X} = [\boldsymbol{x}_1, \boldsymbol{x}_2, \dots, \boldsymbol{x}_n]^{\mathrm{T}} \in \mathbb{R}^{n \times d}$ is the feature matrix and $\boldsymbol{y} = [y_1, y_2, \dots, y_n]^{\mathrm{T}} \in \mathbb{R}^n$ is the target vector, such that each $i$-th observation $\boldsymbol{x}_i = [x_{i1}, x_{i2}, \dots, x_{id}] \in \mathbb{R}^d$ is a row of input features with corresponding output $y_i \in \mathbb{R}$. Again, $n$ is the number of observations, $d$ is the number of features and $y$ is the target variable. For each individual observation, $\boldsymbol{x}_i = [x_{i1}, x_{i2}, \dots, x_{id}]$, a trained model $f$ gives a prediction:

$$\hat{y}_i = f(\boldsymbol{x}_i) \tag{6}$$

Where in this equation, $f$ is either the DTR or XGB regressor we built from the available datasets and whose explicit expressions are given in equations (2) and (5), respectively. Shapley value analysis decomposes this prediction into a sum of feature contributions:

$$f(\boldsymbol{x}_i) = \phi_0 + \sum_{j=1}^{d} \phi_j^{(i)} \tag{7}$$

Where $\phi_0$ is the baseline prediction (often the mean prediction over the training set), and $\phi_j^{(i)}$ is the $j$-th Shapley value, i.e., the contribution of the $j$-th feature to the prediction for the $i$-th *instance*. The Shapley value calculates the average marginal contribution of a feature across all possible subsets of other features for each $i$-th instance. Effectively, it is the average change in the model output when feature $j$ is added to all possible subsets of other features. Mathematically, for feature $j$, the Shapley value for $\boldsymbol{x}_i$ is:

$$\phi_j^{(i)} = \sum_{S \subseteq \mathcal{F} \setminus \{j\}} \frac{|S|!(d-|S|-1)!}{d!} \left[ f_{S \cup \{j\}}(\boldsymbol{x}_i) - f_S(\boldsymbol{x}_i) \right] \tag{8}$$

Where $\mathcal{F} = \{1,2, \ldots, d\}$ is the full set of features, $S$ is a subset of features excluding the $j$-th, and $f_S(\boldsymbol{x}_i)$ is the model's prediction when only features in $S$ are known; the weighting factor in front ensures fairness over all permutations.

The Shapley value analysis thus allows us, for DTR and XGB, to add numerical, per-feature attribution to predictions, effectively helping us quantify how much each feature contributed for each individual sample. This can be seen explicitly in equation (7) where the prediction $f(\boldsymbol{x}_i)$ is decomposed in a sum of terms, $\phi_0$ and $\phi_j^{(i)}$, where each term $\phi_j^{(i)}$ has a numerical value (calculated using eq. (8)) that can be compared to that of any other in the sum. The quantitative Shapley value analysis applied in this study is discussed in detail in § 3.3.

## 3. Results and discussion

In this section, we present and discuss the effectiveness of our DTR and XGB algorithms in predicting target variables from input features. The analysis is organized with respect to the *Tasks I* and *II* we identified for each technique in §§ 2.1.1–2.1.4 and Table ST1 (SI, § S1).

### 3.1. Task I: material synthesis

To measure the effectiveness of the DTR and XGB predictors quantitatively, we use traditional statistical estimators. Training and testing of the models are performed using $k$-fold cross validation (see § 2.3.2). Accuracy and precision of the predictors are determined from the errors, which are calculated based on the differences between the true and predicted values. The true values of the target variables are those reported in the experimental data of the database. The predicted values are the values of the target variables estimated by the DTR and XGB predictors. We quantify the performance of the predictors by measuring their coefficient of determination ($R^2$), mean squared error (MSE) and mean absolute error (MAE).

Figures 2–5 show the main results for *Task I* of each technique organized by diamond color center (columns) and by predictor (rows). The $R^2$, MSE and MAE values for each technique are displayed directly in the figure subpanels both for the test (circles) and training sets (triangles); the values of the latter are merely used as control. The values of $R^2$, MSE and MAE for the DTR/XGB predictors for each synthesis method are also listed in the SI, § S5, Table ST7. In Figures 2–5, the $x$-axis is the true value of the target variable and the $y$-axis the predicted one. The diagonal is thus the *line of perfect accuracy*: a point on this line indicates that the DTR/XGB predictors estimated a target value from the input features coinciding exactly with the true experimental value. Points above and below the line correspond to the model over- and under-predicting the true value, respectively. The distance of the points from this line is thus a measure of the predictors' accuracy (for clarity, in Figs. 2–5, we also use color-coding: the higher the accuracy, the darker the color), while the MSE/MSA values are indicative of their precision.

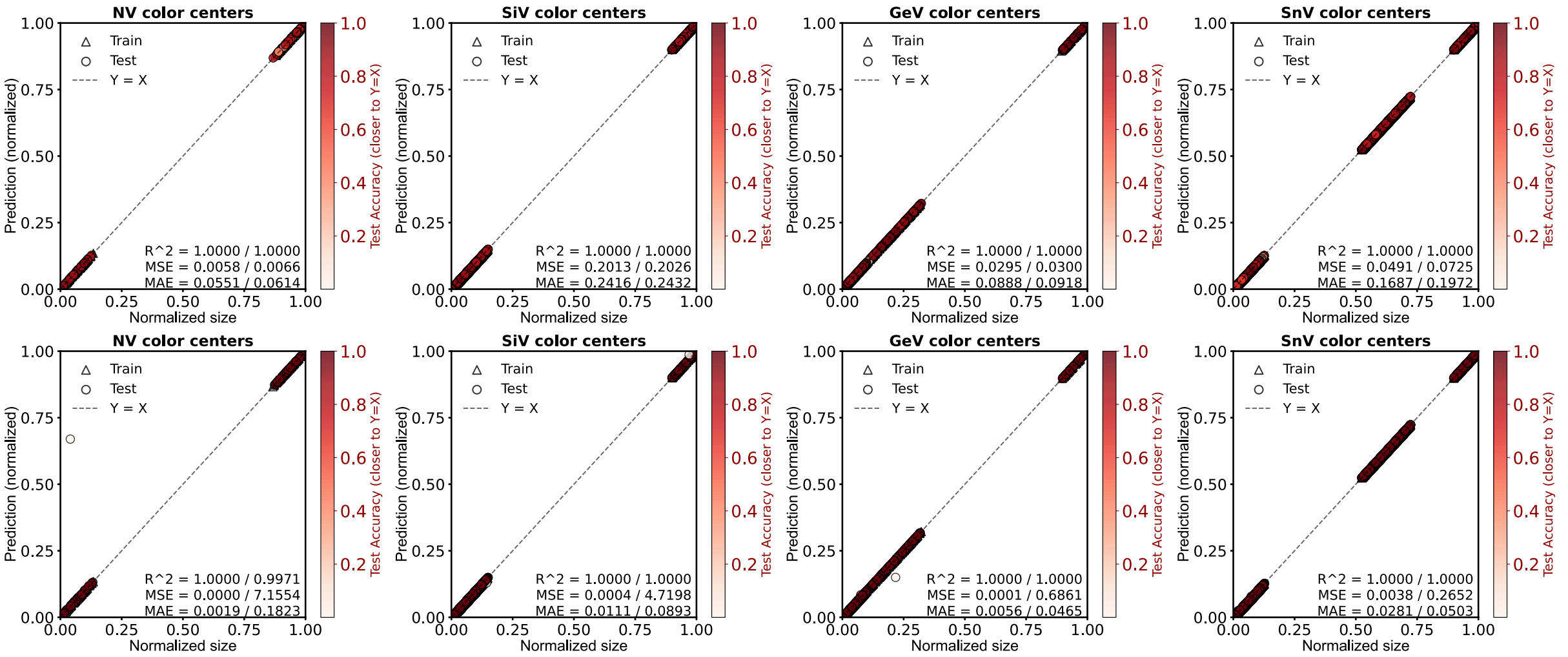

**Figure 2**. HPHT, *Task I* summary. DTR (top row) and XGB (bottom row) predictions of the average HPHT diamond diameter (target variable) from given values for chamber pressure, chamber temperature, and run time (input features), organized by color centers (columns). The values of $R^2$, MSE, and MAE for the training (triangles) and test (circles) sets are listed in the bottom-right corner of each subpanel.

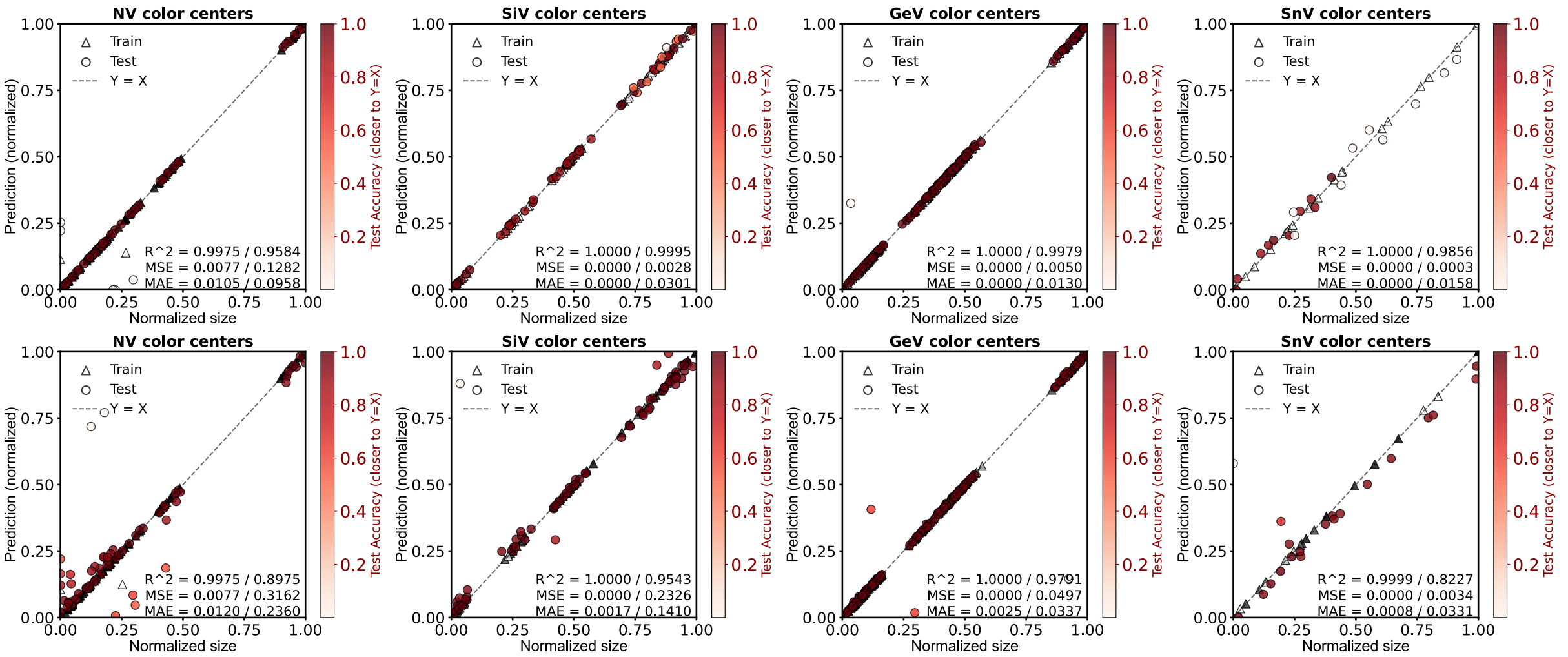

**Figure 3**. CVD/MPCVD, *Task I* summary. DTR (top row) and XGB (bottom row) predictions of the MW plasma power (target variable) from given values for chamber pressure and substrate temperature (input features), organized by color centers (columns). The values of $R^2$, MSE, and MAE for the training (triangles) and test (circles) sets are listed in the bottom-right corner of each subpanel.

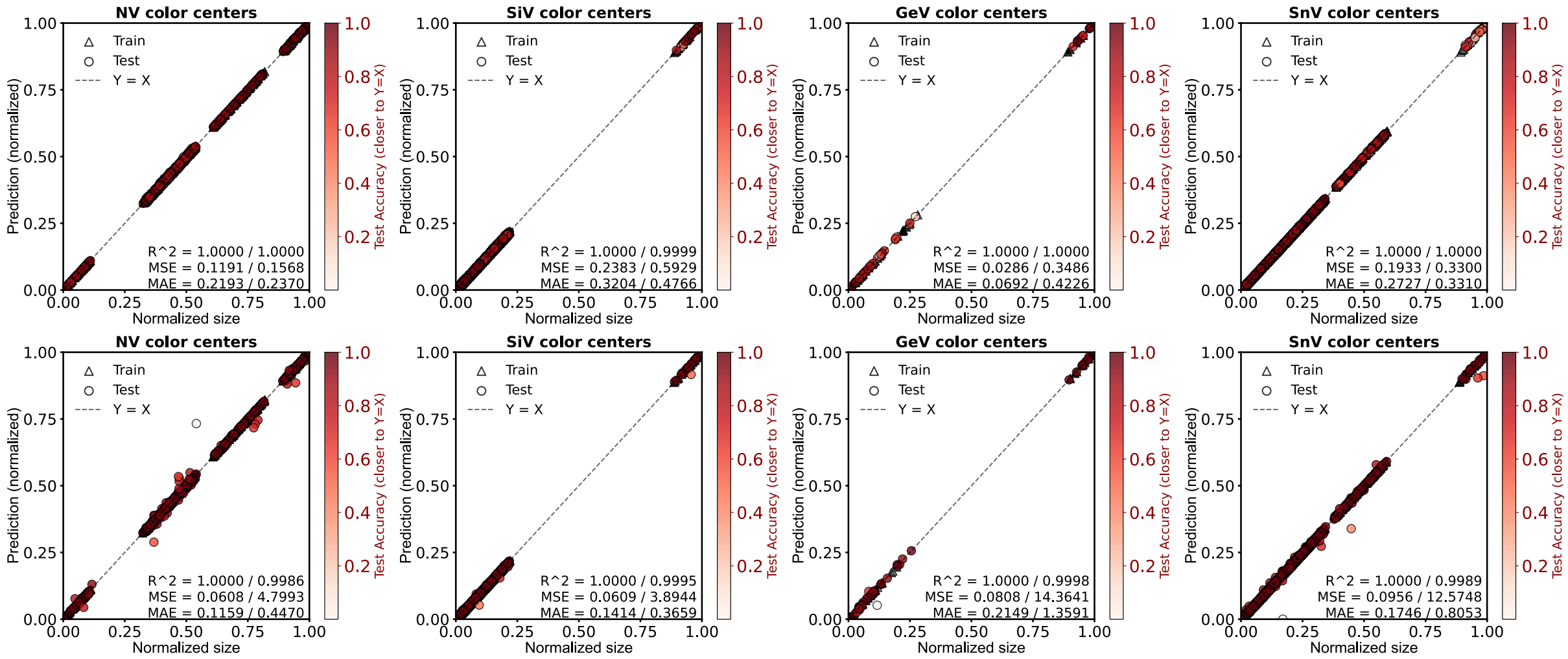

**Figure 4**. Ion implantation, *Task I* summary. DTR (top row) and XGB (bottom row) predictions of the annealing time (target variable) from given values for implantation energy, fluence and annealing temperature (input features), organized by color centers (columns). The values of $R^2$, MSE, and MAE for the training (triangles) and test (circles) sets are listed in the bottom-right corner of each subpanel.

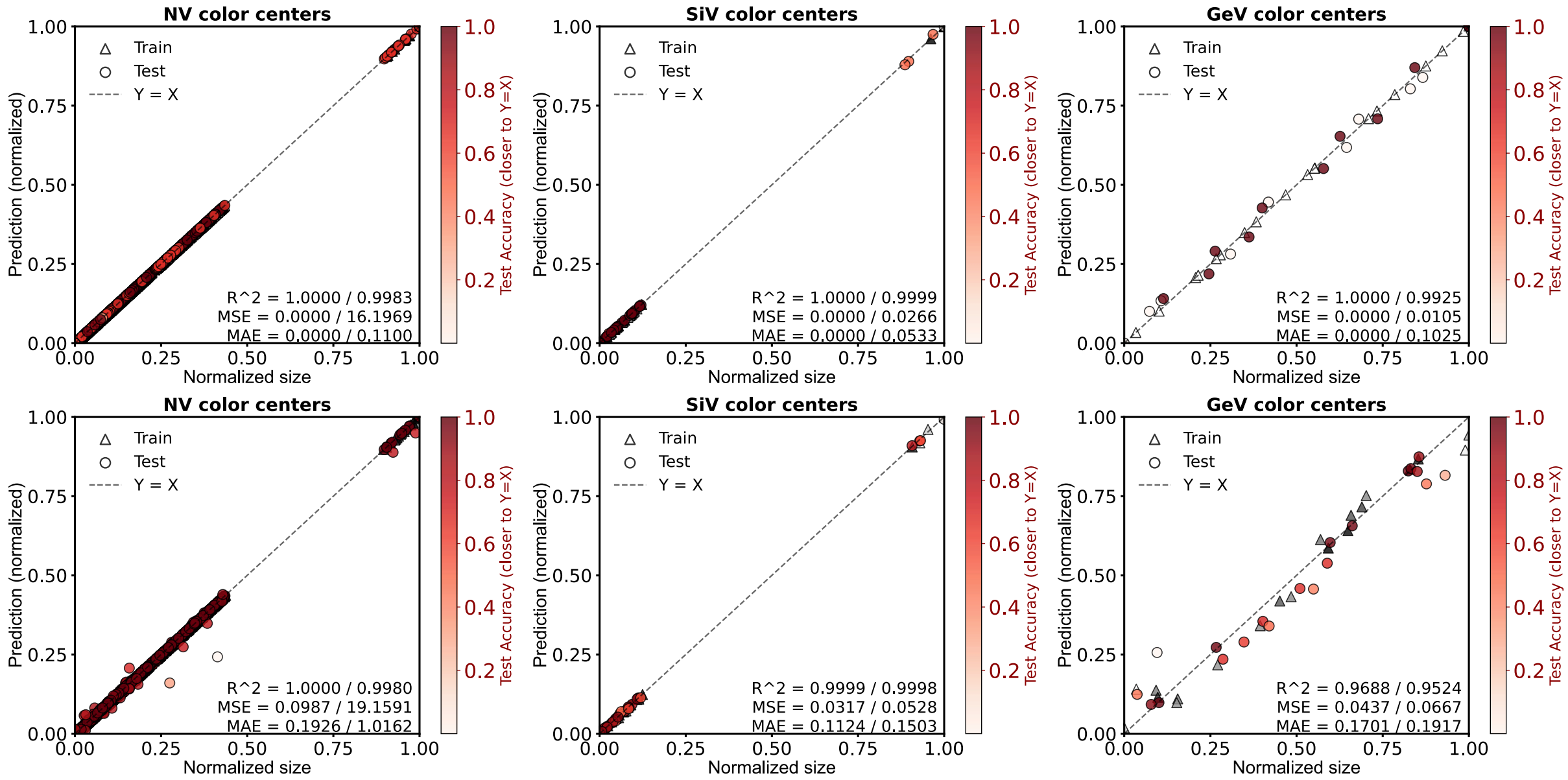

**Figure 5**. Electron/ion irradiation, *Task I* summary. DTR (top row) and XGB (bottom row) predictions of the annealing time (target value) from given values for irradiation energy, fluence and annealing temperature (input features), organized by color centers (columns). The values of $R^2$, MSE, and MAE for the training (triangles) and test (circles) sets are listed in the bottom-right corner of each subpanel.

The numerical results and the plots of our analysis for the DTR and XGB predictors warrant a series of observations.

Overall, the values of the statistical estimators, specifically $R^2$ >0.9, and the relatively small values for MSE and MAE (both for the training and test sets) indicate that the DTR and XGB models are good generalizations. Interestingly, when comparing directly the performance of the DTR and XGB algorithms, we found that DTR is a better predictor than XGB as, in most cases, it has values for $R^2$ closer to 1, and smaller values (roughly by a factor ×5–10) for MSE and MAE. This is likely due to several factors.[98,99] In terms of the number of input features, our data is considered of low dimensionality (only 3–4 features) compared to scenarios of medium ($10^1$–$10^2$ features) or high ($10^2$–$10^3$ features) dimensionality. In low-dimensionality problems, DTR models can learn patterns quickly and the ensemble averaging effects characteristic of XGB can become negligible. These effects are further emphasized by the fact that in our analysis, the features/data-size ratio is relatively small (for us, this ratio is in the low range of 0.004–0.03, as we have 3–4 input features and 128–680 rows of data, depending on the database used; see SI, § S2, Tables ST2–ST5). Finally, a foreshadowing glance at the Shapley value analysis (see §§ 3.3 and Figs. 11–18 for details) indicates that in all four cases (HPHT, CVD/MPCVD, Ion implantation and Electron/ion irradiation), one feature tends to outweigh the others in explaining the variance of the target variable. When this occurs, DTR can efficiently exploit this dominance by making the first split of the tree on that feature and by focusing most of the optimization on it, while using the other features for fine tuning. Conversely, XGB builds one tree at a time and such that each subsequent tree learns to fix the residuals of the previous ones. When one feature is dominant, XGB may dilute its effect by spreading influence across more, smaller steps (especially if the learning rate is low or the depth of the tree is small). This is usually desirable for generalization, but it may slow down convergence or—relevantly here—underuse dominant feature(s) signals.

Analysis of the plots in Figures 2–5, also reveals that for each technique there are certain areas of the parameter space that are unexplored or, in other words, there are no data points corresponding to certain combinations of parameters. This is due to a series of factors including the specific goals of the authors in the original studies, technical limitations, or the fact that, while relatively large (~60 papers), the number of studies we reviewed is just a subset of the existing literature. This is important because, in general, the reliability of trained models should not be trusted beyond the ranges spanned by the training datasets. However, our plots show reliable results, consistently, for different non-contiguous ranges of parameters. This is favorable, for it indicates that the interaction between the various input features and target variables is predictable and mostly devoid of interactions that our models are unable to capture. We thus expect our models to be robust predictors even with parameters in the empty areas of the parameter space—at least within the lowest and highest extremes of the explored ranges.

Finally, we note that for the cases of Ion implantation and Electron/ion irradiation (Figs. 4 and 5), we opted to preprocess the data by grouping the entries in the database by type of ion source as we observed that neglecting to do so led to relatively larger test prediction errors.

### 3.2. Task II: Full width at half maximum of the zero-phonon line

We now want to generalize our approach and apply the DTR and XGB algorithms simultaneously to all synthesis methods, with the intent of drawing direct comparisons. The goal of *Task II* of our analysis is to train the DTR/XGB models to predict the value of a target variable common to all methods from the input features of each technique. The target variable of choice is the full width at half maximum (FWHM) of the zero-phonon line (ZPL) of the various DVCCs, which we chose based on the arguments of § 1. In diamond color centers, the FWHM of the zero-phonon line quantifies the spectral linewidth of the purely electronic transition between the ground and excited states. It serves as an indicator of the local electronic environment experienced by the defect and thus as a proxy for diamond quality with respect to strain, disorder, and decoherence pathways. The ZPL linewidth is predominantly influenced by lattice strain, electric- and magnetic-field fluctuations from nearby defects or impurities, and phonon-mediated dephasing, all of which broaden or destabilize the electronic transition energy.

Ultimately, our intent is to combine the results from *Tasks I* and *II* for the DTR/XGB predictors and identify the synthesis techniques and their parameters that yield diamond with desired material properties and DVCCs quality. A narrow FWHM is usually desirable for optical and quantum applications based on diamond color centers as it improves photon indistinguishability, photon extraction, and the coupling efficiency to optical cavities and photonics structures.[27,100,101]

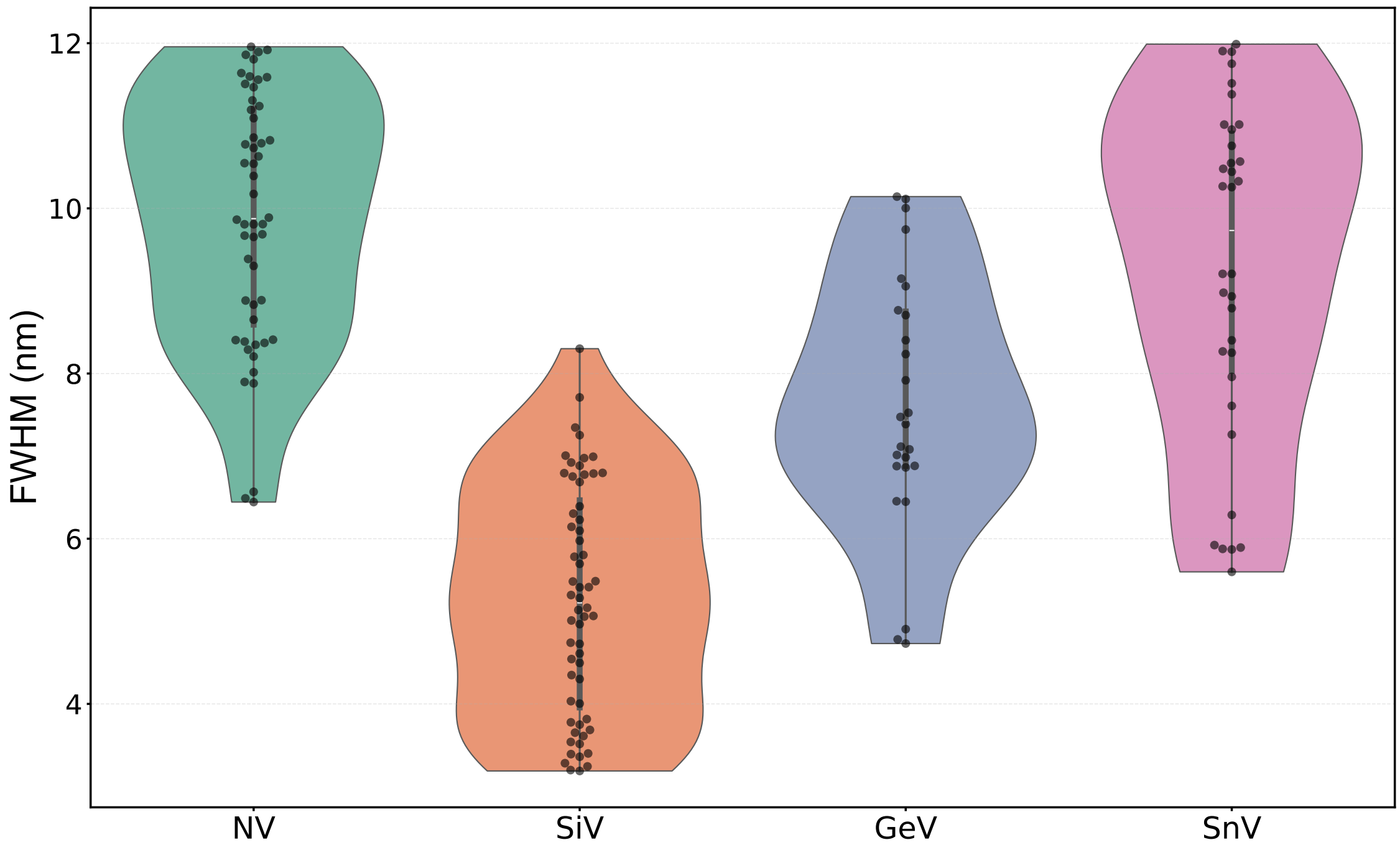

**Figure 6.** FWHM of NV, SiV, GeV and SnV. Each data point (black dot) represents the measured FWHM value. The vertical spread indicates the distribution of FWHM values reported in the literature, and the horizontal width of each plot indicates how many times a specific FWHM value is reported across studies.

To train and test the performance of our DTR and XGB algorithms we use the same approach carried out for *Task I*. We organized the information from the literature in four databases (SI, § S2, Table ST2–ST5) and use $k$-fold cross validation for the analysis. For all the synthesis techniques, the selected target value is the ZPL FWHM, whereas the input features differ for each method (SI, § S1, Table ST1). The performance of the models is based on the difference between the true experimental values and those produced by the DTR/XGB predictors (see § 3.1) and it is assessed by estimating the quantities $R^2$, MSE and MAE. In our analysis, we use the reported values for the FWHM when they are available and otherwise extract them directly from the published spectra. Figure 6 shows a *violin* chart summarizing the experimental values of the ZPL FWHM from the literature, organized by color center. The vertical and horizontal spreads of each plot are indicative of the dispersion of the measured FWHM values and their corresponding observation frequencies, respectively.

The results of the training and testing of the DTR and XGB algorithms for *Task II* are displayed in Figures 7–10 for the HPHT, CVD/MPCVD, Ion implantation and Electron/ion irradiation techniques, respectively. The estimated values for $R^2$, MSE and MAE are displayed in the bottom-right corner of each subpanel both for the test and training data (the latter are merely used as control); they are also summarized in the SI, § S5, Table ST8.

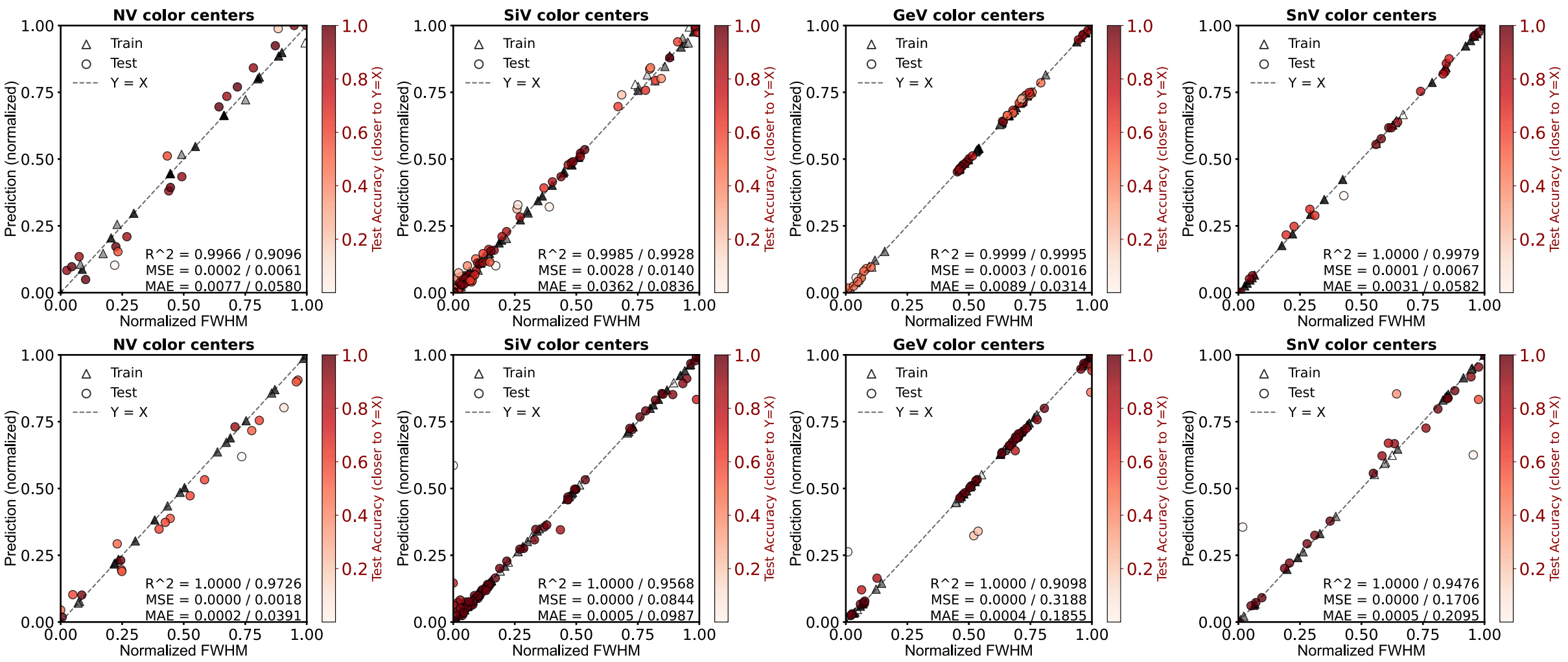

**Figure 7**. HPHT, *Task II* summary. DTR (top row) and XGB (bottom row) predictions of the ZPL FWHM (target variable) from given values for pressure, temperature, run time and average size of the diamond particles (input features), organized by color centers (columns). The values for $R^2$, MSE, and MAE for the training (triangles) and test (circles) sets are listed in the bottom-right corner of each subpanel.

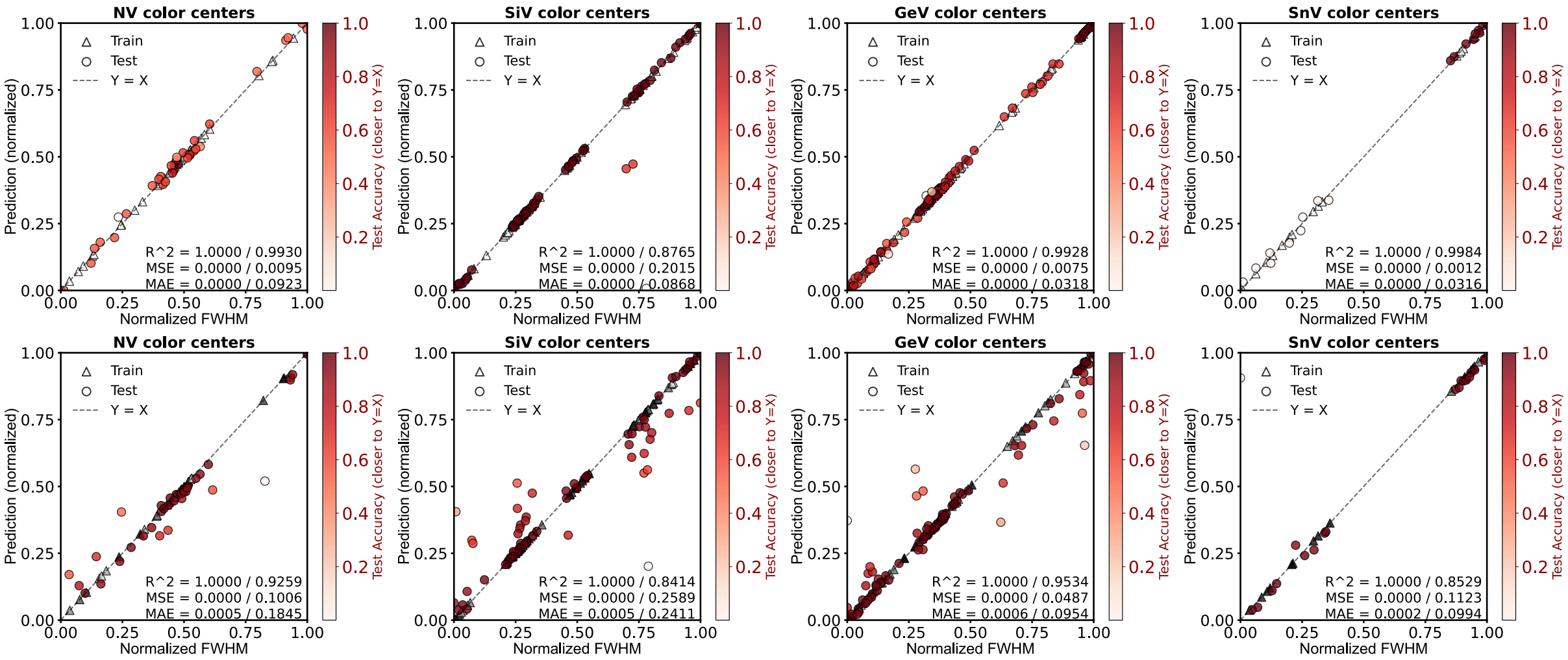

**Figure 8**. CVD/MPCVD, *Task II* summary. DTR (top row) and XGB (bottom row) predictions of the ZPL FWHM (target variable) from given values for chamber pressure, substrate temperature and MW plasma power (input features), organized by color centers (columns). The values for $R^2$, MSE, and MAE for the training (triangles) and test (circles) sets are listed in the bottom-right corner of each subpanel.

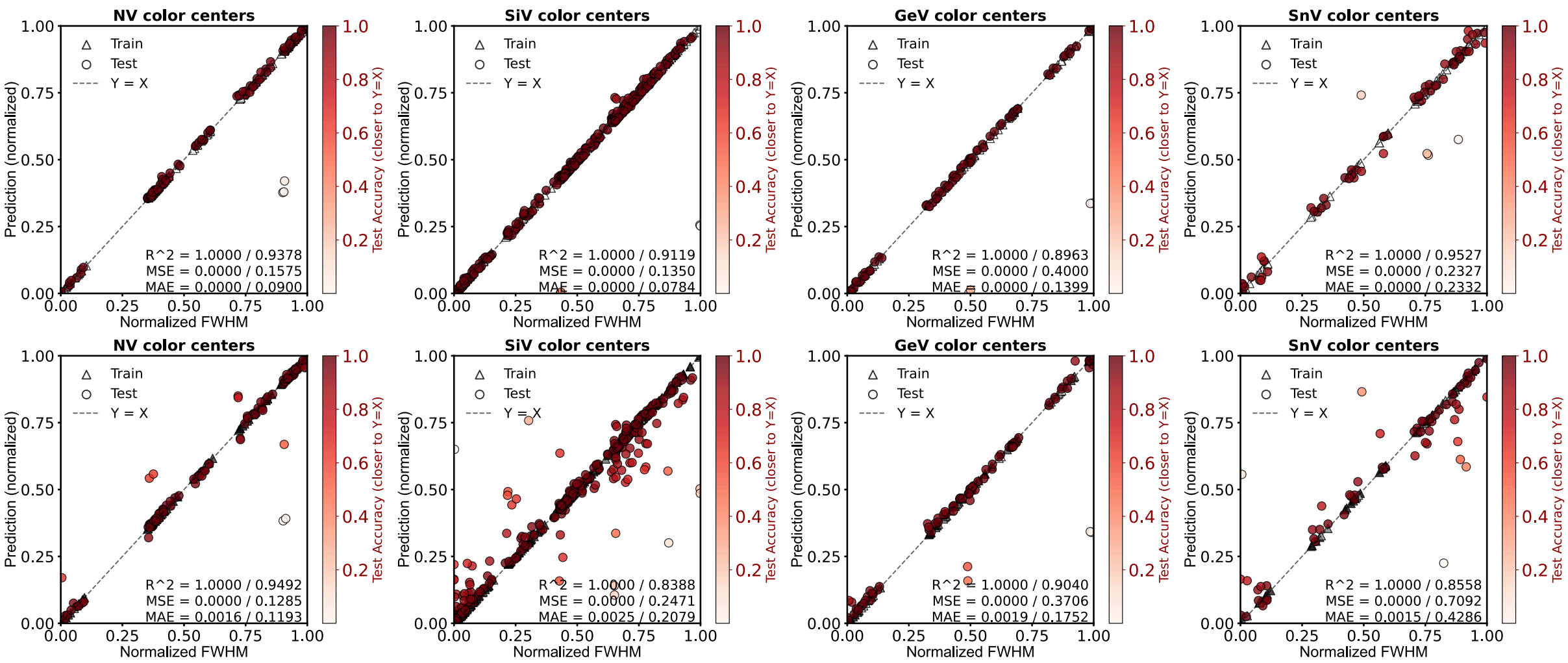

**Figure 9**. Ion implantation, *Task II* summary. DTR (top row) and XGB (bottom row) predictions of the ZPL FWHM (target variable) from given values for implantation energy, fluence, annealing temperature and annealing time (input features), organized by color centers (columns). The values for $R^2$, MSE, and MAE for the training (triangles) and test (circles) sets are listed in the bottom-right corner of each subpanel.

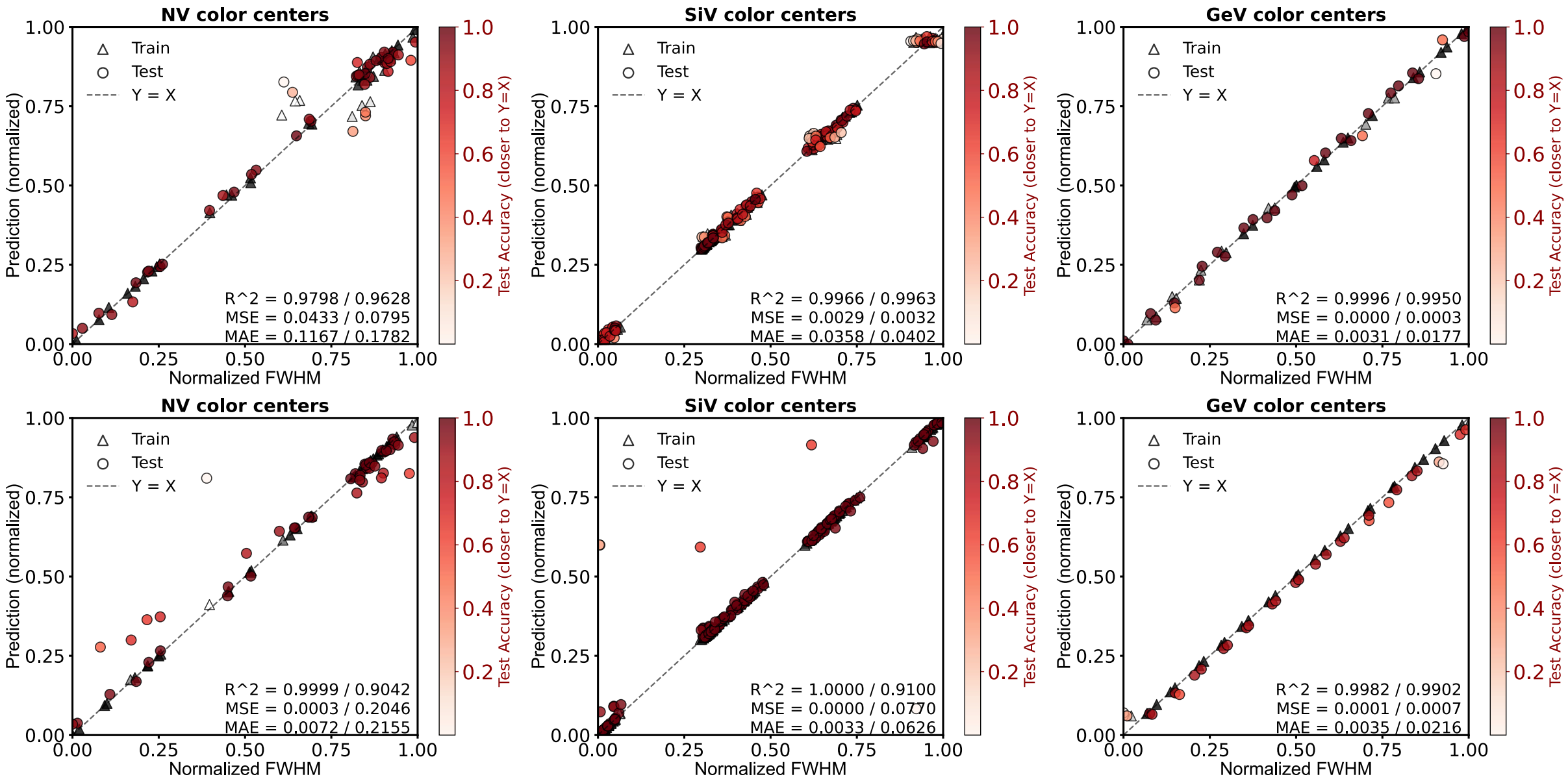

**Figure 10**. Electron/ion irradiation, *Task II* summary. DTR (top row) and XGB (bottom row) predictions of the ZPL FWHM (target variable) from given values for irradiation energy, fluence, annealing temperature and annealing time (input features), organized by color centers (columns). The values for $R^2$, MSE, and MAE for the training (triangles) and test (circles) sets are listed in the bottom-right corner of each subpanel.

Similar to the case of *Task I*, we can make a series of general observations related to the results shown in Figures 7–10 for *Task II*. Overall, the DTR and XGB models can predict the ZPL FWHM value quite well: the estimates for $R^2$ are consistently >0.9 and often close to ~1, with just a few exceptions for the XGB predictor where $R^2 \leq 0.85$. The value of MSE and MAE were also desirably small and consistently $\lesssim$10% of the average target value, in fact regularly within just a few percent points of it. We hypothesize that some of the larger deviations are due to our models not including categorical features such as the substrate materials or the ion source (see § 2.3.2). However, we note that this occurs in a very small minority of cases, underscoring the overall robustness of the predictive ability of our algorithms. We also found that the DTR and XGB predictors were comparable in terms of performance, with DTR slightly outperforming XGB in most cases (see § 3.1). Here, we also applied the DTR and XGB algorithms to predict the value of the Debye-Waller factor (DWF) of the DVCCs, the DTR/XGB models can predict the DWF value relatively well (see § S6, Figs. SF6-10) and summarize the value of $R^2$, MSE and MAE (Table ST9-10).

### 3.3. Shapley value analysis

In this section, we discuss the results of the Shapley value analysis for *Tasks I* and *II*, based on the theoretical framework introduced in § 2.3.3. Shapley value analysis is a particularly valuable tool for assessing and interpreting black-box predictive models. It is model agnostic, and it helps quantify explicitly the relative interdependence between input features and target variables across all observations. Shapley values represent the marginal contribution of individual features and of individual instances to the predicted value. It measures how much each feature, $\phi_j^{(i)}$ in eq. (7), in every instance $i$ (i.e. an individual entry in the database) contributes to return a predicted value for the target variable, $f(\boldsymbol{x}_i)$, above or below the baseline value, $\phi_0$ in eq. (7). In general, the relative weights of the features in our DTR/XGB models do not provide the complete picture. They show the relative importance of each feature, but not how a feature affects directly the prediction $\hat{y}_i = f(\boldsymbol{x}_i)$ of any one particular $i$-th database entry. Conversely, Shapley values show how much and in what direction each feature moves the prediction for an individual instance above or below the mean value, $\phi_0$.

*3.3.1. Shapley value analysis: Task I – material synthesis*

The results of the Shapley value analysis for the DTR model are shown in Figures 11–14 for the HPHT, CVD/MPCVD, Ion implantation and Electron/ion irradiation synthesis methods, respectively. Each figure is organized by color center (columns) and the results are summarized and arranged in rows of subpanels labelled A, B and C, and explained below.

Subpanels A shows the *feature importance bar* plots. In these plots, the features are listed vertically, sorted from most to least important; the ranking is based on the mean absolute Shapley value for the features over all the instances. The $x$-axis shows the average absolute Shapley value of each feature across the entire dataset. It indicates how much, on average, each feature contributed to changing the prediction (regardless of whether it increased or decreased it). A longer (shorter) bar indicates that a feature is more (less) influential in affecting the model's predictions across instances. For example, Figure 11A shows that in the HPHT process for the case of the GeV center, across all the instances, the chamber temperature was significantly more important (+1458) in causing changes in the predicted size of the synthesized diamond materials than the chamber pressure (+235); the run time rarely had any influence (+24).

Subpanels B shows the *violin* plots. The violin plot contains several layers of information. The features are listed in order of importance from top to bottom. The lateral spread shows how variable the contribution of one feature is across observations. If a feature has positive (negative) Shapley values, it means that for those observations/instances the feature increased (decreased) the calculated predicted value with respect to the mean value, $\phi_0$, of the prediction. The plot also carries information through color encoding: red (blue) colors on the positive side indicates that a high (low) feature value pushed the prediction above the average, $\phi_0$, while red (blue) colors on the negative side indicate that a high (low) feature value pushed the prediction below the average, $\phi_0$. The vertical spread is simply indicative of the density of data (more observations of the same Shapley value translate into more points stacked). We can use again the case of the GeV center in Figure 11B as an example. Here, high values (red) of the feature *Time* contributed to the output target variable *Size* by reducing (being to the left of '0') its value below the average. This occurred consistently across all instances, as shown by the fact that all the data concentrates around the same few negative values and are thus spread out vertically.

Subpanels C shows the *heatmap* plots. In these plots, the $x$-axis shows the individual instances (i.e. the entries in the database) sorted by Shapley value magnitude. The $f(\boldsymbol{x})$ graph is the model prediction (eq. (6)) showing the model's output through the individual instances. The red (blue) colors indicate a positive (negative) Shapley value that pushes the prediction above (below) the average value, $\phi_0$; the white color indicates no or negligible contribution. The color changes align with the step changes in $f(\boldsymbol{x})$ indicating why predictions increased or decreased. We can use again the case of the GeV center in Figure 11C as an example. The instances are ordered for convenience to show the predicted values, $f(\boldsymbol{x})$, increasing monotonically. For each value of $f(\boldsymbol{x})$ the heat plot shows the positive (red), negative (blue) or neutral (white) contribution of each feature that, added, produced that value for $f(\boldsymbol{x})$.

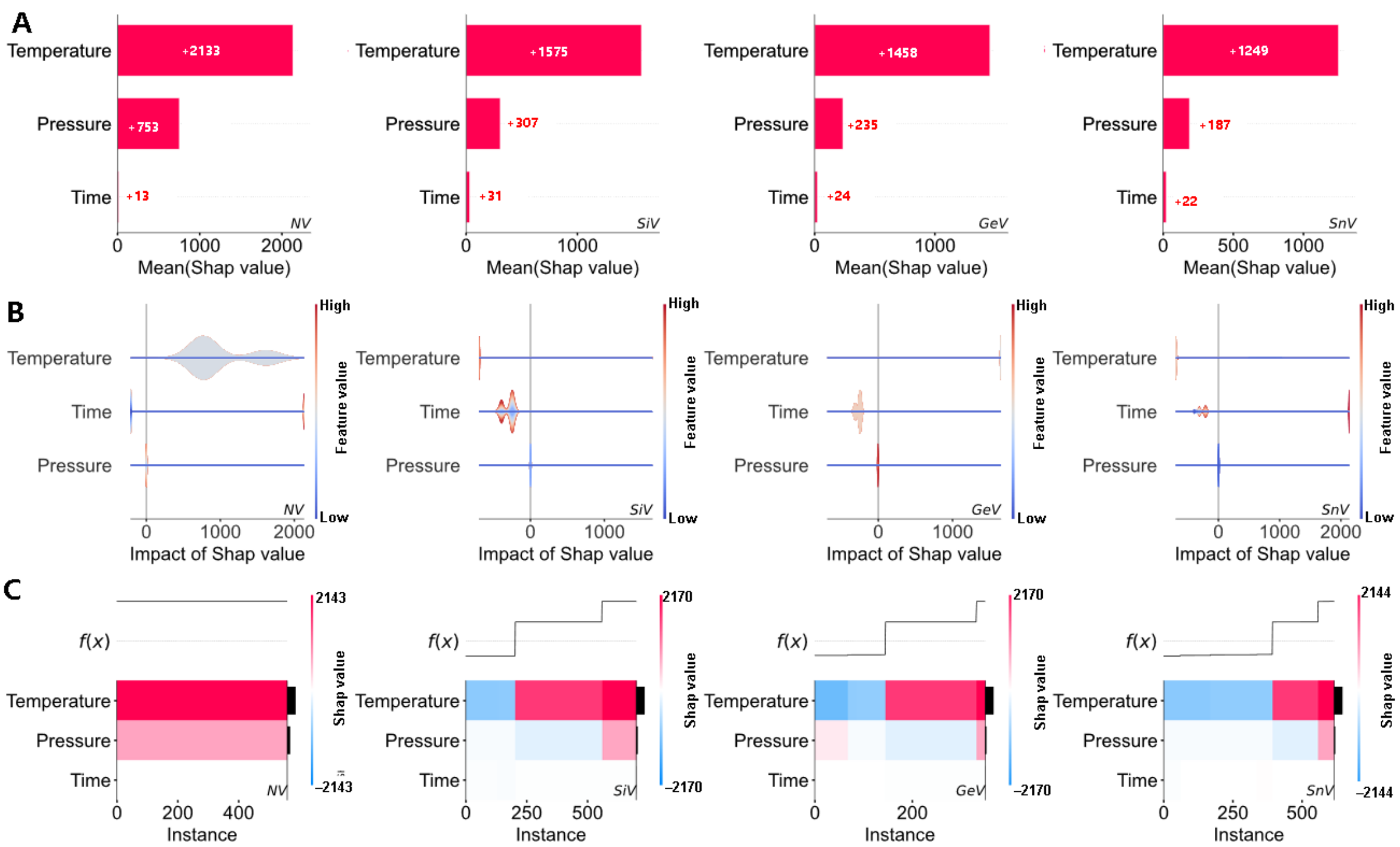

**Figure 11**. Shapley value analysis of the DTR model for HPHT synthesis. **A)** Feature importance bar plot. **B)** Violin plot. **C)** Heatmap plot. See main text for plot details.

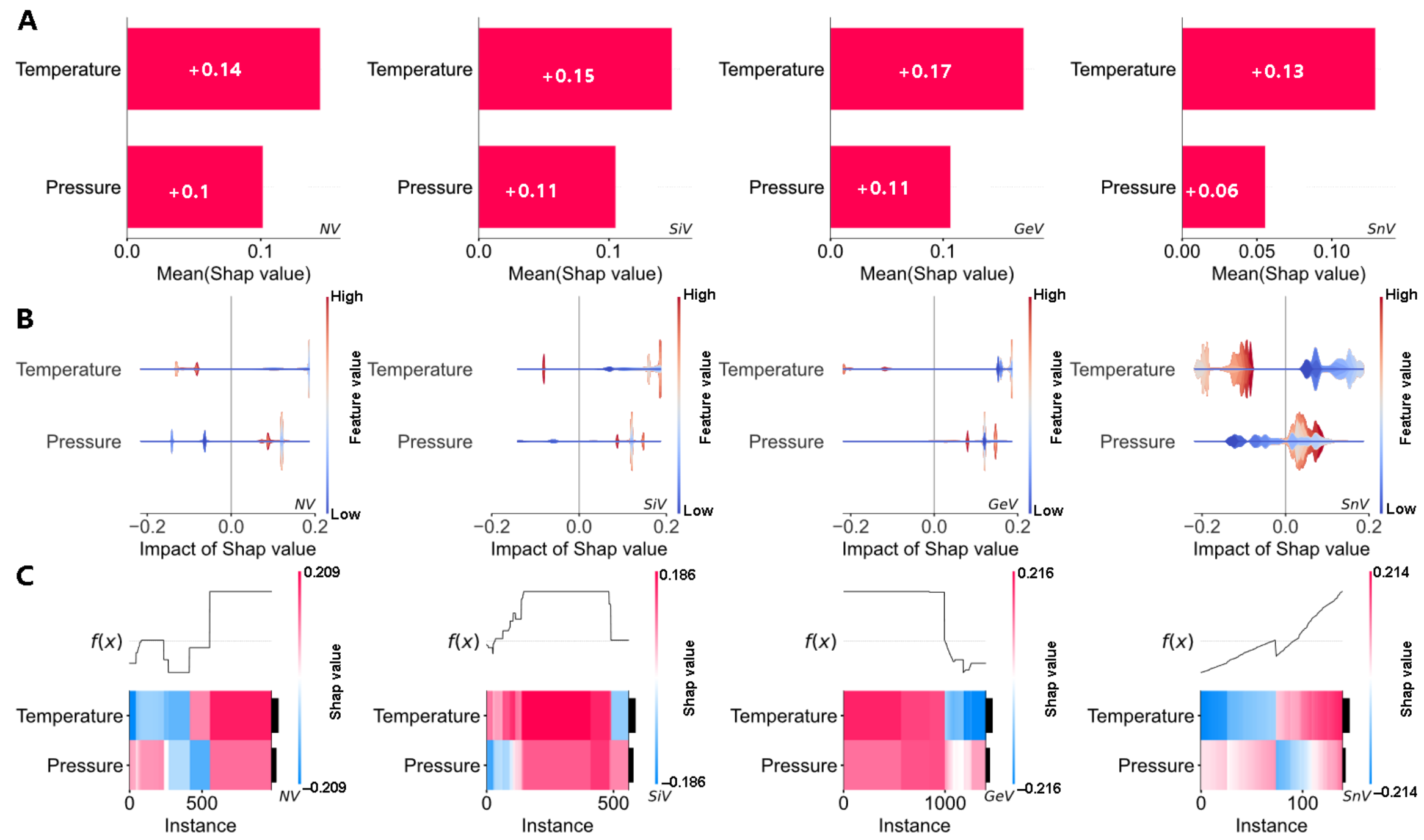

**Figure 12**. Shapley value analysis of the DTR model for CVD/MPCVD synthesis. **A)** Feature importance bar plot. **B)** Violin plot. **C)** Heatmap plot. See main text for plot details.

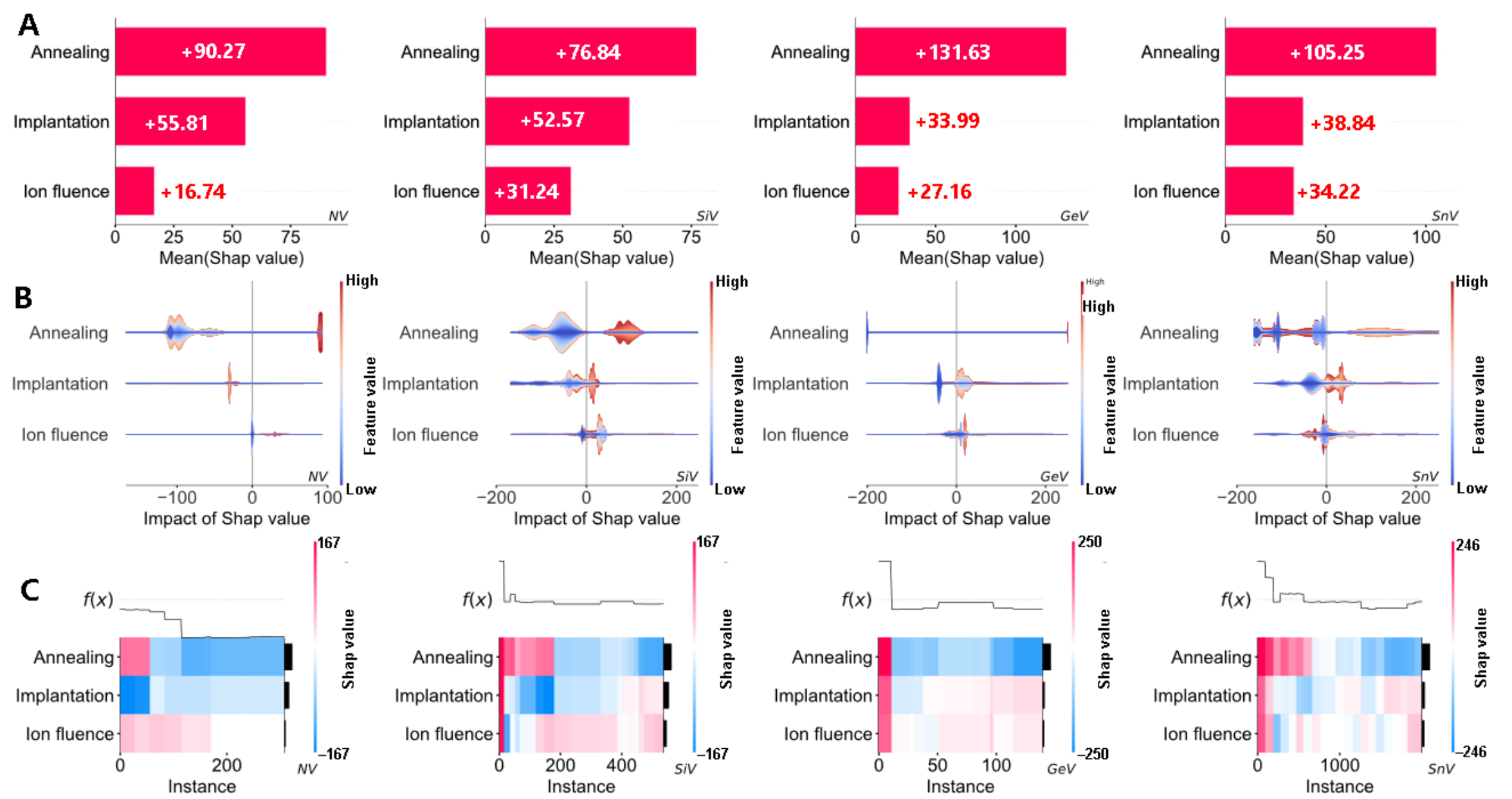

**Figure 13**. Shapley value analysis of the DTR model for Ion implantation. **A)** Feature importance bar plot. **B)** Violin plot. **C)** Heatmap plot. See main text for plot details.

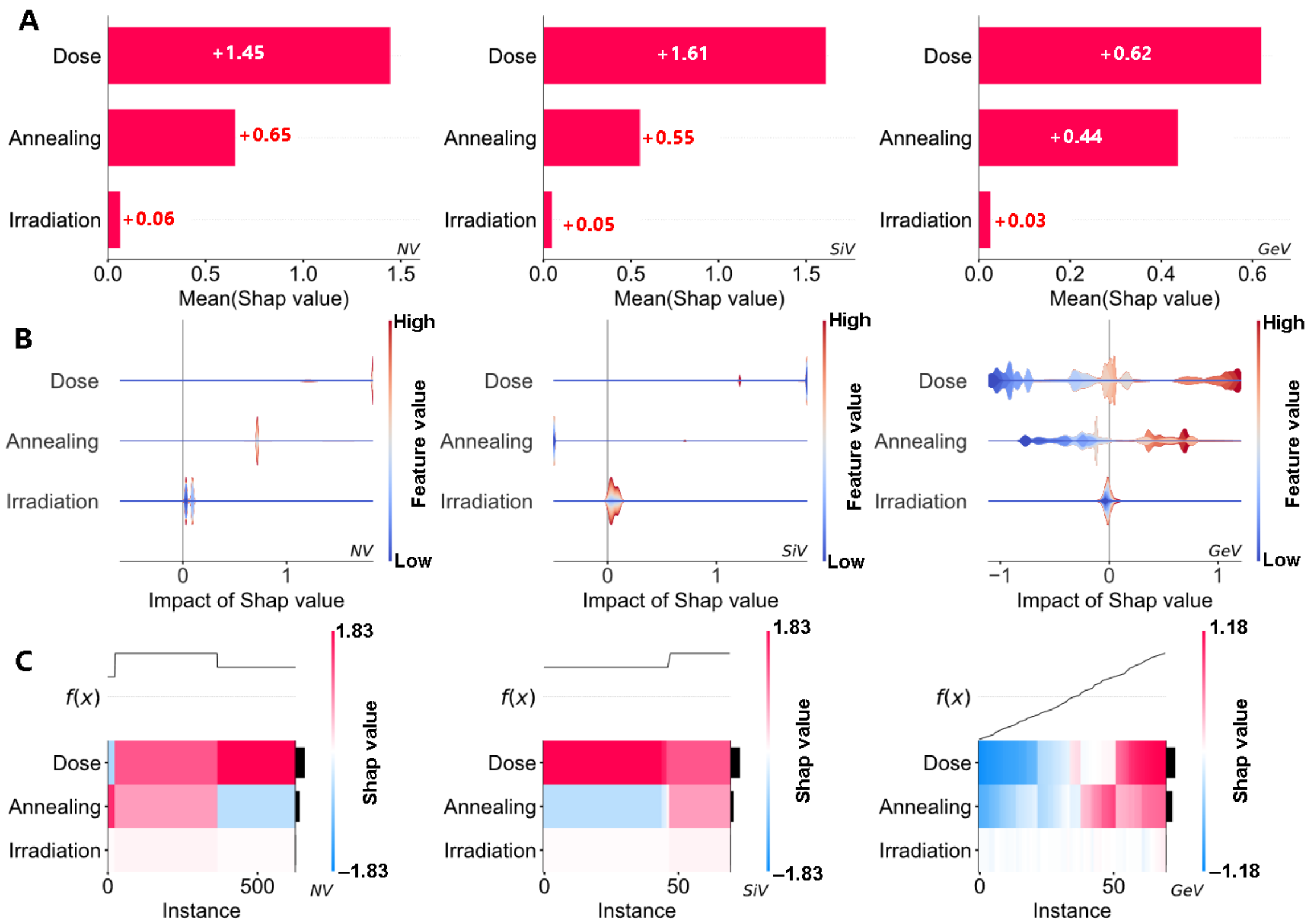

**Figure 14**. Shapley value analysis of the DTR model for Electron/ion irradiation. **A)** Feature importance bar plot. **B)** Violin plot. **C)** Heatmap plot. See main text for plot details.

The analysis of Figures 11–14 provides detailed information about any feature-output interdependence of interest. Here, we briefly highlight the main trends. However, before delving in the analysis, we reiterate that the Shapley value analysis is an instance-dependent explanation not an absolute account on relative weights in the model. The Shapley value of the $j$-th feature explains for any one $i$-th database entry, $\boldsymbol{x}_i$, how much this $j$-th feature shifts the prediction, $\hat{y}_i$, away from the baseline (expected mean output, $\phi_0$). Therefore, if a feature did not vary much in the dataset, its Shapley value would be small, regardless of how large or small its weight is in the model, and vice versa (see e.g. the case of the feature *Time* for HPHT). The Shapley analysis measures the impact on the prediction, not on the model parameters.

For HPHT synthesis, our results suggest that variations in temperature, followed by variations in pressure are the leading cause for deviations from the expected average size of the synthesized diamond materials. Conversely, time does not seem to have a significant effect. These results broadly align with the expectations based on the underlying thermodynamic and kinetic mechanisms. In HPHT diamond synthesis, temperature plays a central role as the primary driver of crystallization. In catalyst-assisted growth, precise control of the temperature—and the associated gradients—govern the carbon solubility, diffusion, and vacancy mobility within the metal-carbon system. These processes are essential to ensuring that the catalyst melts, becomes supersaturated with carbon, and ultimately enables diamond nucleation and growth.[102] Higher temperatures can accelerate vacancy-dopant pairing kinetics (e.g., Ge-V, Si-V), but temperatures that exceed the thermodynamic stability range of diamond can instead promote graphitization or dislocation formation.[103]

Pressure is similarly important. It stabilizes the diamond phase by suppressing graphitization and promoting $sp^3$ bonding, which is consistent with its Shapley value ranking immediately below temperature.

The synthesis time also influences both the crystal size and, to some extent, quality, as longer durations can support more uniform and stable growth. However, its Shapley value is expected to be relatively small. Reported HPHT synthesis times in the literature vary only modestly (typically ±10–20%), giving this feature limited spread. Because the Shapley values quantify the contribution of each feature through variation around its mean, this low variability naturally results in a small Shapley value, regardless of how strongly the model might weigh the feature.

For CVD/MPCVD, chamber pressure and substrate temperature have comparable contributions—with the latter being marginally more influential—on driving shifts in predicted MW power with respect to the expected output. The comparable Shapley importance found for the chamber pressure and substrate temperature mirrors the delicate plasma-surface equilibrium characteristic of gas-phase diamond synthesis. The substrate temperature regulates the surface reconstruction and hydrogen desorption rate, determining the effective incorporation of dopant species (e.g., Si, Ge, and Sn) and the resulting optical quality of the DVCCs. Small variations in substrate temperature can shift the dominant growth regime from step-flow to polycrystalline, thereby altering the optical homogeneity of the color centers.[104] Similarly, chamber pressure modulates the mean free path and plasma energy distribution, influencing both growth rate and defect incorporation efficiency.

For both Ion implantation and Electron/ion irradiation, annealing temperature is the feature that mostly affects differences in annealing time. Perhaps expectedly, higher values of the ion dose and fluence correlate with higher annealing temperatures and annealing times, as—intuitively—the latter are both conducive (or necessary) to the effective repair of damage in the material caused by the irradiation process. The relative importance of fluence, annealing temperature, and annealing duration is a consequence of the fact that these parameters directly determine the balance between defect generation and recovery in the material. Moderate fluence values maximize single-vacancy creation while avoiding amorphization, and post-annealing enables the migration and stabilization of dopant-vacancy complexes such as $NV^-$, $SiV^-$, and $GeV^-$ centers.[105] Excessive fluence or prolonged irradiation, however, can induce clustering and charge instability, which degrades luminescence and coherence properties.[106]

*3.3.2. Shapley value analysis: Task II – FWHM*

Figures 15–18 show the Shapley value analysis for the DTR model applied to the FWHM of the color centers' ZPL. Specific contributions can be drawn from direct analysis of the

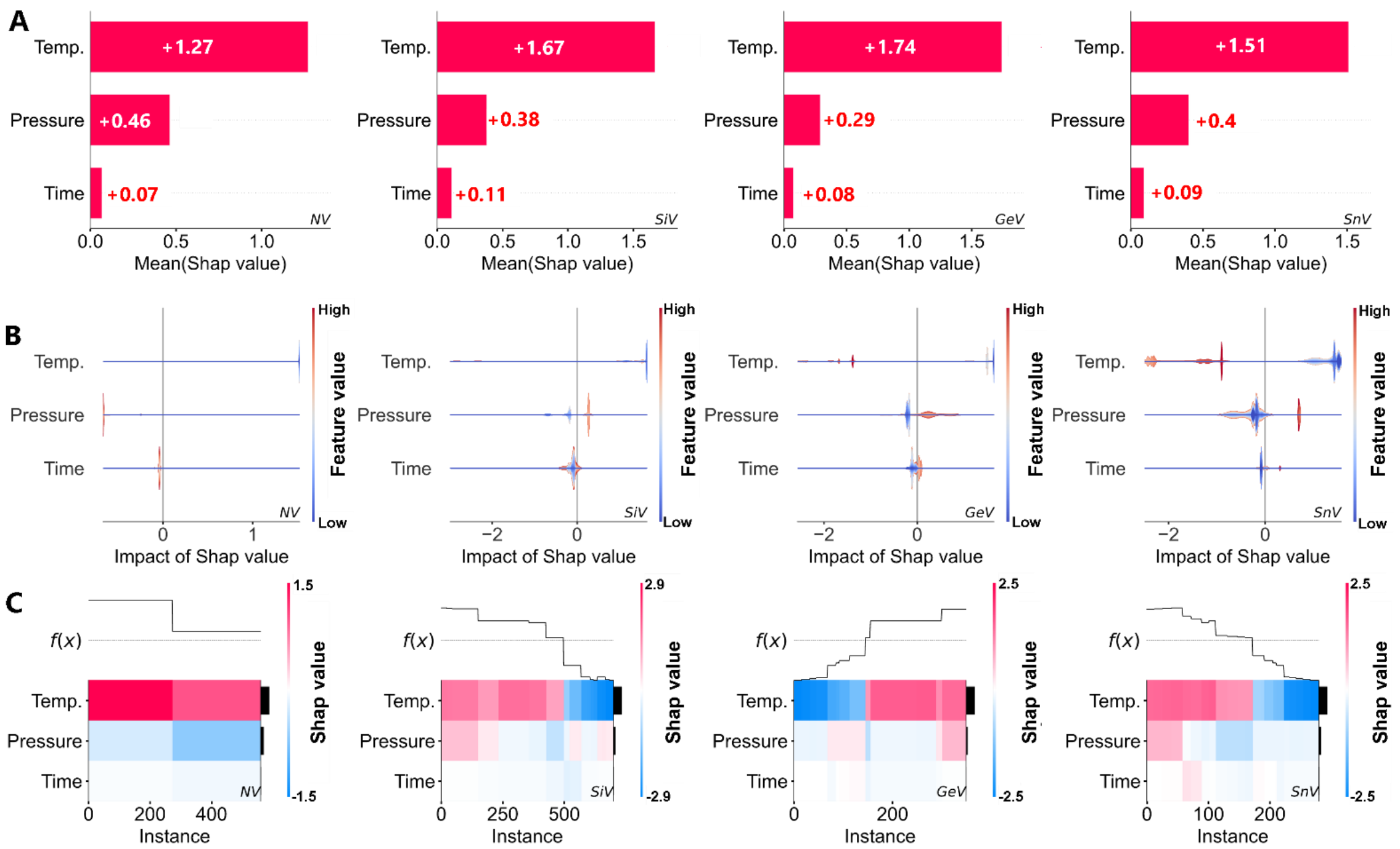

**Figure 15**. Shapley value analysis of the DTR model for the FWHM of the centres' ZPL fabricated by HPHT synthesis. **A)** Feature importance bar plot. **B)** Violin plot. **C)** Heatmap plot.

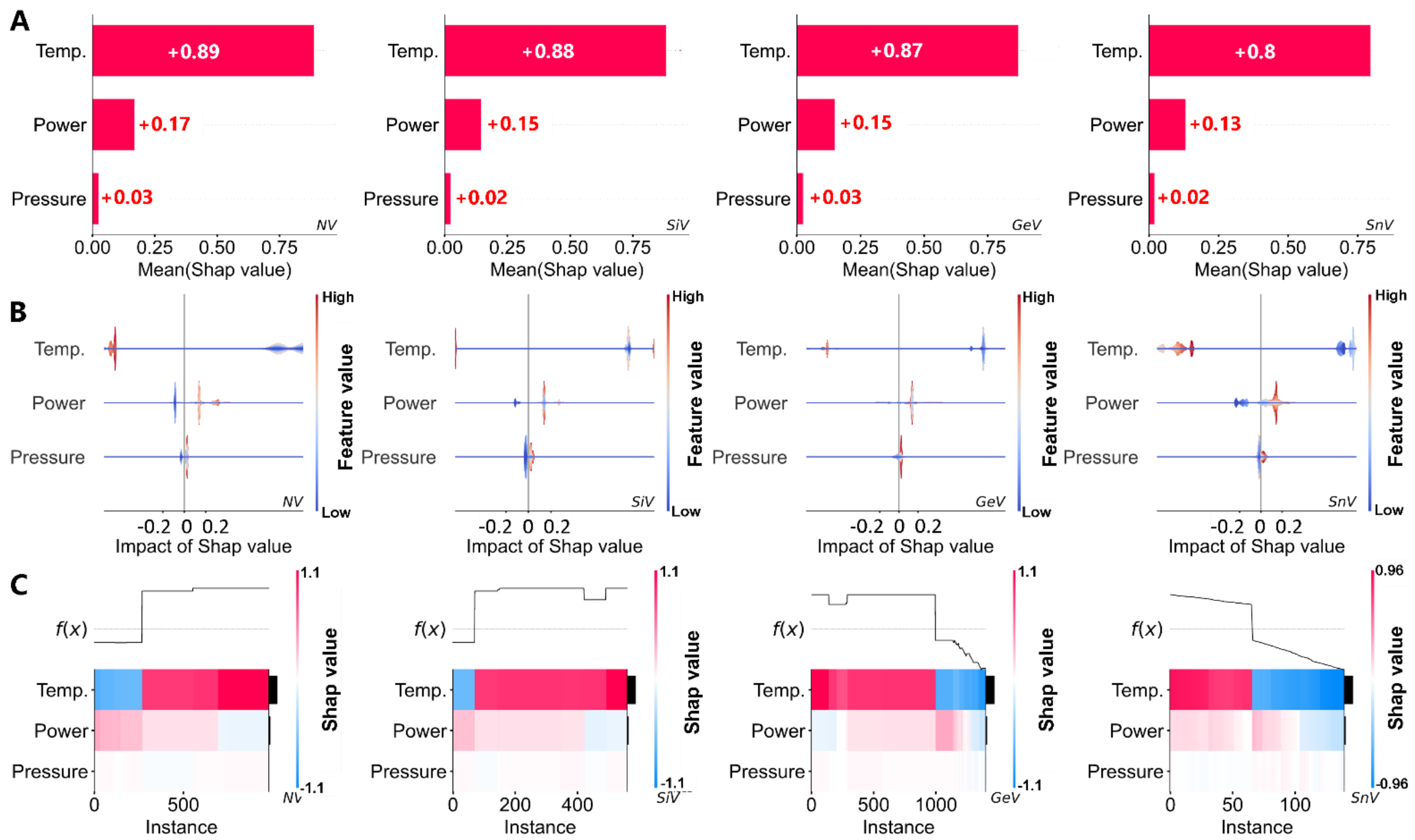

**Figure 16**. Shapley value analysis of the DTR model for the FWHM of the centres' ZPL fabricated by CVD/MPCVD synthesis. **A)** Feature importance bar plot. **B)** Violin plot. **C)** Heatmap plot.

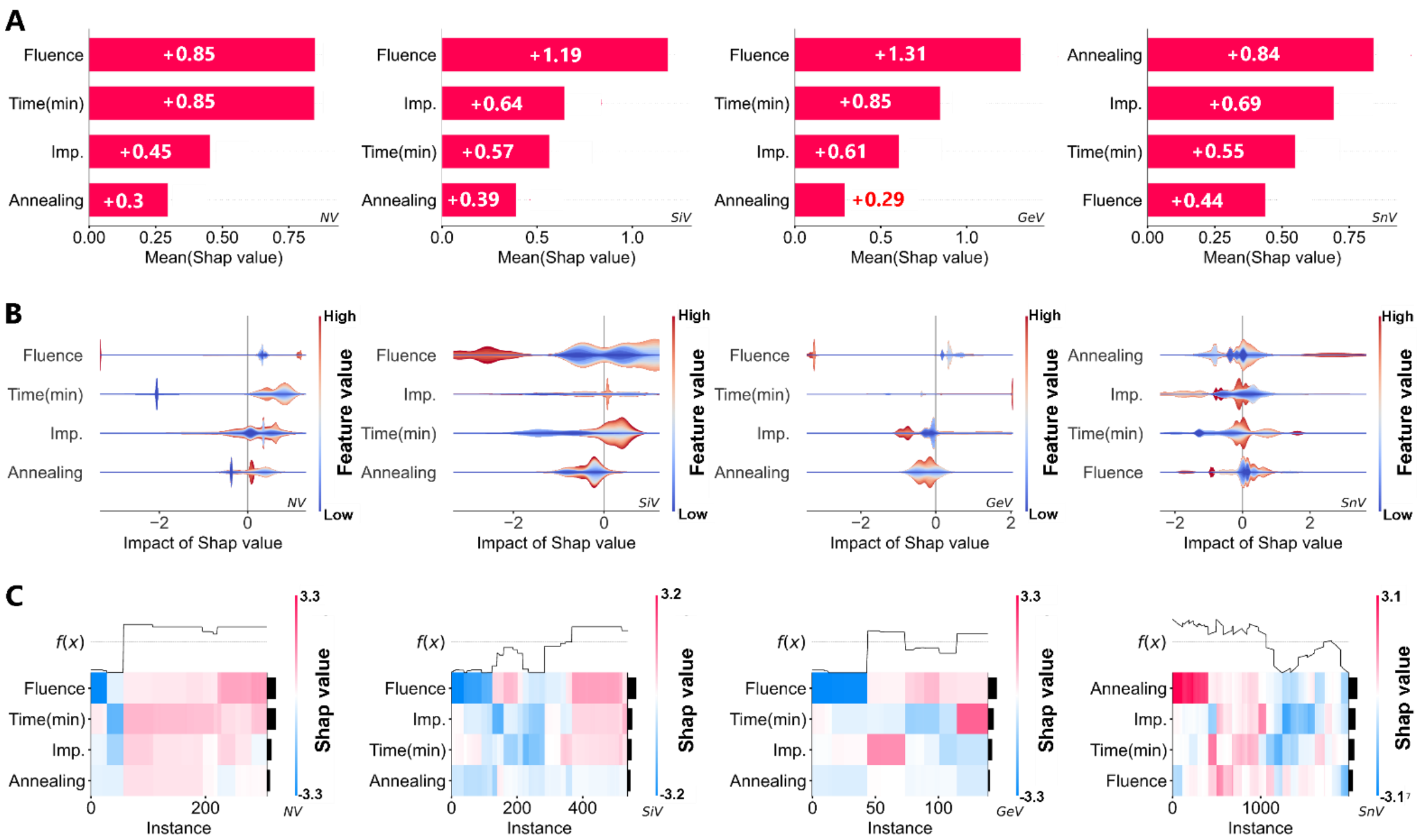

**Figure 17**. Shapley value analysis of the DTR model for the FWHM of the centres' ZPL fabricated by Ion implantation. **A)** Feature importance bar plot. **B)** Violin plot. **C)** Heatmap plot.

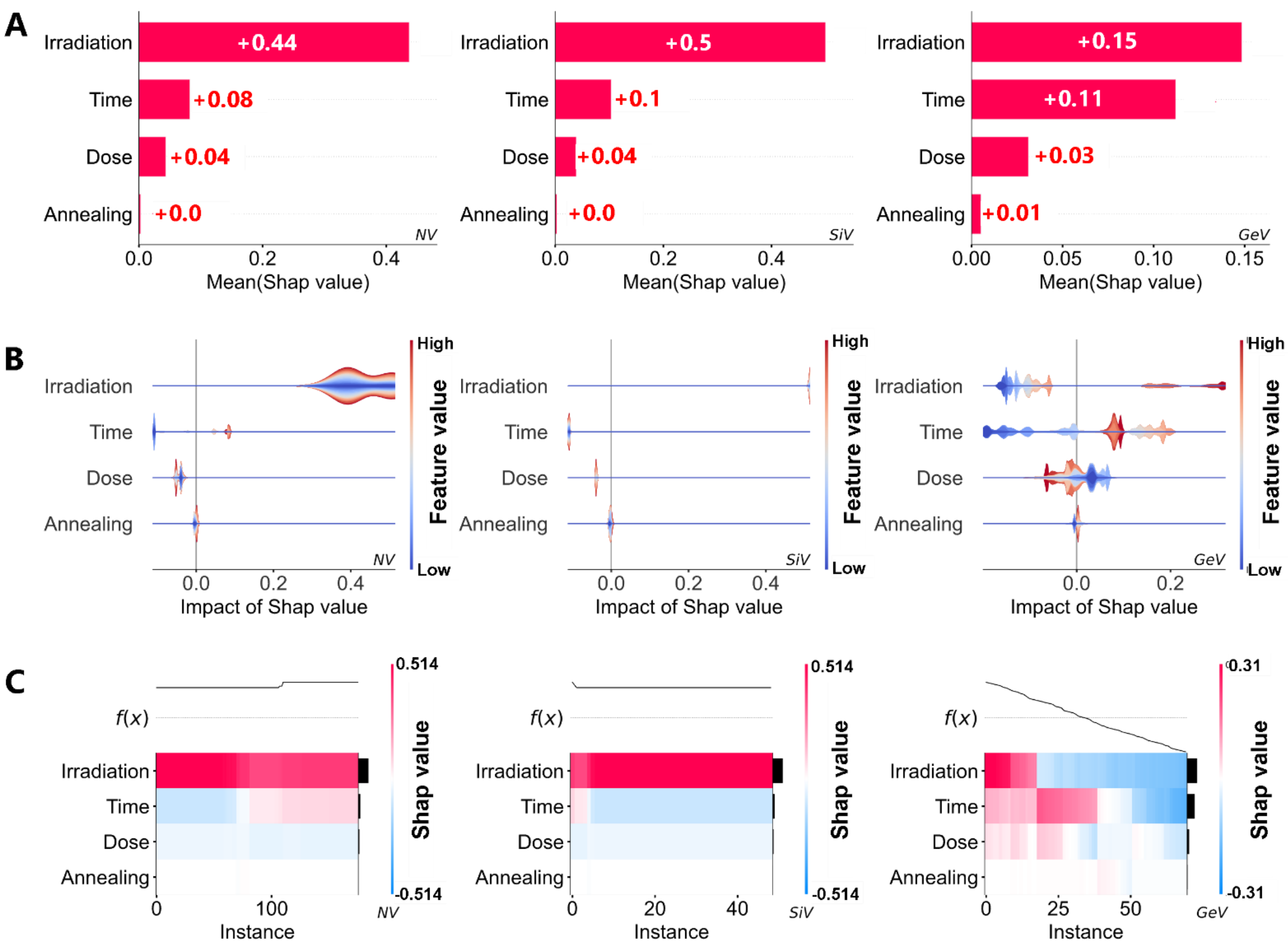

**Figure 18**. Shapley value analysis of the DTR model for the FWHM of the centres' ZPL fabricated by Electron/ion irradiation. **A)** Feature importance bar plot. **B)** Violin plot. **C)** Heatmap plot.

individual graphs. We highlight here a few general results. Figures 15 and 16 show that for equilibrium or quasi-equilibrium growth processes such as HPHT and MPCVD, the dominant Shapley contributor to variations of the FWHM is temperature, with pressure having a weaker—yet still relatively important—effect. This is consistent with the observations made in § 3.3.1 about the underlying thermodynamic and kinetic mechanisms involved in HPHT and MPCVD synthesis. It also indicates that control over temperature and pressure dictates the quality of both the material and of the DVCCs' optical and spectral properties.

Figures 17 and 18 show that, for Ion implantation and Electron/ion irradiation, fluence and irradiation energy are the parameters that most strongly influence the FWHM of the centres' ZPL. This result is physically expected, as both quantities correlate with residual strain and the density of intrinsic defects in the material, which in turn lead to narrower or broader ZPL linewidths for the DVCCs depending on the extent of crystalline damage. Subpanels A capture this trend by showing that the absolute SHAP values associated with fluence and irradiation energy are among the largest relative to other features, indicating their dominant contribution to the variance of the FWHM. However, drawing direct causal conclusions from these results is non-trivial. The violin plots in Subpanels B do not consistently exhibit the intuitive trend that lower fluences or irradiation energies contribute to reduced FWHM values. This behavior reflects the fact that SHAP values encode feature contributions conditioned on the joint distribution of all variables and therefore capture nonlinear effects and inter-feature correlations. In practical terms, this supports the intuition that achieving narrow ZPL linewidths in the DVCCs cannot be reduced to optimizing a single parameter, but instead requires coordinated control of multiple processing variables—for example, longer annealing times that compensate for higher implantation doses or irradiation energies. This multi-parameter interplay is further evidenced in Subpanels A of Figure 17, where several features exhibit comparable absolute SHAP values.

Despite these limitations, Shapley value analysis remains valuable here as it provides a clear, model-agnostic measure of the relative importance of different parameters in determining the variance of the FWHM.

Overall, the statistical relationships captured by the Shapley values of both *Tasks I* and *II* align well with the established physical mechanisms governing the synthesis methods studied. This consistency suggests that the ML models can learn and reflect key kinetic and thermodynamic principles, even without explicitly embedding the underlying physics of diamond and DVCCs synthesis. This ability to recover meaningful physical trends from data—while acknowledging the inherent limits of such general-purpose algorithms[39]—underscores the value of the data-driven framework proposed here.

## 4. Conclusions

In this work we performed a thorough review and meta-analysis of studies involving the synthesis of diamond materials via HPHT, CVD/MPCVD, Ion implantation, and Electron/ion irradiation, over the last decade. Data from approximately 60 published articles was organized into four databases, including a total of 170 data sets and 1692 entries. These were used to train two ML algorithms: DTR and XGB, with the goal of using these models to select diamond synthesis techniques and parameters that would yield materials and DVCCs with the desired characteristics, properties and quality. We organized our analysis in two streams: one aimed at optimizing diamond material synthesis (*Task I*), and one aimed at optimizing the yield of high-quality color centers in the fabricated diamond (*Task II*). We characterized the predictive performance of our DTR/XGB models using traditional statistical indicators, namely $R^2$, MSE, and MAE, and performed a Shapley value analysis to complement their interpretation.

Overall, our results show that the proposed DTR and XGB models are robust predictive algorithms that can leverage knowledge of input features (i.e. synthesis parameters) to make accurate predictions about the specific properties and quality (target variables) of synthesized diamond materials (§ S7, Figure SF11). These models can also be effectively used in reverse to tweak and control—optimally and efficiently—the value of specific synthesis parameters to yield materials with ad hoc characteristics. Thus, the data-driven framework we built is a powerful addition to the tools available to researchers and material scientists involved in the synthesis of diamond materials and diamond color centers.

## 5. Supporting Information

Input features and target variables used in training/testing the algorithms (§ S1, Table ST1), and schematic representation of XGB (§ S1, Figure SF1). Databases used to train the DTR/XGB algorithms (§ S2, Tables ST2–ST5). Hyperparameters and their value ranges used in the models (§ S3, Table ST6). Handling of categorical variables in our algorithms (§ S4, Figures SF2–SF5). Summary of the statistical estimators used to measure the performance of our models for *Tasks I* and *II* (§ S5, Tables ST7 and ST8). Debye-Waller factor analysis (§ S6, Figures SF6–10, Tables ST9 and ST10). Worldwide trends in diamond synthesis (§ S7, Figure SF11).

## 6. Contributions and Acknowledgements

Collection and meta-analysis of the data was performed by Z. J. Theoretical modeling and data analysis were performed by Z.J., C.B and G.G. The analysis and interpretability of the data rationale were provided by M.P. and C.B. All authors contributed to writing the manuscript. The authors declare that they have no conflicts of interest or competing interests related to the research presented in this study.

The Horizon Europe (HORIZON), HORIZON-MSCA-2023-PF-01, Grant101149632–GDSNL, project UID 00481 Centre for Mechanical Technology and Automation (TEMA) and project CarboNCT, 2022.03596.PTDC (DOI: 10.54499/2022.03596.PTDC) are acknowledged.
The Natural Sciences and Engineering Research Council of Canada (DGECR-2021-00234) and the Canada Foundation for Innovation (John R. Evans Leaders Fund #41173) are acknowledged.

## 7. Conflict of Interest

The authors declare no conflict of interest.

## 8. Data Availability

The Supporting Information file contains additional details about the databases, models and their performance. The data and code of the DTR and XGB algorithms used in this study are available in a public repository (linked below).
https://github.com/howgoods/Diamond-color-centers-python-code

# Supplementary Information

## Data-Driven Review and Machine Learning Prediction of Diamond Vacancy Center Synthesis

Zhi Jiang,[1,2] Marco Peres,[3,4,5] Carlo Bradac,[6] and Gil Gonçalves[1,2*]

1. Centre for Mechanical Technology and Automation (TEMA), Mechanical Engineering Department, University of Aveiro, 3810-193 Aveiro, Portugal
2. Intelligent Systems Associate Laboratory (LASI), 4800-058 Guimarães, Portugal
3. Instituto de Engenharia de Sistemas e Computadores - Microssistemas e Nanotecnologia, Rua Alves Redol 9, 1000-029 Lisboa, Portugal
4. IPFN, Instituto Superior Técnico, University of Lisbon, Av. Rovisco Pais 1, 1049 001 Lisbon, Portugal
5. DECN, Instituto Superior Técnico, University of Lisbon, Estrada Nacional 10 (km 139.7), 2695 066 Bobadela, Portugal
6. Department of Physics and Astronomy, Trent University, 1600 West Bank Drive, Peterborough, ON, K9L 0G2, Canada

* Corresponding author: Gil Gonçalves, ggoncalves@ua.pt

## S1. Task features and target values; XGB algorithm

Table ST1 below summarizes the input features and target variables for each diamond synthesis method.

**Table ST1.** Synthesis input features and target variables

| | *Task I* | | *Task II* | |
|---|---|---|---|---|
| | **Target Variable** | **Input Features** | **Target Variable** | **Input Features** |
| **HPHT** | Avg. particle diameter | Chamber pressure<br>Chamber temperature<br>Run time | FWHM of the DVCCs' ZPL | Chamber pressure<br>Chamber temperature<br>Run time<br>Avg. particle diameter |
| **CVD/MPCVD** | MW power | Chamber pressure<br>Substrate temperature | FWHM of the DVCCs' ZPL | Chamber pressure<br>Substrate temperature<br>MW plasma power |
| **Ion implantation** | Annealing time | Implantation energy<br>Fluence<br>Annealing temperature | FWHM of the DVCCs' ZPL | Implantation energy<br>Fluence<br>Annealing temperature<br>Annealing time |
| **Electron/Ion irradiation** | Annealing time | Irradiation energy<br>Fluence<br>Annealing temperature | FWHM of the DVCCs' ZPL | Irradiation energy<br>Fluence<br>Annealing temperature<br>Annealing time |

**Table ST1.** List of target variables and input features organized by *Task* and synthesis method (HPHT, CVD/MPCVD, Ion implantation, Electron/ion irradiation).

Figure SF1 illustrates schematically the application of the XGB algorithm to the synthesis of diamond material via HPHT, used here as an example. Pressure, temperature and run time of every different row in the HPHT database (§ S2, Table ST2) are set as the inputs Fig. SF1, input layer – green left panel). The XGB (or DTR) algorithm produces and optimizes the predictor (Fig. SF1, middle layer – orange middle panel). For a new, unknown set of pressure, temperature and run time input values, the optimized predictor then makes a prediction (main text, eq. (2) for DTR and eq. (5) for XGB) for the average size of the synthesized diamond material, which is the output of interest (Fig. SF1, output layer – cyan right panel).

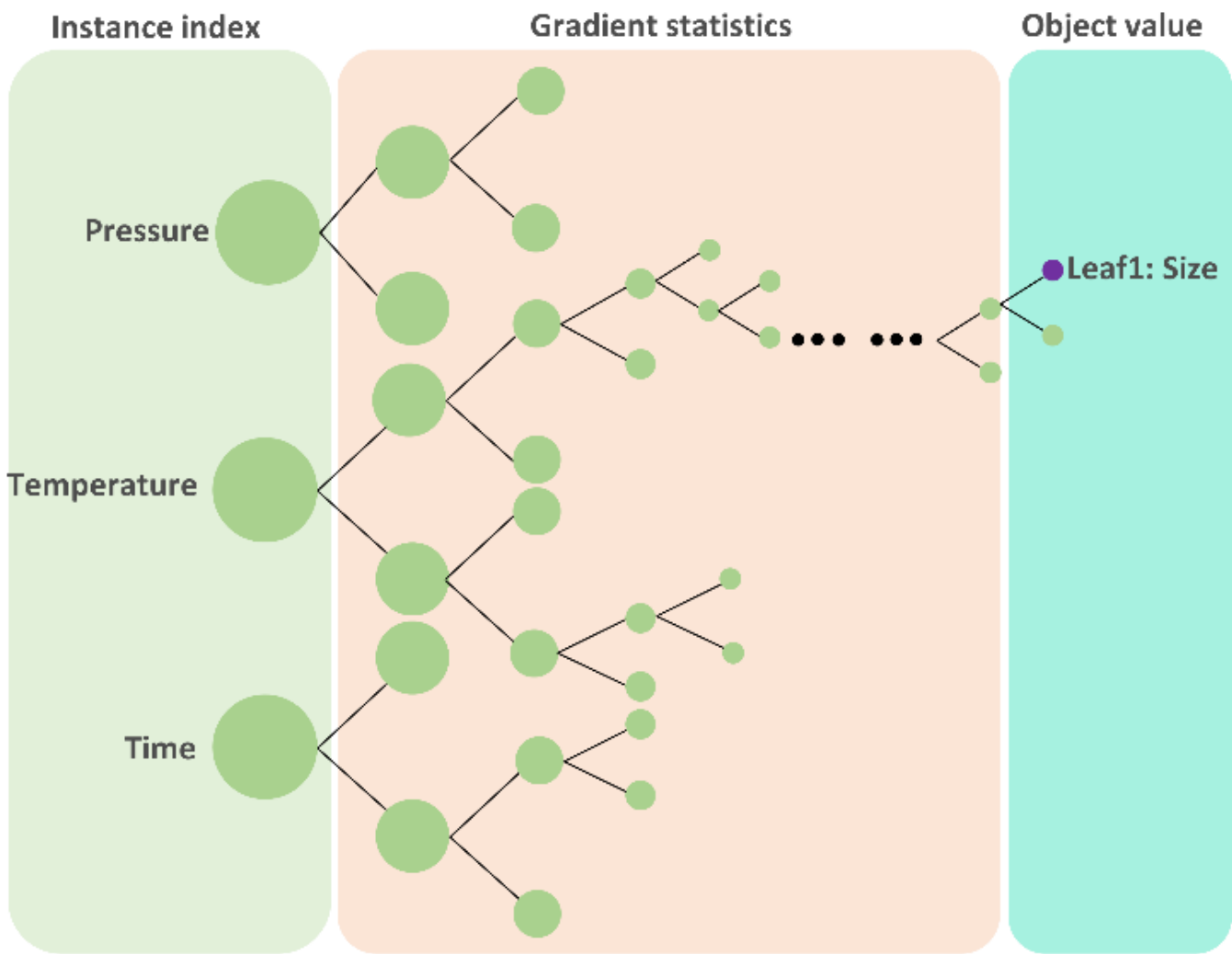

**Figure SF1**. Schematic representation of the Extreme Gradient Boosting (XGB) algorithm for HPHT diamond synthesis.

## S2. Database

To train/test our algorithms we compiled a database from 60 experimental studies. The database has 170 data-sets and 1692 entries organized in Tables ST1–ST4. The tables are arranged by type of color center and show the values and specifications for the critical synthesis parameters discussed in the main text (§§ 2.1.1–2.1.4).

Selection and curation of the data was carried out using the following criteria:

***Data selection***

In the **HPHT** dataset, many secondary parameters (such as catalyst composition or reaction medium) can influence the formation and size of diamond crystals. However, only the key synthesis parameters that are explicitly reported in the literature to affect crystal size and color-center formation—namely temperature, pressure, and process duration—were extracted and analyzed. This was done for clarity and to allow drawing comparisons between methods.

For **MPCVD** growth, most experimental studies emphasize substrate type, deposition time, and gas composition. Although gas mixing ratios and flow rates may also affect color-center density, the parameters most consistently discussed and quantitatively reported across the literature are substrate temperature, chamber pressure, and microwave power.

Similarly, for **Ion implantation** and **Electron/ion irradiation**, once the substrate type is established (typically HPHT- or CVD-grown diamond), the dominant variables are ion fluence, irradiation energy, substrate temperature, and annealing duration. Records that lacked quantitative synthesis conditions, contained duplicate entries, or only reported qualitative experimental descriptions (e.g., substrate thickness, general morphology) were excluded from the database.

***Data normalization and consistency***

All input features were converted into consistent physical units before model training. Temperatures were expressed in degrees Celsius (°C), pressures in kilopascals (kPa), durations in minutes (min), and implantation or irradiation energies in mega-electronvolts (MeV). In cases where publications used approximate or categorical expressions (e.g., "high pressure," "short annealing"), representative numeric values were assigned based on comparable setups reported by the same research group or cross-verified references. This approach ensured both unit consistency and physical interpretability.

***Handling of missing or inconsistent data***

Any dataset with missing values in one or more of the main synthesis parameters was excluded from model training, but retained in the raw database for completeness. When datasets showed large internal variance under nominally similar conditions—for example, between conventional furnace annealing and laser-assisted annealing—data normalization was applied to reduce systematic bias, while preserving underlying physical significance.

***Treatment of uncertainty***

Data transformation was carried out through averaging or range normalization. For experimental data without explicitly reported uncertainty bars, reference-based inference was applied—local variance under similar experimental conditions within the same study was used to estimate the uncertainty. This approach allows us to maintain data balance while minimizing bias caused by inconsistencies in reporting methods.

Overall, although intrinsic variations in experimental setups across research groups cannot be fully eliminated, this multi-step curation and normalization workflow ensured that the final dataset remains physically meaningful, internally consistent, and reproducible.

**Table ST2.** HPHT sub-database

| Year | Carbon source | Vacancy source | Color centers | Pressure range (GPa) | Temperature range (°C) | Time (min) | Size (µm) | References |
|---|---|---|---|---|---|---|---|---|
| 2023 | Graphite | TiN, Fe-Co-Ti | NV | 5.6 – 6.0 | 1645 – 1655 | 3595 – 3605 | 398 – 402 | [78] |
| 2023 | Graphite | $NaN_3$/$Ba(N_3)_2$, Fe-Al-Ti | NV | 5.2 – 5.8 | 1345 – 1355 | 4795 – 4805 | 492 – 502 | [38–44] |
| 2019 | $C_{10}H_{16}$ | $C_{12}H_{36}Si_5$+$C_{12}H_{20}Si$ | SiV | 9.3 – 9.7 | 895 – 905 | 0.43 – 0.47 | 5995 – 6005 | [79] |
| | | | | 8.3 – 8.7 | 1795 – 1805 | 58 – 62 | 29 – 31 | |
| 2019 | $CH_2O$+HCl+ $C_2H_3N$ | $Si(OC_2H_5)_4$ | SiV | 18 – 22 | 1495 – 1505 | 28 – 32 | 0.08 – 0.12 | [96] |
| | | | | 23 – 27 | 1695 – 1705 | | 0.18 – 0.22 | |
| 2019 | $C_{10}H_8$ | $C_{12}H_{36}Si_5$ | SiV | 7.8 – 8.2 | 1295 – 1305 | 0.1 – 0.2 | 0.01 – 0.12 | [97] |
| 2023 | $C_{10}H_8$ +CF1.1 | $C_{12}H_{36}Si_5$ | SiV | 7.8 – 8.2 | 1395 – 1405 | 0.18 – 0.22 | 0.08 – 0.12 | [98] |
| 2024 | SiC | $Fe_{70}Ni_{30}$ | SiV | 7.5 – 7.9 | 1795 – 1805 | 8 – 12 | 198 – 202 | [99] |
| | | | | 8 – 8.4 | 1995 – 2005 | 28 – 32 | 298 – 302 | |
| | | | | 4.8 – 5.2 | 1414 – 1425 | 8 – 12 | 198 – 202 | |
| | | | | 5 – 5.4 | 1495 – 1505 | 8 – 12 | 298 – 302 | |
| 2015 | $C_{10}H_8$ | Germanium | GeV | 7 – 8 | 1295 – 1305 | 0.1 – 0.12 | 10 – 15 | [64] |
| | | | | 8 – 9 | 1595 – 1605 | | 0.03 – 0.1 | |
| 2019 | $C_{10}H_{16}$ | $C_{24}H_{20}Ge$ | GeV | 8.3 – 8.7 | 1895 – 1905 | 235 – 245 | 98 – 102 | [79] |
| 2021 | $C_{10}H_8$ | $C_{24}H_{20}Ge$ | GeV | 7.8 – 8.2 | 1295 – 1305 | 0.8 – 1.2 | 0.1 – 0.5 | [101] |
| 2021 | Graphite and 0.5 mm Diamond | Germanium | GeV | 6.8 – 7.2 | 1795 – 1805 | 3590 – 3605 | 448 – 452 | [102] |
| 2019 | Graphite | $Sn_{20}Mg_{80}$ | SnV | 6.1 – 6.5 | 1415 – 1425 | 5910 – 4005 | 3995 – 4005 | [101] |
| | | $Sn_{20}Mg_{80}$ | | 7.3 – 7.7 | 1895 – 1905 | 235 – 245 | 89 – 102 | |
| | | $Sn_{50}Mg_{50}$ | | | 1995 – 2005 | 475 – 485 | 118 -122 | |
| | | $Sn_{70}Mg_{30}$ | | | 1845 – 1855 | 58 – 62 | 9 -11 | |
| | | $Sn_{20}Mg_{80}$ | | | 1895 – 1905 | 235 – 245 | 98 – 102 | |
| | | $Sn_{50}Mg_{50}$ | | | 1795 – 1805 | 118 -122 | 49 – 51 | |
| | | $Sn_{70}Mg_{30}$ | | | 1695 – 1705 | 475 – 485 | 2498 – 2502 | |
| | | $Sn_{20}Mg_{80}$ | | | 1895 – 1905 | 175 – 185 | 98 – 102 | |
| | | $Sn_{50}Mg_{50}$ | | | 1795 – 1805 | 118 – 122 | 49 -51 | |
| | | $Sn_{70}Mg_{30}$ | | | 1695 – 1705 | 475 – 485 | 2498 – 2502 | |
| 2021 | Graphite | Sn-C system | SnV | 9.2 – 9.6 | 1695 – 1705 | 475 – 485 | 2498 – 2502 | [104] |

**Table ST2.** High pressure high temperature (HPHT). Relevant information includes the carbon/color center sources and experimental conditions such as chamber pressure, temperature, and run time reported in the literature for NV, SiV, GeV and SnV. Pressure and temperature values are in the range 5–9.7 kPa and 895–1,995 °C, respectively. The size of the synthesized diamond NPs is in the range 0.010–6,005 µm. The HPHT database includes a total of 680 data points.

**Table ST3.** CVD/MPCVD, sub-database

| Year | Carbon source | Dopant source | Substrate | Color centers | Pressure range (kPa) | Temperature range (°C) | Power (kW) | References |
|---|---|---|---|---|---|---|---|---|
| 2019 | $CH_4$:$H_2$ (8:192 *sccm*) | $N_2$(2 *sccm*) | CVD SCD on Si | NV | 9.3 – 9.7 | 895 – 905 | 0.43 – 0.47 | [55] |
| 2020 | $CH_4$:$H_2$ (1:100 *sccm*) | $N_2$(2-10 *sccm*) | CVD SCD on *p*-type Si | NV | 3.9 – 4.1 | 745 – 755 | 0.7 – 0.9 | [106] |
| 2022 | $CH_4$ (20 *sccm*) | $N_2O$(480 *sccm*) | Ib-type HPHT (100) SCD | NV | 19 – 21 | 995 – 1005 | 2.9 – 3.1 | [29] |
| 2022 | $CH_4$:$H_2$ (6.8:180 *sccm*) | $N_2$(2 *sccm*) | Carbon films with SCDNs on (100) Si | NV | 9.4 – 10.2 | 895 – 905 | 1.8 – 2.2 | [60] |
| 2023 | $CH_4$:$H_2$ (4: 200 *sccm*) | $N_2$(5 *sccm*) | Ib-type HPHT (100) SCD | NV | 10 – 12 | 1085 – 1095 | 1.3 – 1.5 | [78] |
| 2024 | $CH_4$:$H_2$ (3: 300 *sccm*) | $C_{20}H_{19}N$ | Tantalum | NV | 2.6 – 2.7 | 745 – 755 | 1.2 – 1.4 | [105] |
| 2024 | $CH_4$ (9 *sccm*) | $N_2$ (150 *sccm*) | (001) CVD diamond | NV | 0.9 – 1.1 | 945 – 955 | 5.8 – 6.2 | [44] |
| 2017 | $CH_4$:$H_2$ (24: 375 *sccm*), | $SiH_4$(1 *sccm*) | HPHT thinfilm | SiV | 17 – 18 | 845 – 855 | 3.1 – 3.5 | [109], [111] |
| 2017 | $CH_4$:$H_2$ (20: 378 *sccm*) | $SiH_4$(0.8-2.4 *sccm*) | (100) CVD SCD d or Ib-type MPCVD SCD | SiV | 2.9 – 3.3 | 345 – 355 | 0.3 – 0.5 | [56] |
| 2019 | $CH_4$:$H_2$ (8:192 *sccm*) | Si substrate | CVD SCD on Si | SiV | 9 – 10 | 895 – 905 | 0.4 – 0.5 | [55] |
| 2021 | $CH_4$:$H_2$ (4: 400 *sccm*) | $SiH_4$(5-30 *sccm*) | Si doped diamond or SiC | SiV | 2.8 – 3.2 | 655 – 665 | 5 – 7 | [108] |
| 2021 | $CH_4$:$H_2$ (20: 400 *sccm*) | $Si(CH_3)_4$(1 – 30 *sccm*) | (100) Si | SiV | 5.5 – 6.5 | 855 – 865 | 5.8 – 6.2 | [110] |
| 2022 | $CH_4$:$H_2$ (8: 192 *sccm*) | Carbon films with SCDNs | (100) Si | SiV | 9.6 – 10 | 655 – 665 | 1.8 – 2.2 | [60] |
| 2024 | $CH_4$:$H_2$ (24: 364 *sccm*) | $SiH_4$(0.07 *sccm*) | IIa-type HPHT SCD | SiV | 22 – 23 | 1045 – 1055 | 3.4 – 3.8 | [112] |
| 2024 | $CH_4$:$H_2$ (18: 300 *sccm*) | $SiH_4$(1 *sccm*) | Ib-type MPCVD SCD | SiV | 3.3 – 3.9 | 655 – 665 | 5 – 7 | [113] |
| 2015 | $CH_4$:$H_2$ (2:198 *sccm*) | Ge | Ib-type diamond, Element six | GeV | 5 – 7 | 815 – 825 | 2 – 3 | [70] |
| 2018 | $CH_4$:$H_2$ (1: 100 *sccm*) | Ge/$GeO_2$ | (100) CVD SCD | GeV | 7.8 – 8.2 | 345 – 355 | 0.85 – 0.95 | [118] |
| 2019 | $CH_4$:$H_2$ (8:192 *sccm*) | Ge | CVD SCD on Si | GeV | 9.1 – 9.9 | 895 – 905 | 0.4 – 0.5 | [55] |
| 2019 | $CH_4$:$H_2$ (4: 400 *sccm*) | Ge/$GeO_2$ | SCD on sapphire | GeV | 7.5 – 8.5 | 995 – 1005 | 0.8 – 1 | [119] |
| 2020 | $CH_4$:$H_2$ (1: 100 *sccm*) | $GeO_2$ | Sapphire | GeV | 10 – 11 | 995 – 1005 | 0.8 – 1 | [116] |
| 2021 | $CH_4$:$H_2$ (20: 478 *sccm*) | $GeH_4$ (2 *sccm*) | HPHT diamond | GeV | 11 – 11.6 | 845 – 855 | 2.4 – 2.8 | [120] |
| | | | | | 17 – 18 | 985 – 995 | | |
| 2022 | $CH_4$:$H_2$ (8:192 *sccm*) | Ge | Carbon films with SCDNs on (100) Si | GeV | 9.6 – 10 | 895 – 905 | 1.8 – 2.2 | [60] |
| 2022 | $CH_4$:$H_2$ (15:233 *sccm*) | $GeH_4$ (2 *sccm*) | (100) Si | GeV | 14 – 15 | 795 – 805 | 4.7 – 5.1 | [57] |
| 2023 | $CH_4$:$H_2$ (20:480 sccm) | $GeH_4$ (0.3 *sccm*) | IIa-type diamond Element six-SCD | GeV | 8 – 9 | 845 – 855 | 4.8 – 5.2 | [117] |
| 2020 | $CH_4$:$H_2$ (1: 100 *sccm*) | $SnO_2$ /$SnCl_2$ | Sapphire | SnV | 10 – 11 | 995 – 1005 | 0.8 – 1 | [116] |
| 2020 | $CH_4$:$H_2$ (0.5: 300 *sccm*) | $^{120}Sn^+$ | Electronic-grade diamond, Element Six | SnV | 2.8 – 3.4 | 645 – 655 | 0.9 – 1.3 | [62] |

**Table ST3.** chemical vapor deposition (CVD) and Microwave Plasma Chemical Vapor Deposition (MCVD). Relevant information includes the carbon/color center sources, the type of substrate and experimental conditions such as chamber pressure, substrate temperature, and power of the microwave (MW) plasma reported in the literature for four different DVCCs: NV, SiV, GeV and SnV. Chamber pressure and substrate temperature values are in the range 0.9–18 kPa and 345–1,095 °C, respectively, while the power of the MW plasma is in the range 0.3–7 kW. The CVD/MPCVD-relevant database includes a total of 600 data points.

**Table ST4.** Ion implantation, sub-database

| Year | Ion source | Substrate | Color centers | Implantation energy range(keV) | Fluences range (ions/$cm^2$) | Temperature range (°C) | Annealing time range (min) | References |
|---|---|---|---|---|---|---|---|---|
| 2016 | $He^+$ | IIa-type (001) MPCVD SCD | NV | 3.9 – 4.1 | $9.0\times10^{10} – 1.1\times10^{11}$ | 895 – 905 | 118 – 122 | [146] |
| 2019 | $^{14}N^+$ | IIa- type (111) HPHT diamond | NV | 29 – 31 | $9.0\times10^{10} – 1.1\times10^{11}$ | 745 – 755 | 28 – 32 | [28] |
| 2019 | $^{12}C^+$ | Ib-type SCD | NV | 2.9 – 3.1 | $4.9\times10^{12} – 5.1\times10^{12}$ | 1195 – 1205 | 118 – 122 | [146] |
| 2019 | $N^+$ | Quantum grade diamond, Element Six | NV | 39 – 41 | $9.0\times10^{9} – 1.1\times10^{10}$ | 795 – 805 | 239 – 241 | [122] |
| 2020 | $N^+$ | (001) CVD diamond | NV | 34 – 36 | $2.9\times10^{9} – 3.1\times10^{9}$ | 895 – 905 | 239 - 241 | [123] |
| 2020 | $^{15}N^+$ | Electronic-grade (100) diamond-Element Six | NV | 4 – 6 | $9.0\times10^{11} – 1.1\times10^{12}$ | 995 – 1005 | 178 – 182 | [124] |
| 2021 | $^{14}N^+$ | IIa-type (100) CVD diamond | NV | 29 – 31 | $9.0\times10^{9} – 1.1\times10^{10}$ | 785 – 805 | 118 – 122 | [125] |
| 2022 | $^{12}C^+$ | Electronic grade diamond, Element-Six | NV | 11 – 13 | $9.0\times10^{10} – 1.1\times10^{10}$ | 785 – 805 | 118 – 122 | [76] |
| | | | NV | 49 – 51 | $4.9\times10^{8} – 5.1\times10^{8}$ | 795 – 805 | 118 – 122 | |
| 2023 | $^{15}N^+$ | Electronic grade diamond-Element Six | NV | 5 – 7 | $9.0\times10^{10} – 1.1\times10^{11}$ | 1195 – 1205 | 118 – 122 | [126] |
| 2024 | $N_2^+$ | Ib-type MPCVD and HPHT diamond | NV | 39 – 41 | $1.9\times10^{14} – 2.1\times10^{14}$ | 795 – 805 | 118 – 122 | [127] |
| 2016 | $^{29}Si^+$ | Ib-type CVD diamond Element-Six | SiV | 145 – 155 | $9.0\times10^{9} – 1.1\times10^{10}$ | 1195 – 1205 | 115 – 125 | [128] |
| 2018 | $Si^+$ | CVD SCD, Element-Six | SiV | 395 – 405 | $6.5\times10^{10} – 7.0\times10^{10}$ | 1195 – 1205 | 115 – 125 | [129] |
| 2018 | $Si^+$ | CVD SCD, Element-Six | SiV | 70 – 74 | $3.0\times10^{13} – 3.4\times10^{13}$ | 1095 – 1105 | 59 – 61 | [130] |
| 2020 | $Si^+/Si^{2+}/Si^{3+}$ | P-doped MPCVD diamond on HPHT | SiV | 4.5 – 5.5 | $9.0\times10^{9} – 1.1\times10^{10}$ | 1495 – 1505 | 85 – 95 | [77] |
| | $^{28}Si^+/^{29}Si^+/^{30}Si^+$ | IIa-type electronic grade diamond, Element-Six | | 4.5 – 5.5 | $1.9\times10^{10} – 2.1\times10^{10}$ | 1495 – 1505 | 85 – 95 | |
| | | | | | $4.9\times10^{11} – 5.1\times10^{11}$ | | | |
| | | | | | $4.9\times10^{12} – 5.1\times10^{12}$ | | | |
| 2018 | $Si^{3+}$ | CVD SCD | SiV | 9 – 11 | $9.0\times10^{11} – 1.1\times10^{12}$ | 1045 – 1055 | 115 – 125 | [72] |
| | | | | | $4.9\times10^{12} – 5.1\times10^{12}$ | | | |
| | | | | | $9.0\times10^{12} – 1.1\times10^{13}$ | | | |
| | | | | | $9.0\times10^{13} – 1.1\times10^{14}$ | | | |
| | | | | 19 – 21 | $9.0\times10^{11} – 1.1\times10^{12}$ | 1045 – 1055 | | |
| | | | | | $4.9\times10^{12}– 5.1\times10^{12}$ | | | |
| | | | | | $9.0\times10^{12} – 1.1\times10^{13}$ | 995 – 1005 | | |
| | | | | | $9.0\times10^{13} – 1.1\times10^{14}$ | | | |
| | | | | 49 – 51 | $9.0\times10^{11} – 1.1\times10^{12}$ | 1045 – 1055 | | |
| | | | | | $4.9\times10^{12} – 5.1\times10^{12}$ | | | |
| | | | | | $9.0\times10^{12} – 1.1\times10^{13}$ | 995 – 1005 | | |
| | | | | | $9.0\times10^{13} – 1.1\times10^{14}$ | | | |
| | | | | 99 – 101 | $9.0\times10^{11} – 1.1\times10^{12}$ | 1045 – 1055 | | |
| | | | | | $4.9\times10^{12} – 5.1\times10^{12}$ | | | |
| | | | | | $9.0\times10^{12} – 1.1\times10^{13}$ | 995 – 1005 | | |
| | | | | | $9.0\times10^{13} – 1.1\times10^{14}$ | | | |
| 2023 | $^{29}Si^+$ | CVD SCD, Element-Six | SiV | 65 – 75 | $9.0\times10^{8} – 1.1\times10^{9}$ | 1195 – 1205 | 115 – 125 | [132] |
| 2024 | $^{28}Si^+$ | HPHT and CVD SCD | SiV | 17995 – 18005 | $6.0\times10^{12} – 7.0\times10^{12}$ | 1795 – 1805 | 0.9 – 1.1 | [133] |
| 2024 | $^{28}Si^+$ | CVD SCD, Element-Six | SiV | 130 – 134 | $2.9\times10^{11} – 3.1\times10^{11}$ | 1095 – 1105 | 595 – 605 | [134] |
| 2018 | $Si^+$ | CVD SCD, Element Six | SiV | 95 – 97 | $3.0\times10^{13} – 3.4\times10^{13}$ | 1095 – 1105 | 118 – 122 | [130] |
| | | | | 2995 – 3005 | $6.5\times10^{10} – 7.0\times10^{10}$ | | 115 – 125 | |
| 2018 | $Si^{3+}$ | CVD SCD | SiV | 19 – 21 | $9.0\times10^{7} – 1.1\times10^{8}$ | 1195 – 1205 | 59 – 61 | [131] |
| 2020 | $Si^+/Si^{2+}/Si^{3+}$ | *P*-doped MPCVD diamond on HPHT | SiV | 19 – 21 | $9.0\times10^{11} – 1.1\times10^{12}$ | 1195 – 1205 | 59 – 61 | |
| | | | | | $9.0\times10^{12} – 1.1\times10^{13}$ | | | |
| | | | | | $9.0\times10^{13} – 1.1\times10^{14}$ | | | |
| 2015 | $Ge^+$ | IIa-type (001) SCD Element-Six | GeV | 145 – 155 | $3.0\times10^{8} – 4.0\times10^{8}$ | 795 – 805 | 25 – 35 | [70] |
| | | | | | $5.0\times10^{13} – 6.5\times10^{13}$ | | | |
| | | | | 250 – 270 | $3.0\times10^{8} – 4.0\times10^{8}$ | | | |
| | | | | | $5.0\times10^{13} – 6.5\times10^{13}$ | | | |
| 2021 | $Ge^+$ | MPCVD SCD on Si | GeV | 95 – 105 | $9.0\times10^{13} – 1.1\times10^{14}$ | 895 – 905 | 115 - 125 | [135] |
| 2024 | $^{75}Ge^+$ | (100) SCD-Element-Six | GeV | 25 – 35 | $1.5\times10^{12} – 2.5\times10^{12}$ | 895 – 905 | 8 – 12 | [75] |
| | | | | | $1.5\times10^{13} – 2.5\times10^{13}$ | | | |
| 2024 | $^{74}Ge^+$ | HPHT and CVD SCD | GeV | 17000 – 19000 | $1.4\times10^{13} – 1.6\times10^{13}$ | 595 – 605 | 595 – 605 | [133] |
| | | | | | | 695 – 705 | 595 – 605 | |
| | | | | | | 1700 – 1900 | 0.5 – 1.5 | |
| 2017 | $Sn^+$ | CVD SCD | SnV | 128 – 132 | $1.9\times10^{8} – 2.1\times10^{8}$ | 795 – 805 | 28 – 32 | [136] |
| | | | | 148 – 152 | $1.9\times10^{13} – 2.1\times10^{13}$ | 2095 – 2105 | | |
| 2017 | $Sn^+$ | IIa-type SCD Element-Six | SnV | 58 – 62 | $2.9\times10^{11} – 3.1\times10^{11}$ | 895 – 905 | 28 – 32 | [73] |
| | | | | | $1.9\times10^{10} – 2.1\times10^{10}$ | 945 – 955 | 115 – 125 | |
| | | | | | $4.9\times10^{10}– 5.1\times10^{10}$ | | | |
| | | | | | $9.0\times10^{10} – 1.1\times10^{11}$ | | | |
| | | | | 9995 – 10005 | $4.9\times10^{13} – 5.1\times10^{13}$ | | | |
| 2019 | $Sn^+$ | Quantum grade diamond-Element Six | SnV | 78 – 82 | $2.9\times10^{11} – 3.1\times10^{11}$ | 1195 – 1205 | 475 – 485 | [122] |
| 2019 | $^{120}Sn^+$ | CVD SCD-Element Six | SnV | 365 – 375 | $1.9\times10^{11} – 2.1\times10^{11}$ | 1095 – 1105 | 88 – 92 | [87] |
| 2020 | $^{120}Sn^+$ | Electronic-grade diamond Element Six | SnV | 365 – 375 | $1.9\times10^{13} – 2.1\times10^{13}$ | 1095 – 1105 | 28 – 32 | [62] |
| 2020 | $Sn^+$ | (001) CVD SCD | SnV | 75 – 77 | $9.0\times10^{9} – 1.1\times10^{10}$ | 1195 – 1205 | 235 – 245 | [123] |
| | | | | | $9.0\times10^{10} – 1.1\times10^{11}$ | | | |
| 2020 | $^{121}Sn^+$ | IIa-type (110) Dutch Diamond Technologies | SnV | 58 – 62 | $1.9\times10^{12} – 2.1\times10^{12}$ | 915 – 925 | 8 – 12 | [138] |
| 2020 | $Sn^+$ | Electronic-grade (001) diamond | SnV | 79 – 81 | $1.9\times10^{13} – 2.1\times10^{13}$ | 1195 – 1205 | 235 – 245 | [137] |
| | $^{120}Sn^+$ | Electronic-grade diamond-Element Six | | 695 – 705 | | 2095 – 2105 | 28 – 32 | |
| 2021 | $Sn^+$ | IIa-type (100) diamond | SnV | 58 – 62 | $1.9\times10^{10} – 2.1\times10^{10}$ | 1195 – 1205 | 115 – 125 | [121] |
| 2021 | $Sn^+$ | IIa-type CVD SCD-Element Six | SnV | 345 – 355 | $9.0\times10^{8} – 1.1\times10^{9}$ | 1195 -1205 | 118 – 122 | [65] |
| 2022 | $Sn^+$ | Electronic-grade (001) diamond Element-Six | SnV | 695 – 705 | $9.0\times10^{8} – 1.1\times10^{9}$ | 2095 – 2105 | 115 – 125 | [90] |
| | | | | | $9.0\times10^{11} – 1.1\times10^{12}$ | | | |
| | | | | | $7.9\times10^{13} – 8.1\times10^{13}$ | | | |
| 2024 | $^{117}Sn^+$ | Electronic-grade diamond-Element Six | SnV | 345 - 355 | $9.0\times10^{10} – 1.1\times10^{11}$ | 1195 – 1205 | 715 – 725 | [74] |

**Table ST4.** Ion implantation. Relevant information includes the ion source, substrate, implantation energy, fluence, temperature and annealing time reported for the different DVCCs. Implantation energies and fluences are in the range 3.9–19,000 keV and $1.9\times10^{8}–2.1\times10^{14}$ ions/$cm^2$, respectively; the temperature and annealing time are in the range 595–2,105 °C and 0.9–725 minutes, respectively. The Ion irradiation-relevant database includes a total of 284 data points.

**Table ST5.** Electron/ion irradiation, sub-database

| Year | Ion source | Substrate | Color centers | Implantation energy range (MeV) | Fluences range (ions/cm$^2$) | Temperature range (°C) | Annealing time range (min) | References |
|---|---|---|---|---|---|---|---|---|
| 2017 | Electron beam | CVD SCD, Element-Six | NV | 4.3 – 4.7 | $5.0\times10^{18}$ – $6.0\times10^{18}$ | 1195 – 1205 | 118 – 122 | [85] |
| 2019 | Electron beam | CVD SCD | NV | 1.8 – 2.2 | $4.9\times10^{13}$ – $5.1\times10^{13}$ | 1095 – 1105 | 598 – 602 | [141] |
| 2020 | Electron beam | SCD-Sumitomo Electric | NV | 4.4 – 4.8 | $9.0\times10^{17}$– $1.1\times10^{18}$ | 795 – 805 | 58 – 62 | [140] |
| 2021 | Au$^+$ ion | Ib-type diamond-Element Six | NV | 2.2 – 2.6 | $7.9\times10^{16}$ – $8.1\times10^{16}$ | 1045 – 1055 | 148 – 152 | [95] |
| | | | | | $3.9\times10^{17}$ – $4.1\times10^{17}$ | | | |
| | | | | | $3.9\times10^{18}$– $4.1\times10^{18}$ | | | |
| | | | | | $9.0\times10^{18}$ – $1.0\times10^{19}$ | | | |
| 2022 | Electron beam | MPCVD SCD on Ib-type HPHT (001) diamond | NV | 1.9 – 2.1 | $9.0\times10^{15}$– $1.1\times10^{16}$ | 995 – 1005 | 118 - 122 | [78] |
| | | | | | $1.9\times10^{16}$ – $2.1\times10^{16}$ | | | |
| | | | | | $9.0\times10^{16}$ – $1.1\times10^{17}$ | | | |
| | | | | | $1.9\times10^{17}$– $2.1\times10^{17}$ | | | |
| 2022 | Electron beam | $N_2$ MPCVD SCD on IIa-type HPHT diamond | NV | 0.9 – 1.1 | $9.0\times10^{16}$– $1.1\times10^{17}$ | 995 – 1005 | 118 – 122 | [41] |
| | | | | | $2.9\times10^{17}$– $3.1\times10^{17}$ | | | |
| | | | | | $9.0\times10^{17}$ – $1.1\times10^{18}$ | | | |
| | | | | | $2.9\times10^{18}$ – $3.1\times10^{18}$ | | | |
| 2022 | Electron beam | HPHT (100) diamond | NV | 1.9 – 2.1 | $9.0\times10^{17}$ – $1.1\times10^{18}$ | 995 – 1005 | 118 – 122 | [42] |
| | | | | 1.8 – 2.2 | $9.0\times10^{16}$ – $1.1\times10^{17}$ | | | |
| | | | | | $9.0\times10^{17}$ – $1.1\times10^{18}$ | | | |
| 2022 | Electron beam | Ib-type (100) SCD | NV | 2.1 – 2.5 | $2.9\times10^{18}$ – $3.1\times10^{18}$ | 795 – 805 | 118 – 122 | [29] |
| | | | | 2.0 – 2.6 | | 795 – 805 | 118 – 122 | |
| | | | | | | 98 – 102 | 58 – 62 | |
| | | | | | | 1195 – 1205 | 58 – 62 | |
| 2023 | Electron beam | $N_2$ MPCVD SCD | NV | 9 – 11 | $2.2\times10^{18}$ – $2.6\times10^{18}$ | 845 – 855 | 238 – 242 | [30] |
| 2023 | Electron beam | HPHT SCD-Element Six | NV | 0.18 – 0.22 | $1.4\times10^{18}$ – $1.6\times10^{18}$ | 795 – 805 | 238 – 242 | [142] |
| | | | | | $1.9\times10^{18}$ – $2.1\times10^{18}$ | | | |
| | | | | | $9.0\times10^{18}$ – $1.1\times10^{19}$ | | | |
| | | | | | $4.9\times10^{19}$ – $5.1\times10^{19}$ | | | |
| | | | | | $9.0\times10^{19}$ – $1.1\times10^{20}$ | | | |
| | | | | | $1.9\times10^{20}$ – $2.1\times10^{20}$ | | | |
| | | | | | $4.9\times10^{20}$ – $5.1\times10^{20}$ | | | |
| | | | | 153 – 157 | $1.4\times10^{18}$– $1.6\times10^{18}$ | | | |
| 2024 | Electron beam | Fe-Co-Ni HPHT diamond | NV | 9 – 11 | $9.0\times10^{16}$– $1.1\times10^{17}$ | 795 – 805 | 58 – 62 | [31] |
| 2021 | Electron beam | 5nm DNDs | SiV | 1.8 – 2.2 | $4.9\times10^{18}$– $5.1\times10^{18}$ | 1095 – 1105 | 59 – 61 | [144] |
| 2023 | Electron beam | IIa-type HPHT SCD | SiV | 1.8 – 2.2 | $1.9\times10^{18}$ – $2.1\times10^{18}$ | 1595 – 1605 | 59 – 61 | [86] |
| 2024 | He$^+$ ion | MPCVD SCD | SiV | 1.5 – 1.9 | $2.0\times10^{15}$ – $3.0\times10^{15}$ | 795 – 805 | 118 – 122 | [112] |
| 2021 | Ge$^+$ ion | MPCVD SCD on (100) Si | GeV | 0.08 – 0.12 | $9.0\times10^{14}$ – $1.1\times10^{15}$ | 895 – 905 | 118 – 122 | [135] |

**Table ST5.** Electron/ion irradiation. Relevant information includes the ion source, substrate, implantation energy and fluence, as well as temperature and annealing time reported in the literature for three different DVCCs: NV, SiV and GeV. Implantation energies and fluences are in the range 0.08–157 MeV and $4.9\times10^{13}$–$5.1\times10^{20}$ ions/cm$^2$, respectively; the temperature and annealing time are in the range 98–1,605 °C and 58–602 minutes, respectively. The Ion irradiation-relevant database includes a total of 128 data points.

## S3. Hyperparameters of the DTR and XGB models

Table ST6 lists the hyperparameters with their value ranges, for the DTR and XGB models, including the subsample ratio instance, tree depth, sampling, and learning rate of full training. The hyperparameters alpha (L1) and lambda (L2) are regularization terms that appear in the regularization function $\Omega(f_t)$ in eq. (4) of the Main text and are applied to control the complexity of the model and to avoid overfitting.

**Table ST6.** Hyperparameters for the DTR/XGB predictors

| Model | Hyperparameter | Search range | Optimum value/feature | Model | Hyperparameter | Search range | Optimum value/feature |
|---|---|---|---|---|---|---|---|
| Decision-Tree Regressor | Criterion | mse friedman_mse mae | friedman_mse | XGBoosting | booster | gbtree gblinear | gbtree |
| | splitter | best random | random | | objective | reg, rank, binary, survival multi … | reg: linear |
| | max_features | aqrt, auto, None | None | | learning_rate | 0 – 1 | 0.06 |
| | max_depth | 10 – 100 | 10 | | Eval_metric | logloss | rmse |
| | min_samples_split | 2 – 10 | 2 | | max_depth | - | 6 |
| | min_samples_leaf | 1 – 5 | 1 | | subsample | 0 – 1 | 0.8 |
| | min_weight_fraction_leaf | 0 - 1000 | 0 | | Colsample_bytree | 0 – 1 | 1 |
| | max_leaf_nodes | None, random | None | | lambda | 0 – 1000 | 10 |
| | random_state | None, random | 42 | | alpha | 0 – 1000 | 0 |
| | | | | | seed | 0 – 100 | 66 |

**Table ST6.** List and value ranges of the hyperparameters of the trained DTR and XGB predictors.

## S4. Handling of categorical variables in the models

In this work, *categorical* factors such as the carbon, dopant and ion sources, as well as the substrate have been intentionally excluded. This is to maintain dataset consistency across different fabrication routes and to emphasize the quantitative trends driven by controllable experimental variables, e.g., temperature, pressure, fluence, etc. This is primarily done for practical reasons (see Main text, § 2.3.2). Regardless, while this simplification may introduce certain model biases and limit generalizability, the consistent predictive ability of our models indicates that the trained algorithms are still able to successfully capture principal physical correlations. For context and completeness, below we show the typical substrate, precursor materials and ion sources used across the HPHT, MPCVD, ion implantation, and electron/ion irradiation processes.

**Substrate.** Fabrication protocols largely share similar base materials, typically single- or poly-crystalline diamond substrates obtained from HPHT or CVD growth. The relative consistency in substrate choice and characteristics across methods suggests that the substrate does not play a major role in the observed variability of synthesized materials. The formation of color centers seems to be predominantly governed by differences in process parameters such as temperature, pressure, gas composition, and irradiation conditions.

**Precursors and Dopants.** Figure SF2 summarizes the representative precursors and dopant sources used in HPHT synthesis and color-center formation. The carbon precursors—such as graphite, naphthalene ($C_{10}H_8$), and dipentene ($C_{10}H_{16}$)—provide the fundamental carbon feedstock, while diverse dopant sources including $NaN_3$/$Ba(N_3)_2$, TiN, metallic Ge, tetraethoxysilane, and $Sn_{20}Mg_{80}$ supply the atomic species necessary for forming nitrogen-, silicon-, germanium-, and tin-related vacancy centers. Notably, even when identical precursors are used, variations in annealing temperature or irradiation fluence can profoundly affect vacancy diffusion, dopant-vacancy pairing efficiency, and charge-state stability. The precursors are (chemically) essential to establish and initiate the synthesis process, but the process parameters are the driving factors that determine the variability observed in the physical properties of the material and the different color center.

Figure SF3 shows the substrates and main precursor gases used in CVD/MPCVD growth. The composition of the precursor gas mixture influences both crystal quality and DVCC incorporation. Precursors are needed both for diamond growth (e.g. methane, $CH_4$, to provide the carbon source, and hydrogen, $H_2$, to suppress the formation of graphitic phases and stabilize $sp^3$ surface reconstruction) and for DVCCs' incorporation (e.g., nitrogen, $N_2$; silane, $SiH_4$, and germane, $GeH_4$, for the incorporation of NV, SiV and GeV centers, respectively). The precursor gas chemistry thus governs the overall diamond growth mechanism and potentially provides a tunable parameter for controlling formation pathways, concentrations, and charge states of specific color centers. Yet, beyond precursor composition, the partial pressures of the precursors and the substrate temperature are key parameters affecting both the thermodynamic driving force for incorporation and the kinetics of defect stabilization (for completeness, trace additions of oxygen- or inert-containing gases can further modify vacancy formation energies, etch rates, and surface diffusion dynamics).

**Irradiation.** While different particles—mainly ions or electrons in our analysis—create different damage profiles and efficiencies, defect creation and DVCC yield seems to be mostly determined by irradiation energy, fluence, and annealing temperature and time. Specifically, irradiation energy is key to inducing the formation of vacancies. Fluence allows controlling the concentration of DVCCs, which generally increases monotonically with the dose, up to a saturation point, after which damage aggregation starts to occur and yield is reduced. Annealing temperature and time drive vacancy migration with optimal temperatures promoting DVCC formation and suppressing unwanted defects.

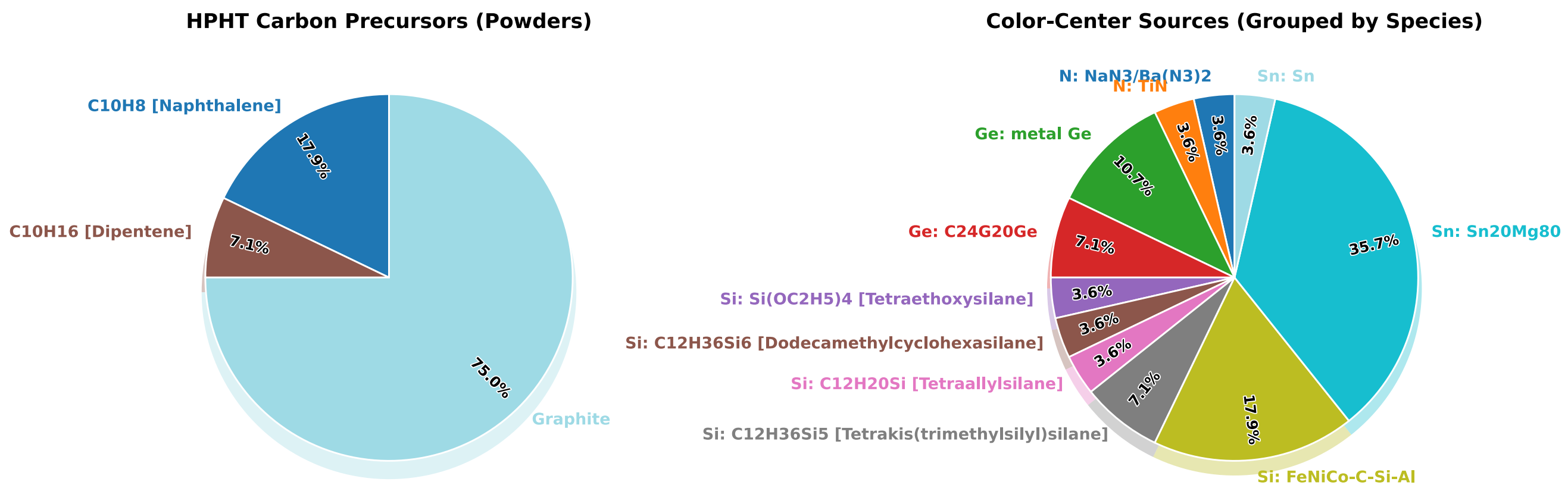

**Figure SF2**. Representative precursors and dopant sources used for HPHT synthesis and DVCC formation. **a)** Composition of carbon precursors employed in HPHT growth, mainly consisting of graphite (21%), naphthalene $C_{10}H_8$ (5%), and dipentene $C_{10}H_{16}$ (2%), which collectively provide both carbon and hydrocarbon species under high pressure high temperature conditions. **b)** Distribution of color-center precursor materials across different dopant types, including nitrogen sources ($NaN_3/Ba(N_3)_2$, TiN), germanium sources (metal Ge, $C_{24}H_{20}Ge$), silicon sources ($Si(OC_2H_5)_4$, $C_{12}H_{36}Si_6$, $C_{12}H_{20}Si$, $C_{12}H_{36}Si_5$), and tin sources ($Sn_{20}Mg_{80}$, Sn), as well as multi-component metallic mixtures (FeNiCo-C-Si-Al).

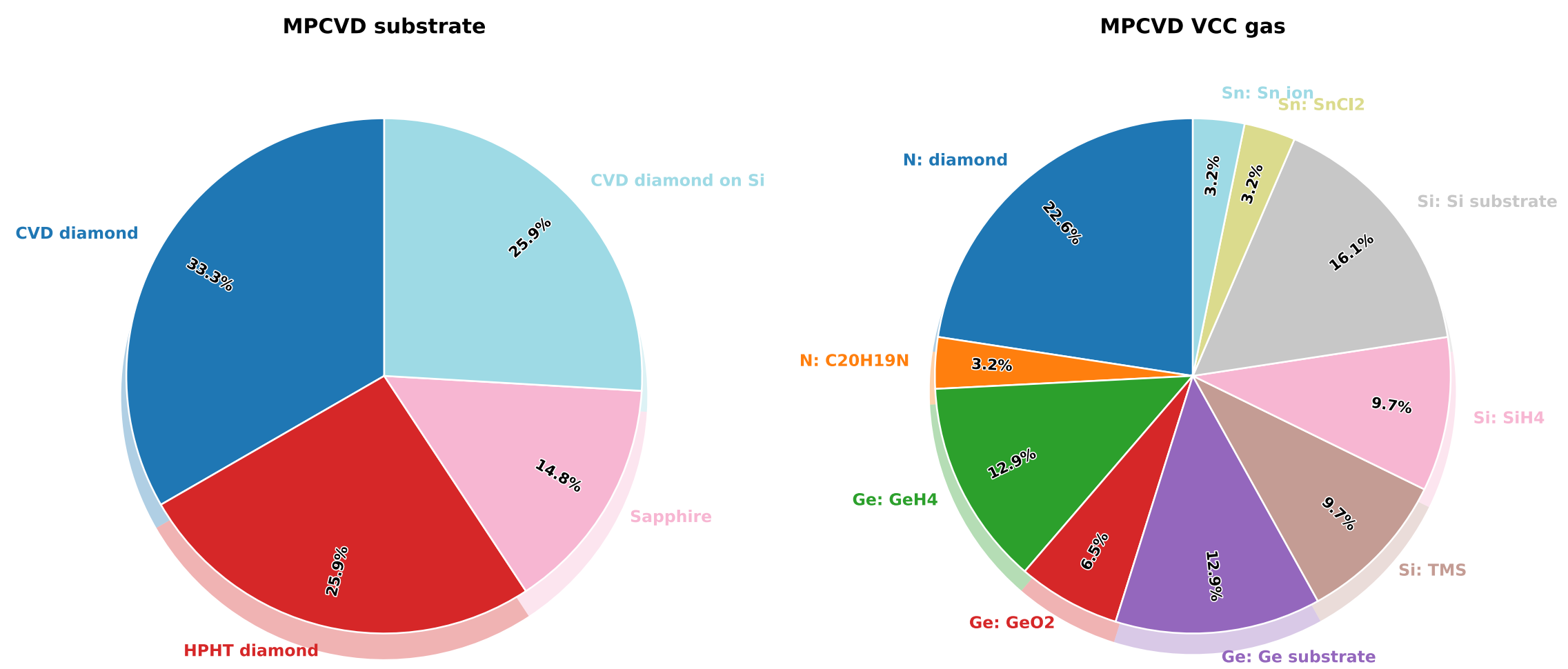

**Figure SF3**. Representative diamond substrate and precursor gas used for MPCVD. **a)** different substrate employed in MPCVD growth, mainly consisting of CVD diamond (33.3%), CVD diamond growth on Si (25.9%), HPHT diamond synthesis (25.9%) and growth of diamond on sapphire (14.8 %). **b)** Distribution of color-center precursor gas across different dopant types, including NV (25.8%), GeV (32.3%), SiV (35.5%), and SnV(6.4%).

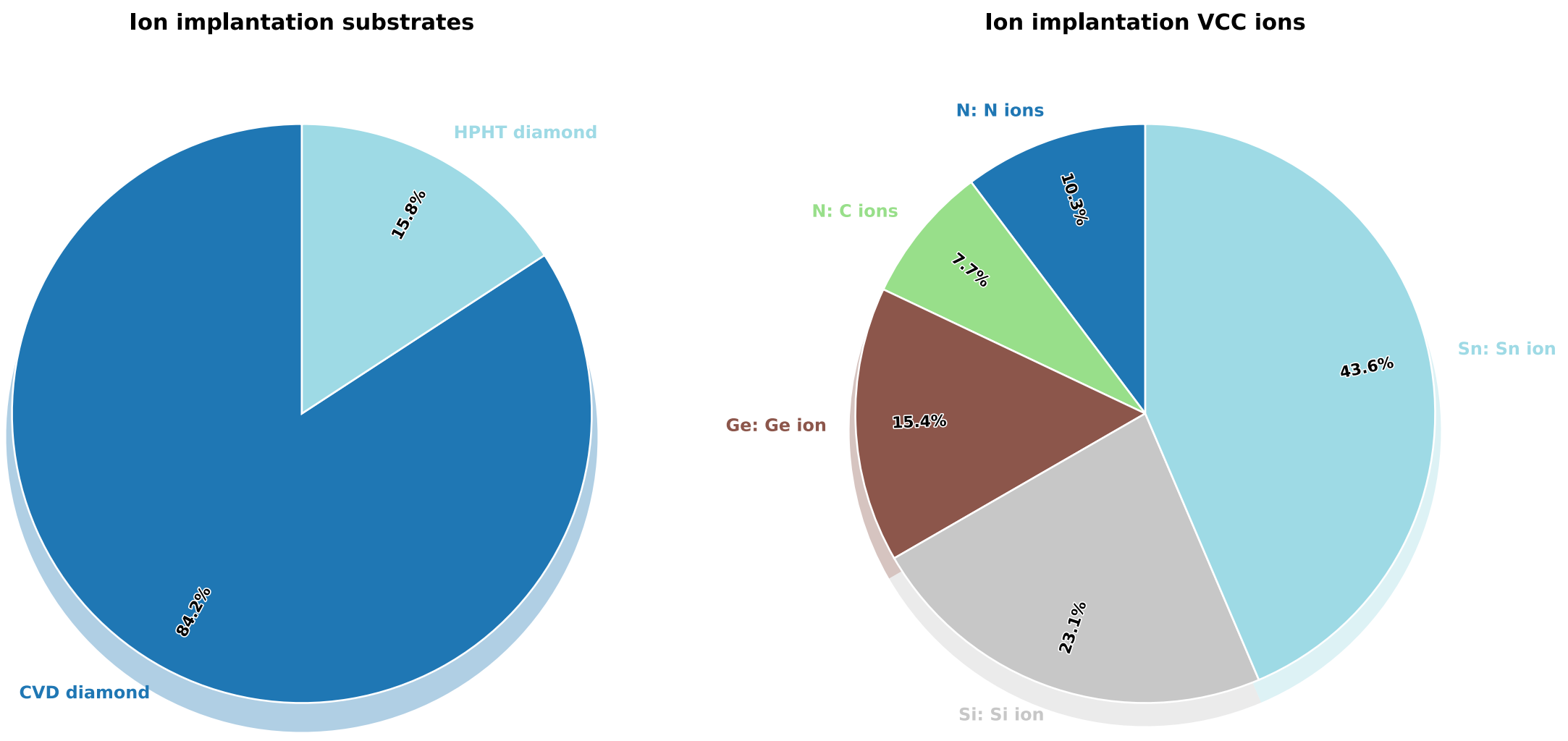

**Figure SF4**. Representative diamond substrate and color center source used for ion implantation. **a)** different substrates employed for ion implantation, mainly consisting of CVD diamond (84.2 %) and HPHT diamond (15.8%). **b)** Distribution of ion sources across different DVCCs: NV (18.0%), GeV (15.4%), SiV (23.1%), and SnV(43.6%).

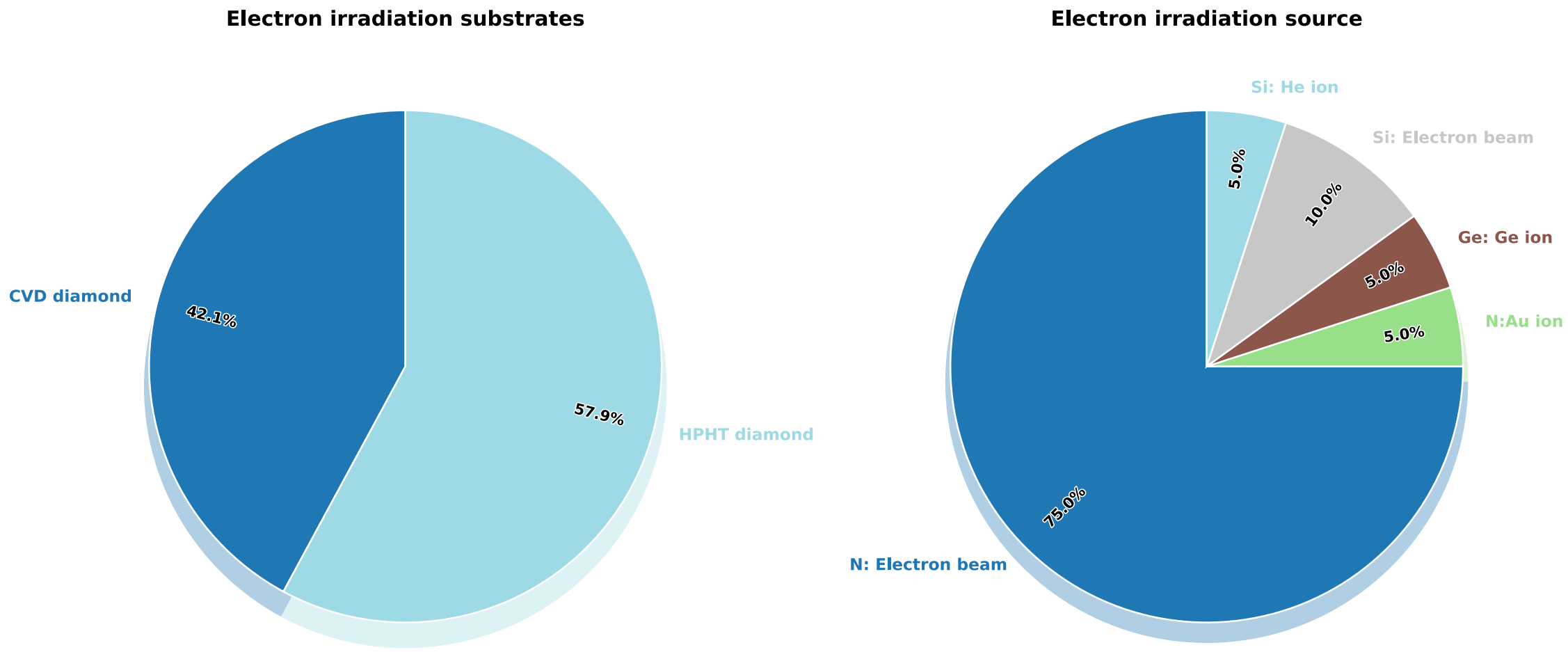

**Figure SF5**. Representative diamond substrates and color center sources used for electron/ion irradiation. **a)** different substrates employed for ion implantation: CVD diamond (42.1 %) and HPHT diamond (57.9%). **b)** Distribution of beam irradiation sources across different DVCCs: NV (80.0%), GeV (5.0%), SiV (23.1%), and SnV(43.6%).

## S5. Performance of the DTR and XGB models

Tables ST7 and ST8 below summarize the values of the statistical indicators, $R^2$, MSE and MAE we used to assess the performance of the DTR and XGB predictors for *Task I* and *II*, for each technique: HPHT, CVD/MPCVD, Ion implantation and Electron/ion irradiation, respectively (see Main text).

**Table ST7.** *Task I*: DTR/XGB performance for each synthesis method

| | | $R^2$ | | MSE | | MAE | |
|---|---|---|---|---|---|---|---|
| | **DVCC** | **DTR** | **XGB** | **DTR** | **XGB** | **DTR** | **XGB** |
| **HPHT** | **NV** | (1.000)<br>1.000 | (1.000)<br>0.9971 | (0.0058)<br>0.0066 | (0.0000)<br>7.1554 | (0.0551)<br>0.0614 | (0.0019)<br>0.1823 |
| | **SiV** | (1.000)<br>1.000 | (1.000)<br>1.000 | (0.2013)<br>0.2026 | (0.0004)<br>4.7198 | (0.2416)<br>0.2432 | (0.0111)<br>0.0893 |
| | **GeV** | (1.000)<br>1.000 | (1.000)<br>1.000 | (0.0295)<br>0.0300 | (0.0001)<br>0.6861 | (0.0888)<br>0.0918 | (0.0056)<br>0.0465 |
| | **SnV** | (1.000)<br>1.000 | (1.000)<br>1.000 | (0.0491)<br>0.0725 | (0.0038)<br>0.2652 | (0.1687)<br>0.1972 | (0.0281)<br>0.0503 |
| **CVD/MPCVD** | **NV** | (0.9975)<br>0.9584 | (0.9975)<br>0.8975 | (0.0077)<br>0.1282 | (0.0077)<br>0.3162 | (0.0105)<br>0.0958 | (0.0120)<br>0.2360 |
| | **SiV** | (1.000)<br>0.9995 | (1.000)<br>0.9543 | (0.0000)<br>0.0028 | (0.0000)<br>0.2326 | (0.0000)<br>0.0301 | (0.0017)<br>0.1410 |
| | **GeV** | (1.000)<br>0.9979 | (1.000)<br>0.9791 | (0.0000)<br>0.0050 | (0.0000)<br>0.0497 | (0.0000)<br>0.0130 | (0.0025)<br>0.0337 |
| | **SnV** | (1.000)<br>0.9856 | (0.9999)<br>0.8227 | (0.0000)<br>0.0003 | (0.0000)<br>0.0034 | (0.0000)<br>0.0158 | (0.0008)<br>0.0331 |
| **Ion implantation** | **NV** | (1.000)<br>1.0000 | (1.000)<br>0.9986 | (0.1191)<br>0.1568 | (0.0608)<br>4.7993 | (0.2193)<br>0.2370 | (0.1159)<br>0.4470 |
| | **SiV** | (1.000)<br>0.9999 | (1.000)<br>0.9995 | (0.2383)<br>0.5929 | (0.0609)<br>3.8944 | (0.3204)<br>0.4766 | (0.1414)<br>0.3659 |
| | **GeV** | (1.000)<br>1.000 | (1.000)<br>0.9998 | (0.0286)<br>0.3486 | (0.0808)<br>14.3641 | (0.0692)<br>0.4226 | (0.2149)<br>1.3591 |
| | **SnV** | (1.000)<br>1.000 | (1.000)<br>0.9989 | (0.1933)<br>0.3300 | (0.0956)<br>12.5748 | (0.2727)<br>0.3310 | (0.1746)<br>0.8053 |
| **Electron/ion irradiation** | **NV** | (1.000)<br>0.9983 | (1.000)<br>0.9980 | (0.0000)<br>16.1969 | (0.0987)<br>19.1591 | (0.0000)<br>0.1100 | (0.1926)<br>1.0162 |
| | **SiV** | (1.000)<br>0.9999 | (0.9999)<br>0.9998 | (0.0000)<br>0.0266 | (0.0317)<br>0.0528 | (0.0000)<br>0.0533 | (0.1124)<br>0.1503 |
| | **GeV** | (1.000)<br>0.9925 | (0.9688)<br>0.9524 | (0.0000)<br>0.0105 | (0.0437)<br>0.0667 | (0.0000)<br>0.1025 | (0.1701)<br>0.1917 |

**Table ST7.** List of the statistical indicators, $R^2$, MSE and MAE for *Task I*, for the different material synthesis techniques (HPHT, CVD/MPCVD, Ion implantation and Electron/ion irradiation), organized by color centers (NV, SiV, GeV and SnV) and by method (DTR and XGB). Values for the test dataset are shown in red, while values for the training dataset are shown in black and within parentheses. The values for the training dataset are merely used for control.

**Table ST8.** *Task II*: DTR/XGB performance for each synthesis method

| | | $R^2$ | | MSE | | MAE | |
|---|---|---|---|---|---|---|---|
| | **DVCC** | **DTR** | **XGB** | **DTR** | **XGB** | **DTR** | **XGB** |
| **HPHT** | **NV** | (0.9966)<br>0.9096 | (1.000)<br>0.9726 | (0.0002)<br>0.0061 | (0.0000)<br>0.0018 | (0.0077)<br>0.0580 | (0.0002)<br>0.0391 |
| | **SiV** | (0.9985)<br>0.9928 | (1.000)<br>0.9568 | (0.0028)<br>0.0140 | (0.0000)<br>0.0844 | (0.0362)<br>0.0836 | (0.0005)<br>0.0987 |
| | **GeV** | (0.9999)<br>0.9995 | (1.000)<br>0.9098 | (0. 0003)<br>0.0016 | (0.0000)<br>0.3188 | (0.0089)<br>0.0314 | (0.0004)<br>0.1855 |
| | **SnV** | (1.000)<br>0.9979 | (1.000)<br>0.9476 | (0.0001)<br>0.0067 | (0.0000)<br>0.1706 | (0.0031)<br>0.0582 | (0.0005)<br>0.2095 |
| **CVD/MPCVD** | **NV** | (1.000)<br>0.9930 | (1.000)<br>0.9259 | (0.0000)<br>0.0095 | (0.0000)<br>0.1006 | (0.0000)<br>0.0923 | (0.0005)<br>0.1845 |
| | **SiV** | (1.000)<br>0.8765 | (1.000)<br>0.8414 | (0.0000)<br>0.2015 | (0.0000)<br>0.2589 | (0.0000)<br>0.0868 | (0.0005)<br>0.2411 |
| | **GeV** | (1.000)<br>0.9928 | (1.000)<br>0.9534 | (0.0000)<br>0.0075 | (0.0000)<br>0.0487 | (0.0000)<br>0.0318 | (0.0006)<br>0.0954 |
| | **SnV** | (1.000)<br>0.9984 | (1.000)<br>0.8529 | (0.0000)<br>0.0012 | (0.0000)<br>0.1123 | (0.0000)<br>0.0316 | (0.0002)<br>0.0994 |
| **Ion implantation** | **NV** | (1.000)<br>0.9378 | (1.000)<br>0.9492 | (0.000)<br>0.1575 | (0.0000)<br>0.1285 | (0.0000)<br>0.0900 | (0.0016)<br>0.1193 |
| | **SiV** | (1.000)<br>0.9119 | (1.000)<br>0.8388 | (0.0000)<br>0.1350 | (0.0000)<br>0.2471 | (0.0000)<br>0.0784 | (0.0025)<br>0.2079 |
| | **GeV** | (1.000)<br>0.8963 | (1.000)<br>0.9040 | (0.0000)<br>0.4000 | (0.0000)<br>0.3706 | (0.0000)<br>0.1399 | (0.0019)<br>0.1752 |
| | **SnV** | (1.000)<br>0.9527 | (1.000)<br>0.8558 | (0.0000)<br>0.2327 | (0.0000)<br>0.7092 | (0.0000)<br>0.2332 | (0.0015)<br>0.4286 |
| **Electron/ion irradiation** | **NV** | (0.9798)<br>0.9628 | (0.9999)<br>0.9042 | (0.0433)<br>0.0795 | (0.0003)<br>0.2046 | (0.1167)<br>0.1782 | (0.0072)<br>0.2155 |
| | **SiV** | (0.9966)<br>0.9963 | (1.000)<br>0.9100 | (0.0029)<br>0.0032 | (0.0000)<br>0.0770 | (0.0358)<br>0.0402 | (0.0033)<br>0.0626 |
| | **GeV** | (0.9996)<br>0.9950 | (0.9982)<br>0.9902 | (0.0000)<br>0.0003 | (0.0001)<br>0.0007 | (0.0031)<br>0.0177 | (0.0035)<br>0.0216 |

**Table ST8.** List of the statistical indicators, $R^2$, MSE and MAE for *Task II* for the different material synthesis techniques (HPHT, CVD/MPCVD, Ion implantation and Electron/ion irradiation), organized by color centers (NV, SiV, GeV and SnV) and by method (DTR and XGB). Values for the test dataset are shown in red, while values for the training dataset are shown in black and within parentheses. The values for the training dataset are merely used for control.

## S6. Debye-Waller factor analysis

To further draw direct comparisons between synthesis methods, we applied the DTR and XGB algorithms to predict the value of the Debye-Waller factor (DWF) of the DVCCs. Similarly to the FWHM of the various DVCC's ZPL, the DWF is a *measurable,* physical observable that is *common* to all synthesis techniques, and is *physically meaningful*, making it a good indicator for the diamond's and the DVCC's quality (see Main text, § 1). The DWF quantifies the fraction of photons emitted by an emitter into the zero-phonon line (ZPL) versus those emitted into phonon sidebands (PSB), in an excited vibrational state. It is defined as the integrated luminescence intensity of the ZPL, $I_{\mathrm{ZPL}}$, divided by the total integrated luminescence intensity of the color center $I_{\mathrm{tot}}$[1,2] A high value of the DWF is usually desirable for optical and quantum applications based on diamond color centers as it improves photon indistinguishability, photon extraction, and the coupling efficiency to optical cavities and photonics structures.[3,4,5]

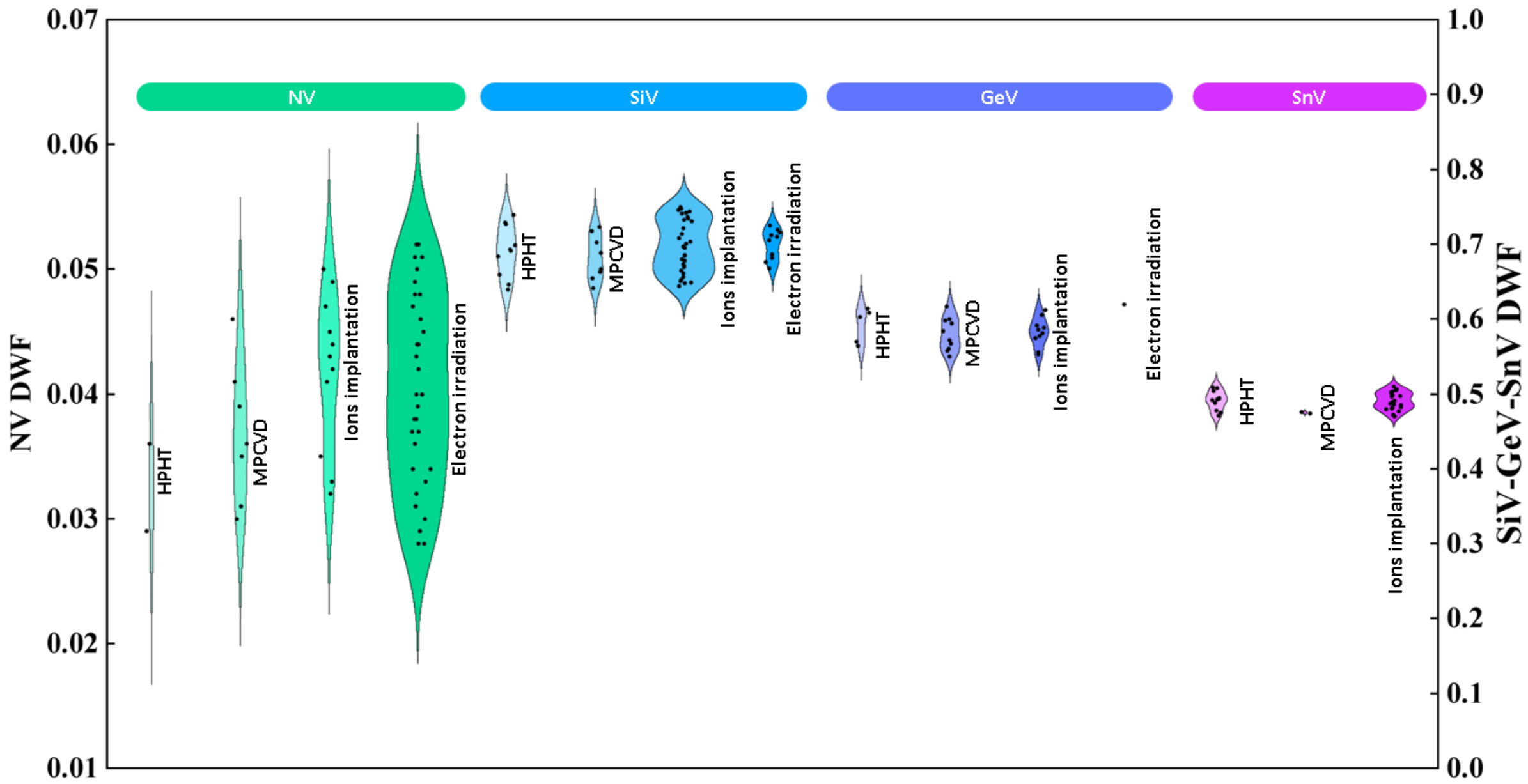

**Figure SF6**. Debye-Waller factor (DWF) for NV, SiV, GeV and SnV, organized by synthesis method. The DWF values were calculated from the photoluminescence spectra available in the published experimental data. Note that the scales are different for the NV center (left $y$-axis) and for the SiV, GeV and SnV centers (right $y$-axis). Each data point (black dot) is a measured DWF value, and the vertical spread indicates the distribution of DWF values found across the literature; the horizontal width of each plot indicates how many times, across studies, the same DWF value was found.

To evaluate the performance of our DTR and XGB algorithms, we use the four databases (§ S2, Tables ST2–ST5) and perform a 70-30 split of the data, where 70% of the entries, at random, are used for training the DTR/XGB predictors and the remaining 30% to test them. For all the synthesis techniques, the selected target value is the DWF, whereas the input features differ for each method analogously to the analysis carried out for *Task II* (see § S1, Table ST1). The performance of the models is based on the difference between the true experimental values and those produced by the DTR/XGB predictors and it is estimated from the quantities $R^2$, MSE and MAE. We remark that the DWF is rarely explicitly reported in any of the experimental studies we reviewed. To determine the DWF values, we extracted the spectrum from each publication and calculated the DWF from the available data. Figure SF6 shows a *violin* chart summarizing the experimental values of the DWF from the literature, organized by DVCC and synthesis method. The vertical and horizontal spreads of each plot indicate the dispersion of the measured DWF values and their observation frequency, respectively.

The results of the training and testing of the DTR and XGB algorithms are displayed in Figures SF7–10 for the HPHT, CVD/MPCVD, Ion implantation and Electron/ion irradiation techniques, respectively. The estimated values for $R^2$, MSE and MAE are displayed in the bottom-right corner of each subpanel both for the test and training data (the latter are merely used as control); they are also summarized in Table ST9.

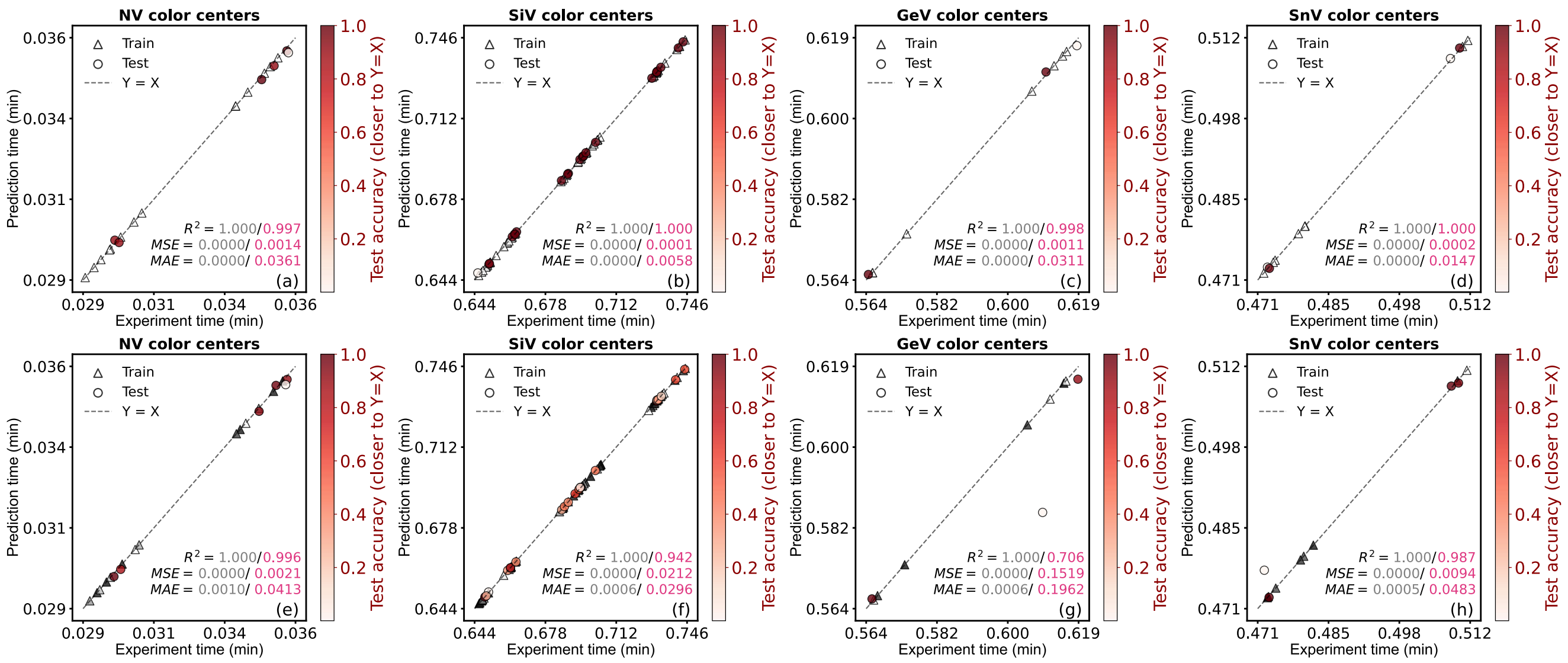

**Figure SF7**. HPHT, DWF summary. DTR (top row) and XGB (bottom row) predictions of the DWF (target variable) from given values for pressure, temperature, run time and average size of the diamond particles (input features), organized by color centers (columns). The values of $R^2$, MSE, and MAE for the training (triangles) and test (circles) sets are listed in the bottom-right corner of each subpanel.

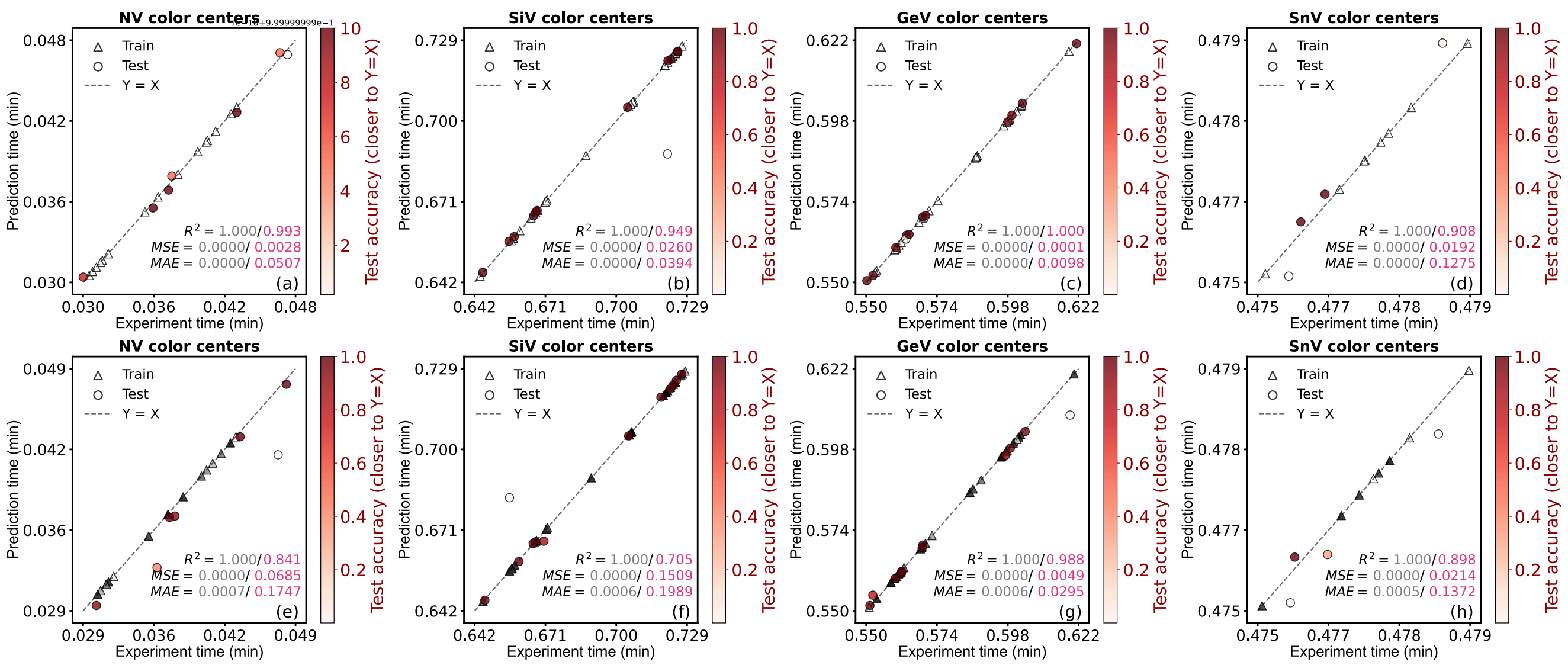

**Figure SF8**. CVD/MPCVD, DWF summary. DTR (top row) and XGB (bottom row) predictions of the DWF (target variable) from given values for chamber pressure, substrate temperature and MW plasma power (input features), organized by color centers (columns). The values of $R^2$, MSE, and MAE for the training (triangles) and test (circles) sets are listed in the bottom-right corner of each subpanel.

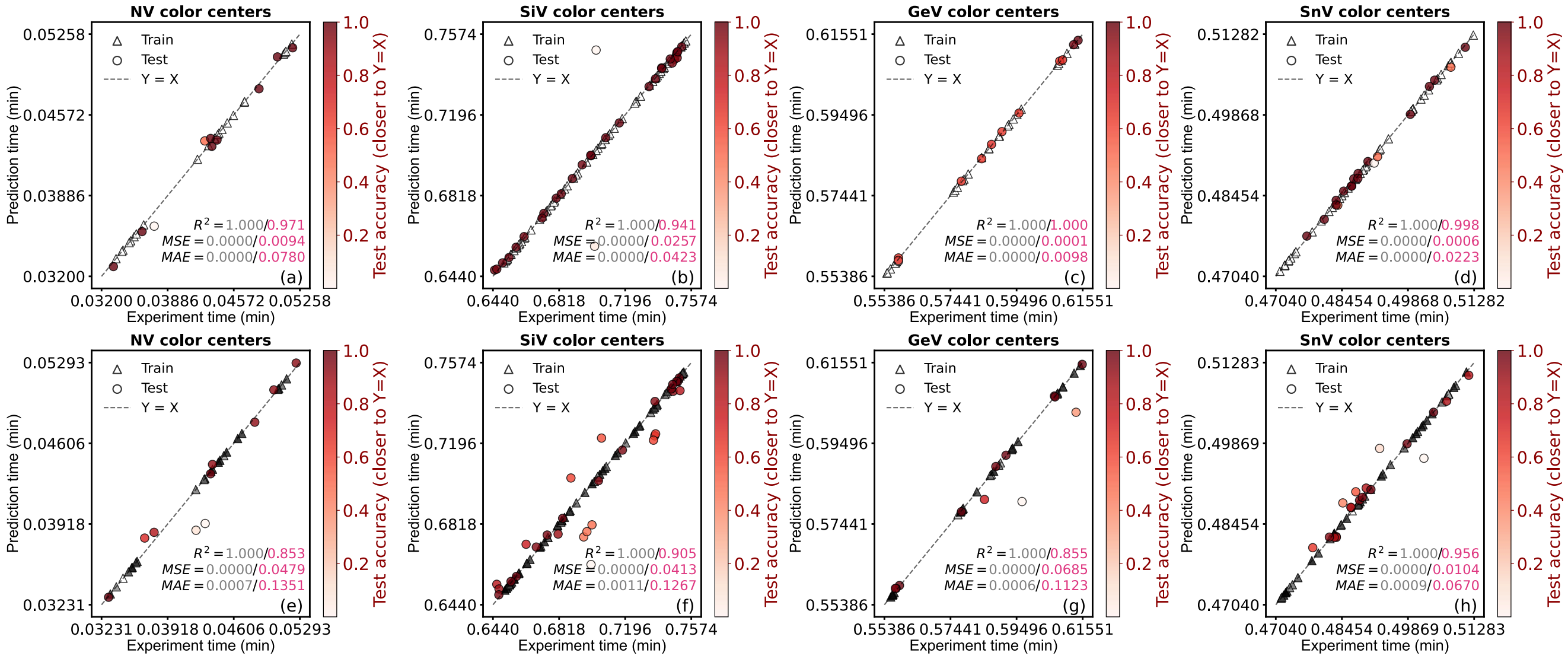

**Figure SF9**. Ion implantation, DWF summary. DTR (top row) and XGB (bottom row) predictions of the DWF (target variable) from given values for implantation energy, fluence, annealing temperature and annealing time (input features), organized by color centers (columns). The values of $R^2$, MSE, and MAE for the training (triangles) and test (circles) sets are listed in the bottom-right corner of each subpanel.

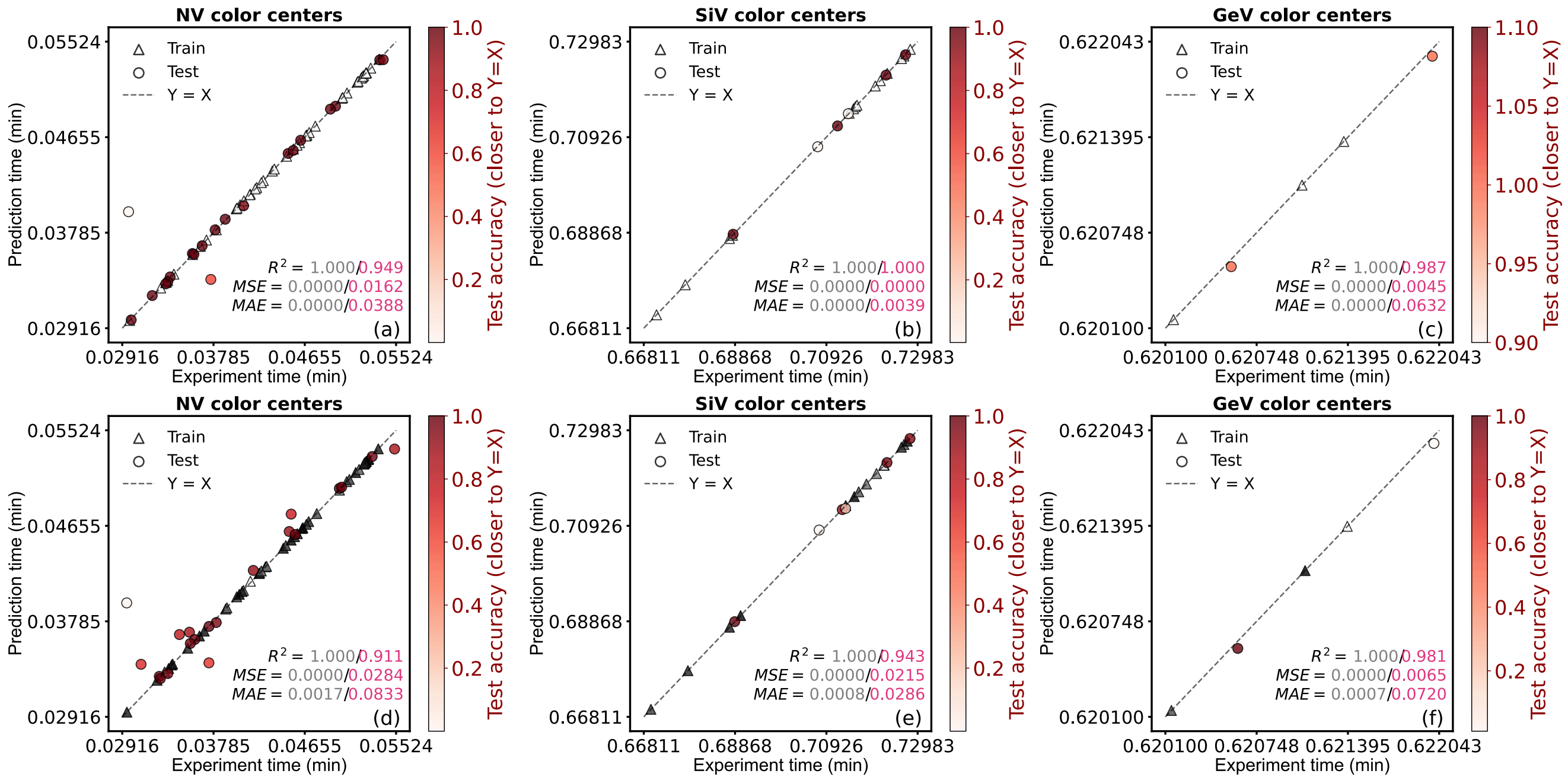

**Figure SF10**. Electron/ion irradiation, DWF summary. DTR (top row) and XGB (bottom row) predictions of the DWF (target variable) from given values for irradiation energy, fluence, annealing temperature and annealing time (input features), organized by color centers (columns). The values of $R^2$, MSE, and MAE for the training (triangles) and test (circles) sets are listed in the bottom-right corner of each subpanel.

The DTR/XGB models can predict the DWF value relatively well: the estimates of $R^2$ are consistently >0.9 and often close to ~1, with just a few exceptions for the XGB predictor where $R^2 \leq 0.85$. The value of MSE and MAE are also desirably small (≲10% of the average target value). Similar to the case of the ZPL FWHM, larger deviations are likely due to not including categorical features in our models (see Main text, § 3.2).

**Table ST9.** *Debye-Waller Factor* prediction: DTR/XGB performance for each synthesis method

| | | $R^2$ | | MSE | | MAE | |
|---|---|---|---|---|---|---|---|
| | **DVCC** | **DTR** | **XGB** | **DTR** | **XGB** | **DTR** | **XGB** |
| **HPHT** | **NV** | (1.000) 0.997 | (1.000) 0.996 | (0.0000) 0.0014 | (0.0000) 0.0021 | (0.0000) 0.0361 | (0.0010) 0.0413 |
| | **SiV** | (1.000) 1.000 | (1.000) 0.942 | (0.0000) 0.0001 | (0.0000) 0.0212 | (0.0000) 0.0058 | (0.0006) 0.0296 |
| | **GeV** | (1.000) 0.998 | (1.000) 0.706 | (0.0000) 0.0011 | (0.0000) 0.1519 | (0.0000) 0.0311 | (0.0006) 0.1962 |
| | **SnV** | (1.000) 1.000 | (1.000) 0.987 | (0.0000) 0.0002 | (0.0000) 0.0094 | (0.0000) 0.0147 | (0.0005) 0.0483 |
| **CVD/MPCVD** | **NV** | (1.000) 0.993 | (1.000) 0.841 | (0.0000) 0.0028 | (0.0000) 0.0685 | (0.0000) 0.0507 | (0.0007) 0.1747 |
| | **SiV** | (1.000) 0.949 | (1.000) 0.705 | (0.0000) 0.0260 | (0.0000) 0.1509 | (0.0000) 0.0394 | (0.0006) 0.1989 |
| | **GeV** | (1.000) 1.000 | (1.000) 0.988 | (0.0000) 0.0001 | (0.0000) 0.0049 | (0.0000) 0.0098 | (0.0006) 0.0295 |
| | **SnV** | (1.000) 0.908 | (1.000) 0.898 | (0.0000) 0.0192 | (0.0000) 0.0214 | (0.0000) 0.1275 | (0.0005) 0.1372 |
| **Ion implantation** | **NV** | (1.000) 0.971 | (1.000) 0.853 | (0.0000) 0.0094 | (0.0000) 0.0479 | (0.0000) 0.0780 | (0.0007) 0.1351 |
| | **SiV** | (1.000) 0.941 | (1.000) 0.905 | (0.0000) 0.0257 | (0.0000) 0.0413 | (0.0000) 0.0423 | (0.0011) 0.1267 |
| | **GeV** | (1.000) 1.000 | (1.000) 0.855 | (0.0000) 0.0001 | (0.0000) 0.0658 | (0.0000) 0.0098 | (0.0006) 0.1123 |
| | **SnV** | (1.000) 0.998 | (1.000) 0.956 | (0.0000) 0.0006 | (0.0000) 0.0104 | (0.0000) 0.0223 | (0.0009) 0.0670 |
| **Electron/ion irradiation** | **NV** | (1.000) 0.949 | (1.000) 0.911 | (0.0000) 0.0162 | (0.0000) 0.0284 | (0.0000) 0.0388 | (0.0017) 0.0833 |
| | **SiV** | (1.000) 1.000 | (1.000) 0.943 | (0.0000) 0.0000 | (0.0000) 0.0215 | (0.0000) 0.0039 | (0.0008) 0.0286 |
| | **GeV** | (1.000) 0.987 | (1.000) 0.981 | (0.0000) 0.0045 | (0.0000) 0.0065 | (0.0000) 0.0632 | (0.0007) 0.0720 |

**Table ST9.** List of the statistical indicators, $R^2$, MSE and MAE of the DWF for the different material synthesis techniques (HPHT, CVD/MPCVD, Ion implantation and Electron/ion irradiation), organized by color centers (NV, SiV, GeV and SnV) and by method (DTR and XGB). Values for the test dataset are shown in red, while values for the training dataset are shown in black and within parentheses. The values for the training dataset are merely used for control.

As a final remark, one of the merits of our meta-analysis is that it reveals that for the same color center, different techniques yield different ranges for DWF values. These are listed explicitly and in full in Table ST10 and show that ion implantation and electron/ion irradiation tend to yield DVCCs with higher DWF values—which is usually desirable for optical, photonic and quantum applications.

**Table ST10.** Measured DWF values for each technique

| | HPHT | CVD/MPCVD | Ion implantation | Electron/ion irradiation |
|---|---|---|---|---|
| **NV** | 0.029 – 0.036 | 0.030 – 0.048 | 0.032 – 0.049 | 0.028 – 0.054 |
| **SiV** | 0.640 – 0.741 | 0.642 – 0.725 | 0.644 – 0.752 | 0.668 – 0.727 |
| **GeV** | 0.564 – 0.616 | 0.550 – 0.619 | 0.553 – 0.613 | 0.620 – 0.622 |
| **SnV** | 0.471 – 0.510 | 0.474 – 0.479 | 0.470 – 0.512 | – |

**Table ST10.** Ranges of measured DWF values organized by synthesis techniques, HPHT, CVD/MPCVD, Ion implantation and Electron/ion irradiation (columns) and organized by color centers, NV, SiV, GeV and SnV (rows). In red we highlight, for each color center, the techniques that yield the largest DWF value, which is usually desirable for optical, photonic and quantum applications.

## S7. Worldwide trends in diamond synthesis

The database we compiled is rich in information that goes beyond process parameters and fabrication protocols. We can leverage this information to produce Sankey diagrams and visualize general trends in diamond synthesis over the last decade, as inferred from the studies we reviewed in our meta-analysis. The diagram (Fig. SF11) has three main nodes: geographical region, synthesis method and year. Specific trends of interest can be deduced directly from the analysis of the diagram; below we highlight a few main notable points.

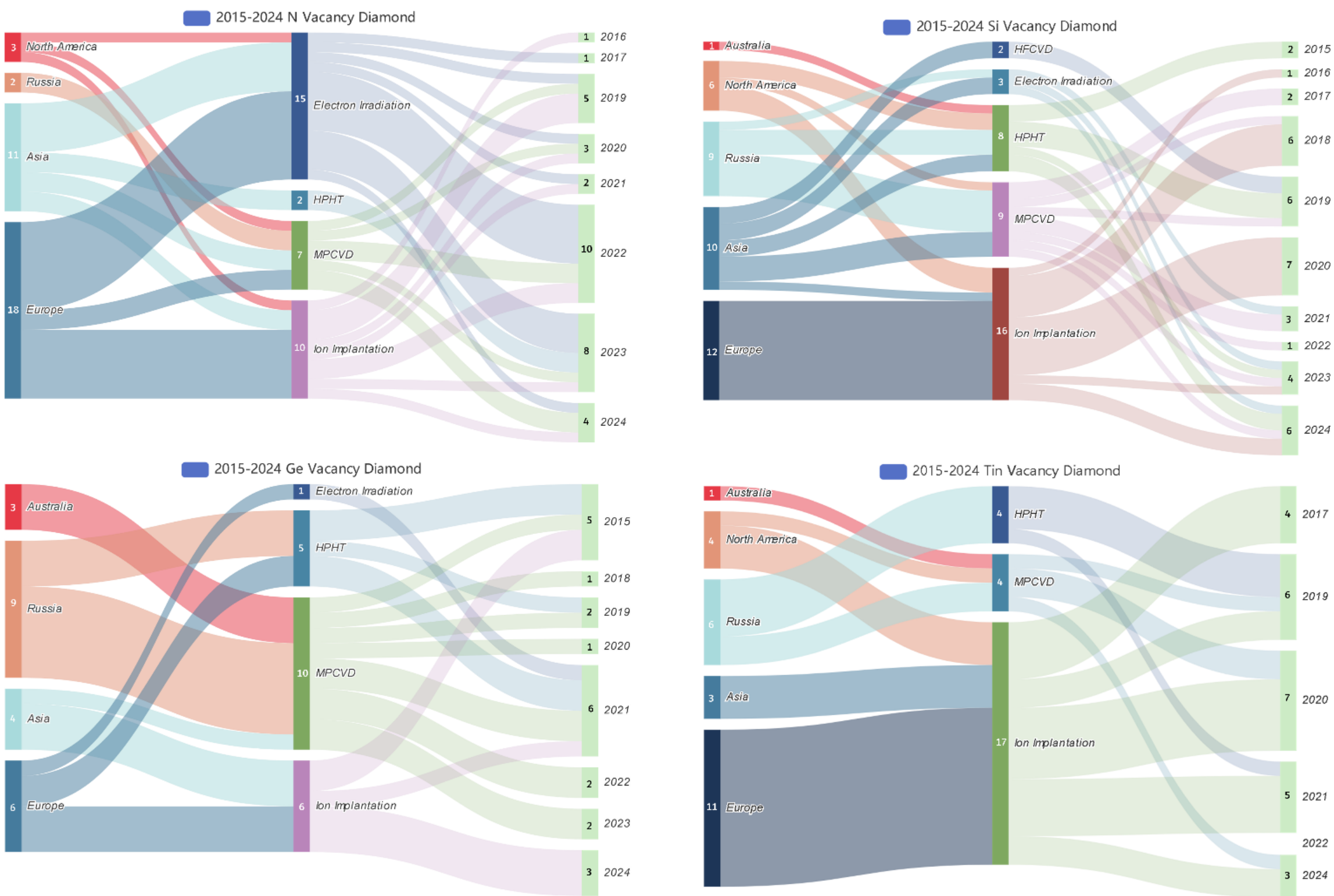

**Figure SF11.** Sankey diagrams for the synthesis of diamond materials containing NV, SiV, GeV, and SnV color centers, organized by geographical region, synthesis method, and year. The diagrams were extracted from a subset of studies surveyed for our meta-analysis, focusing on data from the last decade.

Geographically, most of the studies we gathered were conducted by groups in Europe (47/119, 39%), Asia (28/119, 23%) and Russia (26/119, 22%). Regarding diamond synthesis and DVCC inclusion, ion implantation and electron/ion irradiation are, across the board, widely used for color centers incorporation in diamond, with many of the studies surveyed coming out of Europe (37/68, 54%) and Asia (18/68, 26%). The number of studies we collated was split roughly evenly with respect to the synthesis method, with a slight edge for CVD/MPCVD (30/49, 61%) over HPHT (19/49, 39%). Electron irradiation is more commonly used for incorporation of NV color centers (15/25, 60%), whereas other color centers are mostly created through ion implantation (23/27, 85%). This is expected because nitrogen is abundant in diamonds, and electron irradiation is usually sufficient to create vacancies and—upon annealing—stable NV centers. Conversely, for SiV, GeV, and SnV, the incorporation of the relevant dopant via implantation is either needed or desirable, especially in studies where it is performed deterministically (e.g., to leverage emission from single color centers). We note that MPCVD is also commonly used for incorporation of color centers in diamond directly during growth, especially for GeV (10/22, 45%) and SiV (9/38, 24%). For SnV the preferred incorporation method is ion implantation (17/25, 68%).

## Supplementary References